\documentclass[twocolumn,noshowpacs,prl,superscriptaddress]{revtex4}
\usepackage{graphicx}
\usepackage{amsmath}
\usepackage{amssymb}

\newcommand{\tr}{\text{Tr}} 
\newcommand{\A}{\mathcal{A}} 
\newcommand{\W}{\mathcal{W}} 
\newcommand{\w}{\wedge}
\newcommand{\ket}[1]{\left| #1 \right>} 
\newcommand{\bra}[1]{\left< #1 \right|} 
\newcommand{\braket}[2]{\left< #1 \vphantom{#2} \right| \left. #2 \vphantom{#1} \right>} 

\begin{document}

\title{Quantized Electric Multipole Insulators} 

\author{Wladimir A. Benalcazar}
\affiliation{Department of Physics and Institute for Condensed Matter Theory, University of Illinois at Urbana-Champaign, IL 61801, USA}
\author{B. Andrei Bernevig}
\affiliation{Department of Physics, Princeton University, Princeton, New Jersey 08544, USA}
\author{Taylor L. Hughes}
\affiliation{Department of Physics and Institute for Condensed Matter Theory, University of Illinois at Urbana-Champaign, IL 61801, USA}

\begin{abstract}
In this article we extend the celebrated Berry-phase formulation of electric polarization in crystals to higher electric multipole moments. We determine the necessary conditions under which, and minimal models in which, the quadrupole and octupole moments are topologically quantized electromagnetic observables. Such systems exhibit \emph{gapped} boundaries that are themselves lower-dimensional topological phases. Furthermore, they manifest topologically protected corner states carrying fractional charge, i.e., fractionalization at the boundary of the boundary. To characterize these new insulating phases of matter, we introduce a new paradigm whereby `nested' Wilson loops give rise to a large number of new topological invariants that have been previously overlooked. We propose three realistic experimental implementations of this new topological behavior that can be immediately tested. 
\end{abstract}

\maketitle 

One of the most exciting outcomes of the discovery of topological phases of matter is the observation of sharply quantized physical phenomena. Several important examples are known: (i) the quantization of charge polarization $P_1$ in crystals with either charge conjugation or inversion symmetry~\cite{zak1989berry,hughes2011inversion,turner2012}, (ii) the quantization of the Hall conductance $\sigma_{xy}$ in 2D integer quantum Hall or Chern insulator systems~\cite{klitzing1980,thouless1982,haldane1988}, and (iii) the quantization of the magneto-electric polarizability $P_3$ in 3D time-reversal invariant or inversion symmetric topological insulators~\cite{qi2008,essin2009,hughes2011inversion,turner2012}. These response coefficients can be elegantly expressed as quantized topological invariants in terms of the Berry phase vector potential $\A$:
\begin{align}
P_1&=-\frac{e}{2\pi}\int_{BZ} \tr \left[\A\right]
\label{eq:polarization}\\
\sigma_{xy}&=-\frac{e^2}{2\pi h}\int_{BZ} \tr\left[d \A+i \A\w \A\right]\nonumber\\
P_3&=-\frac{e^2}{4\pi h}\int_{BZ} \tr\left[\A \w d\A + \frac{2i}{3} \A \w \A \w \A \right],
\nonumber
\end{align}
where $BZ$ is the Brillouin zone in one, two, and three spatial dimensions respectively, and $\A$ has components $[\A_{i}({\bf k})]^{mn}=-i\langle u^{m}_{\bf k}\vert\partial_{k_i}\vert u^{n}_{\bf k}\rangle$, where $\ket{u^{n}_{\bf k}}$ is the Bloch function of band $n$, and $m, n$ run only over occupied energy bands. 

We recognize that the forms of $\sigma_{xy}$ and $P_3$ are natural \emph{mathematical} generalizations of $P_1$ to higher dimensions. However, there is no theory for the natural \emph{physical} generalizations of the dipole polarization $P_1$ to higher electric multipole moments in crystalline insulators.  
In this article we provide a generalization of the polarization \eqref{eq:polarization} to higher multipole moments.  These higher moments are also of topological origin, and are hence, quantized by certain crystalline symmetries. We explain this quantization by providing a new paradigm for topological invariants based on nested Wilson loops that opens an unexplored direction in the analysis of topological phases of matter. Our bulk topological invariants determine the properties of \emph{gapped}, but possibly non-trivial \emph{boundaries}, and allow us to use robust topological phenomena to provide a definition of a ``boundary of a boundary." The physical manifestation of our multipole moments is two-fold: (1) it manifests fractional corner charges (but not edge or surface) which are quantized to $\pm e/2$ in the presence of certain symmetries, and (2) it generates surface-localized electric currents as the result of adiabatic deformations that preserve inversion symmetry.
In addition to our theory, we propose physical realizations in cold-atom lattice systems and in photonic crystals that can be demonstrated in current experiments.

\begin{figure}[t]
\centering
\includegraphics[width=\columnwidth]{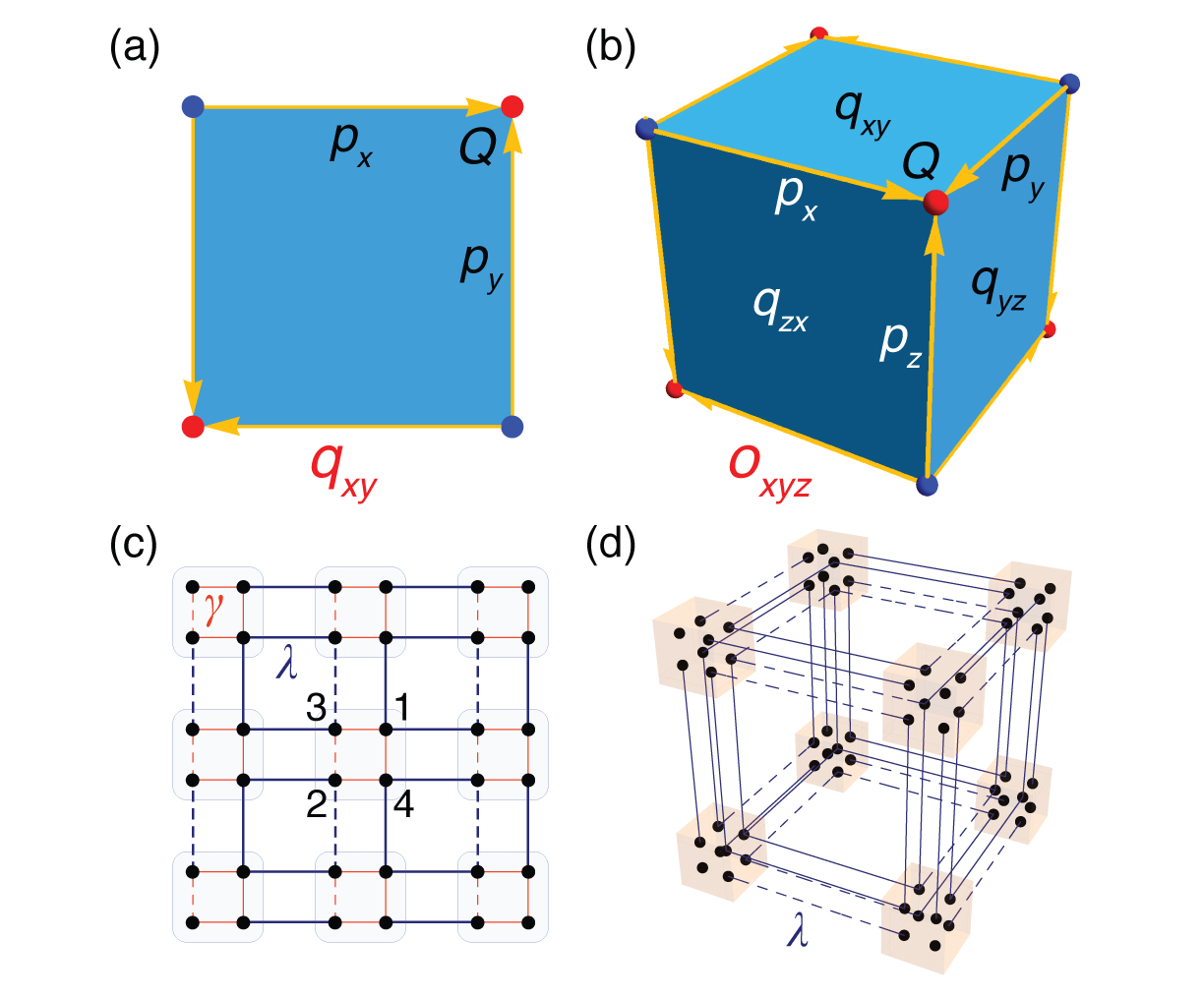}
\caption{\emph{Quadrupole and octupole moments in the classical continuum (a, b) and quantum mechanical crystalline (c, d) theories}. (a) Bulk quadrupole moment $q_{xy}$ and its boundary polarizations $p_i$ and corner charges $Q$. (b) Bulk octupole moment $o_{xyz}$ and its surface quadrupoles $q_{ij}$, hinge polarizations $p_{i}$, and corner charges $Q$. (c) Tight binding model with quantized quadrupole moment $q_{xy}.$ $\gamma,\lambda$ represent two hopping strengths, dashed lines represent hopping terms with negative signs. The topological phase occurs when $|\gamma/\lambda |<1.$ Numbers indicate basis for $\Gamma$-matrices (see Eq.~\ref{eq:quad_hamiltonian}). (d) Tight binding model with quantized octupole moment $o_{xyz}$. Each square plaquette in each direction contains $\pi$-flux, i.e., one of the bonds on each square has a negative sign (see Eq.~\ref{eq:octupole_hamiltonian}). In (d) $\gamma=0$ for easy of presentation.}
\label{fig:classical_multipoles}
\end{figure}

Classically, the \textit{primitive}  dipole, quadrupole, and octupole moments of a continuum, volume charge density $\rho({\bf r})$ are defined as~\cite{raabbook}: $p_i = \int d^3 {\bf r} \rho({\bf r}) r_i$,  $q_{ij} = \int d^3 {\bf r} \rho({\bf r}) r_i r_j$, and $o_{ijk} = \int d^3 {\bf r} \rho({\bf r}) r_i r_j r_k.$ 
In the modern theory of polarization in crystals, the dipole moment $p_i$ is a Berry phase of the bulk electron states  (see Eq.~\ref{eq:polarization}), and manifests itself through the presence of surface charge. More generally, higher moments generate a cascade of observable properties, for which the current theory of  polarization is not sufficient. 
For example, a 2D (3D) insulator with a square (cubic) shape and only a non-vanishing quadrupole moment $q_{xy}$ (octupole moment $o_{xyz}$) has the electromagnetic properties (see Sec. 1 of the Supplementary Material (SM) for a derivation and general formulation): 
\begin{align}
p_{j}^{edge\; \alpha}&=n^{\alpha}_{i}q_{ij},\;\;\; Q^{corner \; \alpha, \beta}=n^{\alpha}_{i}n^{\beta}_{j}q_{ij};\label{eq:qdensities}\\
q_{ij}^{face\; \alpha}&=n^{\alpha}_{i} o_{ijk},\;\;\; p_{k}^{hinge \; \alpha,\beta}=n^{\alpha}_{i}n^{\beta}_{j}o_{ij k},\nonumber\\ Q^{corner\; \alpha,\beta,\gamma}&=n^{\alpha}_{i} n^{\beta}_{j} n^{\gamma}_{k} o_{ijk},\label{eq:odensities}
\end{align} 
where repeated indices are summed over. Here $p_j^{edge\; \alpha} (p_{j}^{hinge\; \alpha,\beta}$)  are the edge (hinge) tangential polarization per unit length on the square (cube), $q_{ij}^{face\; \alpha}$ are the surface quadrupoles per unit area of the cube, and $Q^{corner\; \alpha,\beta,\gamma}$ are the charges localized on the corners. The Greek letters $\alpha, \beta, \gamma= \pm x, \pm y, \pm z$ label the surfaces of the square/cube with outward pointing unit normal vectors $n_{i}^{\alpha}=s_{\alpha}\delta^{|\alpha|}_{i},$ where the sign $s_{\alpha=\pm}=\pm 1$  encodes the direction. Hinges and corners are associated to the intersection of multiple surfaces and hence, have more than one associated surface label and normal vector; the only non-vanishing contributions occur for $\alpha\neq\beta\neq \gamma.$ 
These properties are represented pictorially in Fig.~\ref{fig:classical_multipoles}a,b. 

To illustrate the bulk nature of the quadrupole moment we draw a distinction between systems with edge polarizations induced by a bulk quadrupole moment $q_{xy}$, and those with boundary polarizations in the absence of a bulk quadrupole. For the second case, consider an unpolarized 2D insulator with $q_{xy}=0$ which has tangent \emph{edge} polarizations per unit length following the same pattern as in Fig.~\ref{fig:classical_multipoles}a (but where in general $|p^{edge\; +x}_y| \neq |p^{edge\; +y}_x|$). In this configuration $Q^{corner\; +x,+y}=p^{edge\; +x}_y+p^{edge\; +y}_x$ (notice that all of these quantities have units of charge), which is just the induced charge from the two independent edge polarizations converging to the corner or diverging from it. Now, suppose that instead we have an unpolarized system with $q_{xy}\neq 0$. According to Eq.~\ref{eq:qdensities},  this moment induces tangent edge polarizations and corner charges of \emph{equal} magnitude, $|p^{edge\; \pm y}_{x}|=|p^{edge\; \pm x}_{y}| = |Q^{corner\; \pm x,\pm y}|= |q_{xy}|$, and with the pattern shown in Fig.~\ref{fig:classical_multipoles}a. 
Crucially, the corner charge \emph{is not} the sum of the two boundary polarizations, i.e., it is not the result of edge phenomena alone. Instead, both edge polarizations and corner charge are boundary manifestations of the bulk. In a system without bulk polarization but with both quadrupole moment and (bulk-independent) edge polarizations, the quadrupole moment determines exactly how much the corner charge differs from the sum of the tangent edge polarizations according to $Q^{corner\; +x,+y}-P^{edge\; +x}_{y}-P^{edge\; +y}_{x}=-q_{xy},$, where the upper-case $P^{edge\; \alpha}$ is the total edge polarization on edge $\alpha$ regardless of its physical origin. 

We will now show that the quadrupole and octupole moments can be defined in crystals, and are quantized by the presence of reflection symmetries.\\

\paragraph*{Model with Quantized Quadrupole Moment}
Heuristically, a 2D crystal with a quantized quadrupole moment should have two properties: (i) at least two occupied bands, and (ii) symmetries that quantize the bulk quadrupole moment and bulk dipole moment--the latter made to vanish. We expect (i) since canonical quadrupoles are built from two canceling dipole moments, and two occupied bands are needed to generate each non-vanishing individual dipole, but a total canceling dipole (see SM, Sec. 1). Regarding (ii), moments that are odd under a symmetry must naively vanish, but, in following with the usual quantization paradigm of $P_1$~\cite{zak1989berry,hughes2011inversion,turner2012} and $P_3$~\cite{qi2008,essin2009,hughes2011inversion,turner2012}, these moments can more generally satisfy, e.g., $P_1=-P_1\mod e$ (in 1D), which has a non-vanishing solution $P_1=\pm e/2$ when a lattice is introduced~\cite{King-SmithVanderbilt1993,vanderbilt1993}. For (ii) we will enforce reflection symmetries $M_x:x\to -x$ and $M_y: y\to -y.$

\begin{figure}[t]
\centering
\includegraphics[width=\columnwidth]{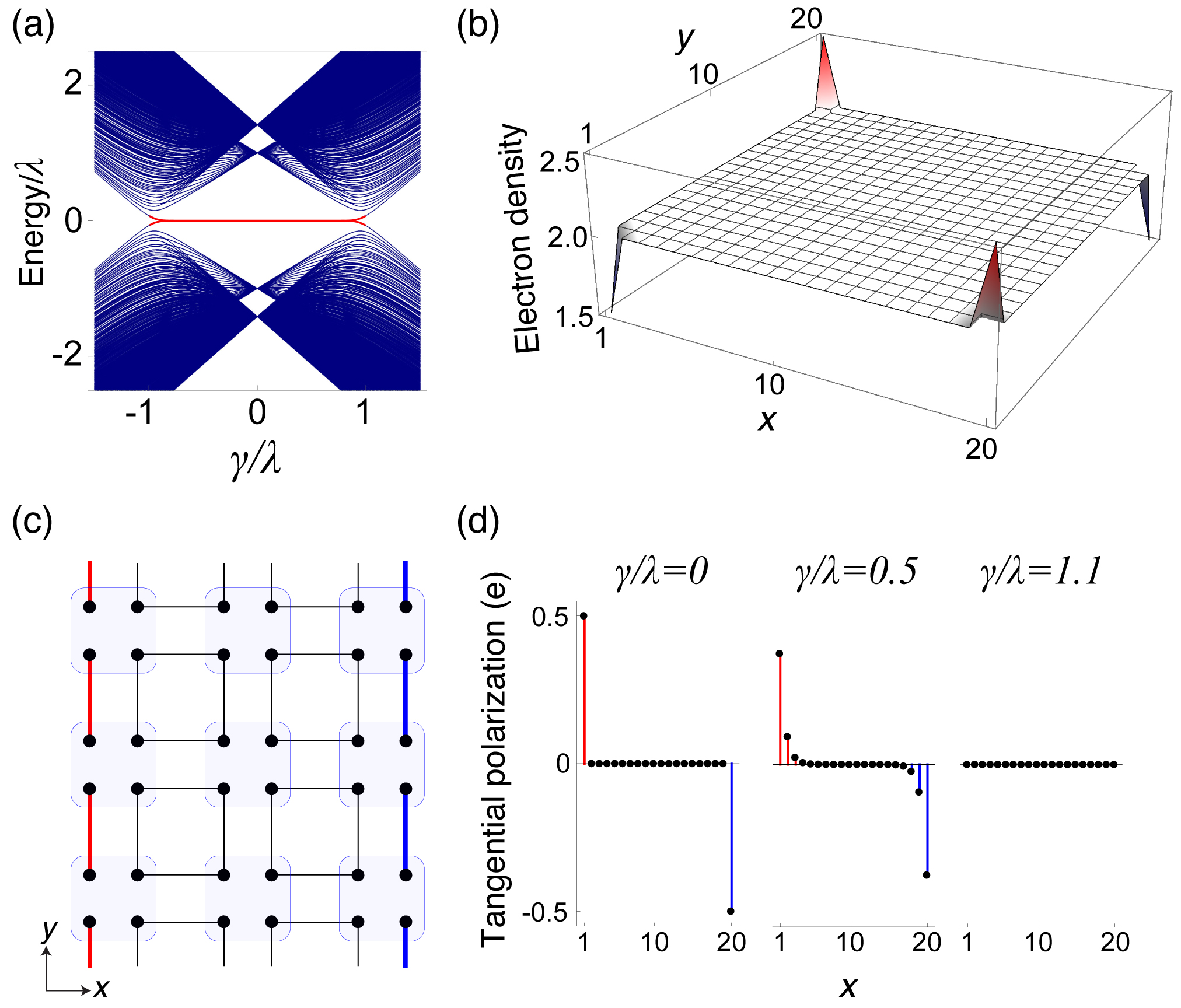}%
\caption{\emph{Properties of quadrupole insulator with the Bloch Hamiltonian in Eq.~\ref{eq:quad_hamiltonian}}. (a) Energy spectrum  for a system with open boundary conditions in $x$ and $y$ as a function of $\gamma/\lambda$. Red lines correspond to four energy-degenerate, corner-localized states. (b) Electronic charge density in the non-trivial phase (here $\lambda=1$, $\delta=\gamma = 10^{-3}$ on a square lattice with 20 sites per side). (c) Lattice with open boundaries along $x$ and periodic boundaries along $y$ at $\gamma=0$. Its edges (with blue and red hoppings) are 1D non-trivial topological insulators. (d) Tangential polarization along $y$ as a function of lattice sites along $x$ for the geometric configuration in (c), but for three values of $\gamma/\lambda$. For $\gamma/\lambda=0.5$ the integrated polarization over half of the lattice width is still $\pm e/2.$}
\label{fig:quadrupole}
\end{figure}

As our first model, let us consider the 4-band insulator with Bloch Hamiltonian
\begin{align}
h^q({\bf k},\delta)&= \left(\gamma + \lambda  \cos(k_x)\right) \Gamma_4 + \lambda \sin(k_x) \Gamma_3 \nonumber\\
&+ \left(\gamma + \lambda \cos(k_y)\right) \Gamma_2 + \lambda \sin(k_y) \Gamma_1+\delta \Gamma_0,
\label{eq:quad_hamiltonian}
\end{align}
which can represent an insulator with a quadrupole moment. Here $\Gamma_0= \tau_3 \sigma_0, \Gamma_k=-\tau_2 \sigma_k, \Gamma_4=\tau_1 \sigma_0$ for $k=1,2,3$; $\tau,\sigma$ are Pauli matrices for the degrees of freedom within a unit cell, and we will choose lattice constants $a_{x,y}=1$ throughout.  This model, shown in Fig.~\ref{fig:classical_multipoles}c along with the basis of the $\Gamma$-matrices, represents spinless electrons on a square lattice, having $\pi$-flux per plaquette and with bond dimerization in the $x$ and $y$ directions. $h^{q}({\bf k},0)$ has energies $E=\pm \sqrt{2\lambda^2+2\gamma^2 + 2\lambda \gamma (\cos(k_x)+\cos(k_y))}$, each of which is two-fold degenerate. An energy gap exists unless $\gamma/\lambda= \pm 1$ (Fig.~\ref{fig:quadrupole}a). Hence, at half-filling, with two electrons per cell, it is an insulator. A phase transition occurs at $\gamma/\lambda=1$ ($\gamma/\lambda=-1$) with a bulk energy gap closing at the ${\bf M}=(\pi,\pi)$ [${\bf \Gamma}=(0,0)$] point of the BZ.  
When $\delta =0$, $h^q$ has reflection symmetries $\hat{m}_i h^q({\bf k}) \hat{m}_i^\dagger= h^q(M_i {\bf k})$ for $i=x,y$, where $\hat{m}_x= \tau_1 \sigma_3$, $\hat{m}_y= \tau_1 \sigma_1$ and $M_x (k_x, k_y) = (-k_x, k_y)$, $M_y (k_x, k_y) = (k_x, -k_y).$ These symmetries quantize both components of the polarization and, as we will see, the quadrupole moment $q_{xy}$. Being topological and symmetry-protected, this quantization is robust with respect to the spatial displacement of the internal degrees of freedom of the unit cell that preserve the reflection symmetries (see SM, Sec. 10). To obtain a non-trivial topological quadrupole, however, it is crucial that $\hat{m}_{x}$ and $\hat{m}_y$ do not commute (see SM, Sec. 8). For our model they satisfy $\{\hat{m}_x,\hat{m}_y\}=0$ because the reflection symmetry is only preserved up to a gauge transformation (see SM, Sec. 9).  Because of its simple form, this model also  preserves time-reversal (TR) and charge-conjugation (CC) symmetries, with $\Theta=K$ and $C=\Gamma_0 K$ the TR and CC operators, as well as $C_4$ symmetry $\hat{r}_4 h^q({\bf k}) \hat{r}_4^\dagger = h^q(R_4{\bf k})$ up to a gauge transformation, with
\begin{align}
\hat{r}_4=\left(
\begin{array}{cc}
0 & \sigma_0\\
-i \sigma_y & 0
\end{array} \right)\nonumber
\end{align}
and $R_4$ the rotation of the momentum by $\pi/2$. However, these symmetries are not necessary for our discussion and could be removed without affecting the conclusions. 
When $\delta \neq 0$, the model breaks charge conjugation, both reflections, and $C_4$ symmetries, but preserves inversion symmetry. This leaves the polarization quantized,  however, it does break the symmetries required to quantize the quadrupole moment. For most of our discussion we will only tune $\delta$ infinitesimally away from $0$ in order to fix the sign of $q_{xy}.$

In the gapped phases of $h^q,$ the polarization and Chern number are zero, and we confirm that generically there are no gapless edge states. However, in the phase $|\gamma/\lambda|<1$, we find that the insulator has a quantized \emph{boundary} polarization, tangent to the edge, of magnitude $\pm e/2,$ and quantized corner charges $\pm e/2$ (Fig.~\ref{fig:classical_multipoles}a and Figs.~\ref{fig:quadrupole}b,d). Together, these are a defining signature of a quantized bulk quadrupole moment $q_{xy}=\pm e/2$ (c.f., Eq.~\ref{eq:qdensities}). On the other hand, for $|\gamma/\lambda|>1$, both the corner charges and boundary polarizations vanish. Thus, we denote the phase $|\gamma/\lambda|<1$ ($|\gamma/\lambda|>1$) as the non-trivial (trivial) quadrupole phase. Associated with the corner charges observed in the non-trivial phase is the existence of four, corner-localized, energy-degenerate, mid-gap modes  when $\delta =0$ (red lines in Fig.~\ref{fig:quadrupole}a). At half filling, the negative energy bulk bands and \emph{two} of the four corner modes are filled. An infinitesimal shift of $\delta$ suffices to split the degeneracy of the corner modes by the on-site potential $\delta\Gamma_0$. The sign of $\delta$ determines whether the corner modes at one diagonal or the other are filled, resulting in the electron density pattern shown in Fig.~\ref{fig:quadrupole}b. 
If we account for a charge of $+2e$ per unit cell due to atomic contributions, the overall charge density is  essentially zero in the bulk,  with $\pm e/2$ fractional charges exponentially localized at the corners in the limit $\delta \to 0.$ 

The edge polarization in a quadrupole is a subtle feature. When it changes in space or time it leads to observables that will be discussed with more detail below. In the mean time, we occupy in the characterization of the \emph{quantized} edge polarization of the symmetry-protected quadrupole phase.
For this purpose, let us consider a cylinder geometry by connecting the top and bottom (left and right) edges of our model. Notice that the remaining x-normal (y-normal) edge preserves $M_y$ ($M_x$) symmetry. As a starting point, we will describe the edge polarization in the simple limit $\gamma=0$. This case is illustrated in Fig.~\ref{fig:quadrupole}c. We see that on each edge there is a localized 1D, two-band, gapped insulator (at half-filling) with reflection symmetry. The 1D polarization of such a system, which lies tangent to the edge, is quantized in units of $e/2$, and signals a boundary topological insulator phase. Remarkably, the edge polarization remains quantized even after departing from the $\gamma=0$ limit. In Fig.~\ref{fig:quadrupole}d we show calculations of the spatially-resolved polarization [see also~\cite{Vanderbilt2015} for a similar characterization],  for several values of $|\gamma/\lambda|$ (we assume the edge unit cells are identical to the bulk and not relaxed/reconstructed). We find that if $\gamma/\lambda=0,$ the quantized polarization of $\pm e/2$ is \emph{exactly} localized on the edge, but as $|\gamma/\lambda|\rightarrow 1$ the polarization, although still exactly quantized, penetrates into the bulk exponentially slowly as determined by the bulk gap. This protection by the bulk gap is clear evidence that the edge polarization is due to the topology of the bulk and not just an edge property. Indeed, we find the edge polarization exactly vanishes after the bulk phase transition, just as we saw the corner-localized modes disappear in a system with full open boundaries. In all cases the \emph{total}  polarization vanishes since opposite edges have opposite polarization.

This nontrivial edge polarization implies the existence of topological corner states.
Because the topology is protected by spatial symmetry, spatial defects such as corners create an effective edge for the boundary topological insulator, and hence form the boundary of the boundary.  The edge phases that converge at a corner are both topological, and, crucially, the corner-localized topological mode is a simultaneous end state of the two boundary topological insulators that meet at that corner, and not a conventional domain wall mode trapped between a topological and trivial phase (see SM, Sec. 2).\\

\begin{figure}[t]
\centering
\includegraphics[width=\columnwidth]{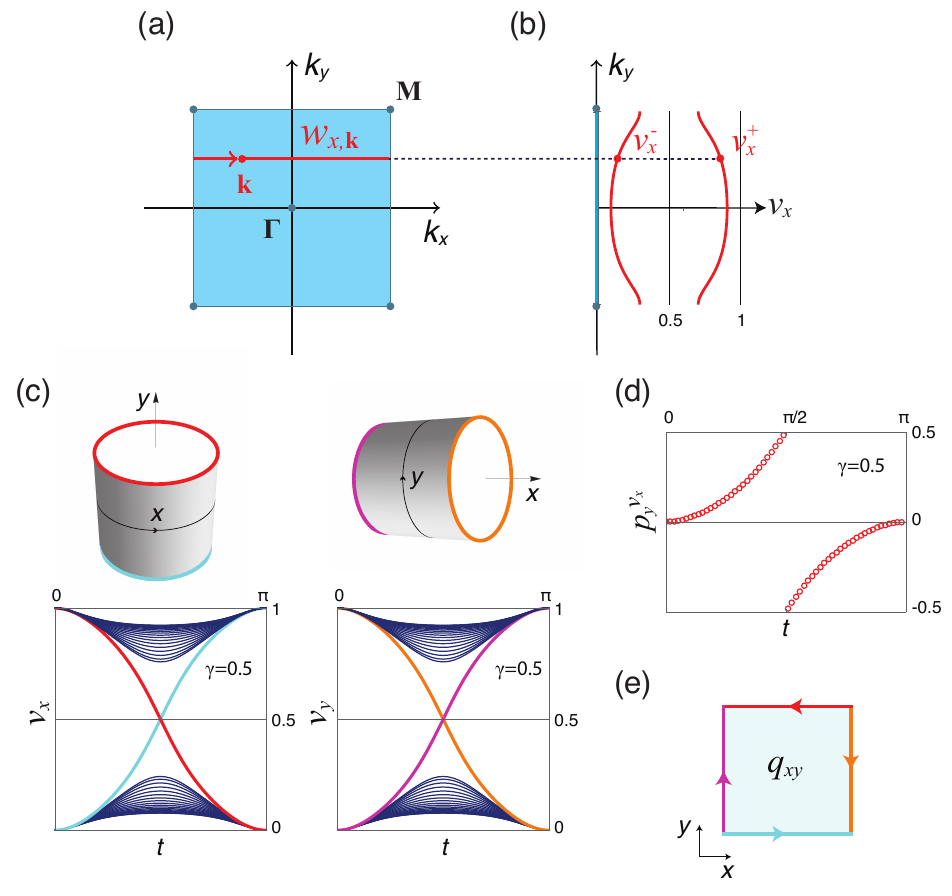}%
\caption{\emph{Wannier bands for quadrupole insulator and quadrupole pumping.} (a) Wilson loop $\W_{x,\bf k}$ along $k_x$ at starting point ${\bf k}$ over the 2-dimensional Brillouin zone. (b) Construction of Wannier bands $\nu^\pm_x(k_y)$ (marked red) for Bloch Hamiltonian in Eq.~\ref{eq:quad_hamiltonian} by means of Wilson loop diagonalization. (c) Flow of Wannier centers during adiabatic pumping. Wannier centers crossing the Wannier gap have wavefunctions localized at edges. (d) Wannier-sector polarization as a function of adiabatic parameter for the pumping in (c). There is no change between $t=\pi$ and $t=2\pi$. (e) Pattern of electric current during this pumping process.}
\label{fig:Wannier_bands}
\end{figure}

\paragraph*{Nested Wilson Loops}
As seen above, a bulk topological quadrupole moment $q_{xy}$ generates a topologically polarized edge when the system is cut parallel to either the $x$ or $y$ directions. To formalize this property we consider the Wilson loop operators in the $x$ and $y$ directions: $\W_{x, \bf k}$, or $\W_{y, \bf k}$, where ${\bf k} = (k_x,k_y)$ is the starting point, or base point, of the loop (Fig.~\ref{fig:Wannier_bands}a). A crystal with $N_{orb}$ degrees of freedom per unit cell and $N_{occ}$ occupied energy bands has the occupied Bloch functions $\ket{u_{{\bf{k}}}^{n}}$, for $n=1\ldots N_{occ}$, with components $[u_{{\bf{k}}}^{n}]^\alpha$ for $\alpha=1\ldots N_{orb}$. We define 
$[F_{x,\bf k}]^{mn} =\braket{u^{m}_{{\bf k}+{\bf \Delta k_x}}}{u^n_{{\bf k}}}$, for ${\bf \Delta k_x}=(2\pi/N_x,0)$. Hence,  $\W_{x, \bf k} = F_{x,{\bf k}+N_x{\bf \Delta k_x}} \ldots F_{x,{\bf k} + {\bf \Delta k_x}} F_{x,\bf k},$ and similarly for $\W_{y, \bf k}.$ We now define a Wannier Hamiltonian $H_{\W_{x}({\bf{k}})}$ of the $x$-edge via
\begin{equation}
\W_{x, \bf k}\equiv e^{iH_{\W_x}(\bf k)},
\end{equation}
which has been shown~\cite{Klich2011} to be adiabatically connected to the physical Hamiltonian of the edge perpendicular to direction $x$, and which we use to characterize the boundary topology.  

$H_{\W_{x}({\bf{k}})}$ has eigenvalues $2\pi \nu^{j}_{x}(k_y),$ $j=1\ldots N_{occ},$ that only depend on the coordinate $k_y$ of the Wilson loop base point ${\bf k}$ (and vice-versa for the eigenvalues of $H_{\W_{y}({\bf{k}})}$ of the $y$-edge). With fully periodic boundary conditions, the Wilson loop \emph{for our model} diagonalizes as:
\begin{align}
\W_{x,\bf k}&=\sum_{j=\pm} \ket{\nu^j_{x,\bf k}} e^{2\pi i \nu^j_x(k_y)} \bra{\nu^j_{x,\bf k}},
\label{eq:Wilson_loop_eigensystem}
\end{align}
where the eigenstates $\ket{\nu^j_{x,\bf k}}$, $j=1,2$, have components $[\nu^j_{x,\bf k}]^n$, $n=1,2$.  It is well-known that the Wannier centers are proportional to the phases $\nu^{j}_x(k_y)$ of these eigenvalues~\cite{kivelson1982,King-SmithVanderbilt1993,vanderbilt1993,marzari1997,coh2009,yu2011,qi2011}. Due to the $x \to-x$  reflection symmetry, $\nu_x^-(k_y) = -\nu_x^+(k_y)$ mod 1~\cite{Alexandradinata2014} (see SM, Sec. 6); for convention we choose $\nu_x^-(k_y)\in[0,1/2].$ Crucially, in our model, unlike previously characterized topological insulators~\cite{coh2009,yu2011,qi2011,Alexandradinata2014,aris2016,wang2016a,aris2016b}, the Wannier centers of the two occupied bands do not touch at any point over the entire range $ k_y \in (-\pi,\pi]$  (red lines in Fig.~\ref{fig:Wannier_bands}b). Thus, $H_{\W_x}$ is gapped, and we refer to the spectra $\nu_x^\pm(k_y)$ as being two \textit{Wannier bands}, and to their separation as the \textit{Wannier gap}. The Wannier bands are defined over the 1D BZ $ k_y\in (-\pi,\pi],$ and they obey the identification $\nu_x^j \equiv \nu_x^j \mod 1.$

The key feature of our construction is that, being gapped, the Wannier bands \emph{can carry their own topological invariants.} These can be evaluated by calculating nested Wilson loops, as follows. Let us start by defining the Wannier band subspaces 
\begin{align}
\ket{w^\pm_{x,\bf k}} = \sum_{n=1,2}\ket{u^n_{\bf k}} [\nu^\pm_{x,\bf k}]^n,
\label{eq:Wannier_basis}
\end{align} 
which provide a natural splitting of our original pair of occupied energy bands, which were necessarily degenerate in energy at the special points in the BZ (due to $\{\hat{m}_x,\hat{m}_y\}=0),$ into two separate single-band subspaces (we generalize this formalism to arbitrary numbers of occupied energy bands and Wannier bands in the SM, Secs. 4, 5, and 6).
If we define $F^\pm_{y,\bf k} = \braket{w^\pm_{x,{\bf k}+{\bf \Delta k_y}}}{w^\pm_{x,\bf k}}$, where ${\bf \Delta k_y}=(0, 2\pi/N_y)$, the nested Wilson loops along $k_y$ in the Wannier bands $\nu^\pm_x$ are 
\begin{align}
\tilde\W^\pm_{y, k_x} = F^\pm_{y,{\bf k}+N_y{\bf \Delta k_y}} \ldots F^\pm_{y,{\bf k} + {\bf \Delta k_y}} F^\pm_{y,\bf k},
\label{eq:nested_Wilson_loop}
\end{align}
and their associated polarizations are
\begin{align}
p^{\nu^\pm_x}_y=-\frac{i}{2\pi} \frac{1}{N_x}\sum_{k_x} \text{Log} \left [\tilde\W^\pm_{y, k_x} \right] = \left\{
\begin{array}{ll}
0 & \frac{|\gamma|}{|\lambda|} >1\\
1/2 & \frac{|\gamma|}{|\lambda|} <1
\label{eq:pumping_spherical}
\end{array}
\right. .
\end{align}
The polarization of $1/2$ along $y$ implies that the Wannier Hamiltonian of the $x$-edge $H_{\W_{x}({\bf{k}})}$ is a topological insulator when $|\gamma/\lambda|<1$, which is in agreement with our findings in the numerical simulation of our model (Fig.~\ref{fig:quadrupole}c,d). Importantly, though the subspace of occupied energy bands has trivial polarization along $y$, each of the separate Wannier band subspaces has non-vanishing polarization.
In the thermodynamic limit the polarization of the Wannier bands is 
\begin{align}
p^{\nu^\pm_x}_y = -\frac{1}{(2\pi)^2} \int_{BZ} \tilde\A^{\pm}_{y, \bf k} d^2\bf k,
\label{eq:polarization_Wannier_sector}
\end{align} 
where
\begin{eqnarray}
\tilde\A^{\pm}_{y, \bf k} &=& -i \bra{w^\pm_{x,\bf k}} \partial_{k_y} \ket{w^\pm_{x,\bf k}}
\end{eqnarray}   
is the Berry potential over Wannier band $\nu_x^\pm$ respectively. In Secs. 4 and 5 of  the SM, we provide more formal details of these expressions, as well as their relation with lattice position operators projected into the energy or Wannier-band subspaces.\\

\paragraph*{Symmetry Constraints and Quantization}
 To obtain a well defined quadrupole moment we require a crystal with a vanishing bulk polarization, hence inversion-symmetric~\cite{zak1989berry}. To quantize the quadrupole moment we want to preserve reflection symmetries as well.
Under reflections  $x \rightarrow -x$ ($M_x$), and $y \rightarrow -y$ ($M_y$), and inversion $(x,y)\to (-x,-y)$ ($\mathcal{I}$) the polarization of the Wannier-band subspace obeys
\begin{align}
p^{\nu^+_x}_y \stackrel{M_x}{=} p^{\nu^-_x}_y, \;\;
p^{\nu^\pm_x}_y \stackrel{M_y}{=} -p^{\nu^\pm_x}_y, \;\; 
p^{\nu^+_x}_y \stackrel{\mathcal{I}}{=} -p^{\nu^-_x}_y \;\;\mbox{mod 1},
\label{eq:symmetry_constraints}
\end{align}
as shown in detail in SM, Sec. 6. 
Hence,  the Wannier-sector polarizations $p^{\nu^\pm_x}_y$ and $p^{\nu^\pm_y}_x$ (calculated from $H_{\W_{x}}$ and $H_{\W_{y}}$ respectively) take quantized values
\begin{align}
p^{\nu^\pm_x}_y, p^{\nu^\pm_y}_x \stackrel{\mathcal{I},M_x,M_y}{=} 0 \mbox{ or } 1/2
\end{align}\noindent when the constraints due to all three symmetries are satisfied. Hence, the most general classification of the Wannier band topology is $\mathbb{Z}_2 \times \mathbb{Z}_2$, with the non-trivial quadrupole state corresponding to the class with symmetry-protected invariants $(p^{\nu^\pm_x}_y, p^{\nu^\pm_y}_x)=(1/2,1/2)$.  The phases with $(p^{\nu^\pm_x}_y, p^{\nu^\pm_y}_x)=(1/2,0)$ or $(0,1/2)$ can be reached  by allowing for separate $x$ and $y$ hopping parameters $\lambda_{x},\lambda_{y}$ in Eq.~\ref{eq:quad_hamiltonian} and tuning one or the other outside the topological regime, i.e., $|\lambda_{i}/\gamma|>1$. A transition between the quadrupole phase (1/2, 1/2) and, e.g., the (1/2, 0) phase is not marked by a closing of the 2D bulk energy gap. Instead the transition occurs in the Wannier bands of $H_{\mathcal{W}_{x}}$, or equivalently through a \emph{bulk-driven} gap closing in the 1D \emph{edge} Hamiltonian parallel to the $y$-axis (a transition on the physical edge alone is not enough as the interior region would still be in the (1/2, 1/2) phase). 
Enforcing $C_4$ symmetry reduces the classification to $\mathbb{Z}_2$, and the quadrupole invariant $q_{xy}$ can be identified as 
\begin{align}
q_{xy} = p^{\nu^+_x}_y= p^{\nu^-_x}_y= p^{\nu^+_y}_x = p^{\nu^-_y}_x = 0 \mbox{ or } 1/2.
\label{eq:quadrupole_invariant}
\end{align} 
Transitions between the $C_4$-symmetric gapped phases do occur along with a closure of the 2D  bulk energy gap, as seen in Fig.~\ref{fig:quadrupole}a, or determined explicitly from the bulk energy bands. 

Finally, we point out that reflection symmetry can be used to compute the invariant in a simpler way. The Wannier band basis  \eqref{eq:Wannier_basis} obeys
\begin{align}
\hat{m}_y \ket{w^\pm_{x,(k_x,k_y)}}=\ket{w^\pm_{x,(k_x,-k_y)}} \alpha^{\pm}(k_x,k_y),
\label{eq:Wannier_basis_reflection}
\end{align}
with a $U(1)$ phase $\alpha^{\pm}(k_x,k_y)$. In particular, at the reflection-invariant momenta $k_y^*=0,\pi$,  $\alpha^{\pm}(k_x, k_y^*)$ are the eigenvalues of the reflection representation of $\ket{w^\pm_{x,\bf k}}$ at $(k_x,k_y^*).$ These can take the values $\pm 1$  $(\pm i)$ for spinless (spinful) fermions. If the representation is the same (different) at $k_y^* = 0$ and $k_y^* = \pi$, the Wannier-sector polarization is trivial (non-trivial). Thus, in reflection-symmetric insulators the quadrupole topology captured in \eqref{eq:polarization_Wannier_sector} can then be easily computed by
\begin{align}
\mbox{exp}\left\{i2\pi p^{\nu^\pm_x}_y\right\} = \alpha^{\pm}(k_x,0) \alpha^{\pm\ast}(k_x,\pi).
\label{eq:invariant_reflection}
\end{align}\\

\paragraph*{Edge currents and quadrupole pumping} 
In crystals, corner charges alone do not imply the existence of a quadrupole moment (see SM, Sec. 11 for a counterexample). There must exist   concomitant edge polarizations  having \emph{the same} magnitude as the corner charge, in agreement with Eq. \eqref{eq:qdensities}. Although the edge polarization itself is not an observable quantity, changes in polarization over space or time are, as they amount to localized charge or charge current, respectively. Thus, even in the absence of corners or other edge inhomogeneities, changes in bulk quadrupole moments over time manifest as inversion-symmetric, edge-localized currents. In contrast, currents due to changes in bulk polarization are distributed in the bulk, and are induced by breaking inversion symmetry (otherwise the bulk polarization is locked).

\begin{figure}%
\centering
\includegraphics[width=\columnwidth]{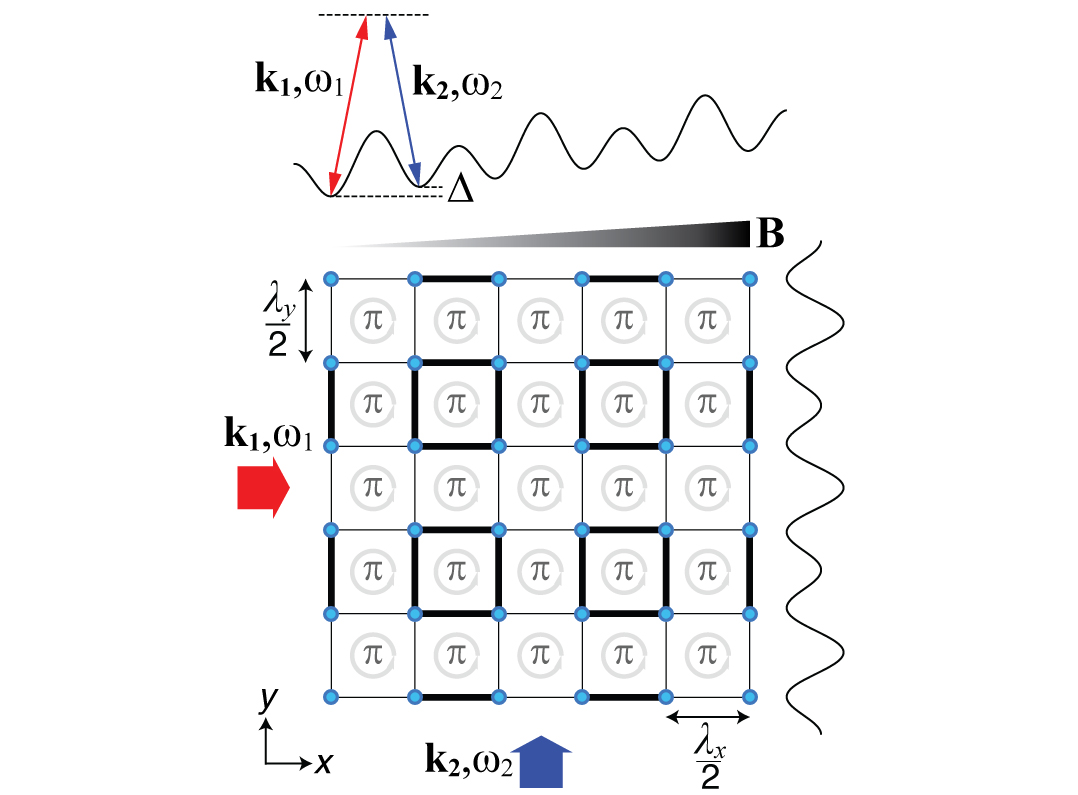}%
\caption{\emph{Proposal for the experimental realization of a crystalline quadrupole using ultra cold atoms in an optical lattice}.  The lattice has a superlattice optical pattern and laser assisted tunneling that induces an effective $\pi$ flux per plaquette.}%
\label{fig:cold_atoms}
\end{figure}

An illustration of such a process is given by the family of Hamiltonians parameterized as $h^{q}_{t}\equiv (\sin t)h^{q}({\bf{k}},0)+(\cos t)\Gamma_0$ for $0 \leq t \leq \pi$ and $h^{q}_t\equiv (\sin t)(\Gamma_2+\Gamma_4)+(\cos t)\Gamma_0$ for $\pi \leq t \leq 2\pi$. The parameter $t$ represents the time over which the instantaneous Hamiltonian is \emph{adiabatically} deformed. Along this entire process inversion symmetry as well as both the bulk and edge energy gaps are maintained. We enforce the first property in order to lock the bulk polarization to vanish, and the second property for adiabaticity. During the first half of the cycle all of the transport occurs (the second half of the cycle has only on-site terms that connect two trivial Hamiltonians and therefore cause no transport). A plot of the Wannier values $\nu_x$ as a function of $t$ for $0 \leq t \leq \pi$ is shown in Fig.~\ref{fig:Wannier_bands}c, left. In this calculation the $x$-direction is periodic, but the $y$-direction is open. At each $t$, we diagonalize the Hamiltonian $h^q_t(k_x,y)$ and use the subspace of $2 N_y$ occupied bands as a function of $k_x$ to perform the Wilson loop along $x$. What we show in Fig.~\ref{fig:Wannier_bands}c, left, are the resulting $2 N_y$ Wannier centers. 
At $t=0$, the Hamiltonian is $\Gamma_0$, which is a trivial insulator, and the Wannier centers are all 0 (mod 1). As $t$ progresses, a pair of Wannier centers with corresponding eigenstates localized at the edges separate from the `bulk' Wannier centers (dark blue lines), approaching the value of $\nu_x=1/2$ at $t=\pi/2$ (i.e. at the symmetry-protected quadrupole phase). Aside from the edge-localized Wannier centers, the bulk Wannier centers remain gapped, in agreement with Fig.~\ref{fig:Wannier_bands}b. For $t>\pi/2$ the edge Wannier centers move away from $1/2$ towards a value of $0$ but \emph{towards its neighboring unit cell}, following the pattern in Fig.~\ref{fig:Wannier_bands}c, left. At $t=\pi$, all Wannier centers become 0 again as the system becomes a trivial insulator. The second half of the cycle then connects the two trivial insulators $h^q_\pi=-\Gamma_0$ with $h^q_0=\Gamma_0$ via on-site, gap-preserving terms that do not induce any transport. Thus, the net effect is that electrons are pumped along the edge by one unit cell to the left on the top edge, and by one unit cell to the right on the bottom edge. For the same cycle, but with open boundaries along $x$ and periodic along $y$, the Wannier centers along $y$ are shown in Fig.~\ref{fig:Wannier_bands}c, right. In both cases, transport is quantized. This quantization is captured by the winding of the Wannier-sector polarization
\begin{align}
\Delta q_{xy} = \int_0^{2\pi} dt \frac{\partial}{\partial t} p^{\nu^\pm_y}_x(t) = 1,
\end{align}
as shown in Fig.~\ref{fig:Wannier_bands}d. Notably, the overall charge transport along horizontal and vertical edges is not that of a circulating current; rather, it follows the pattern shown schematically in Fig.~\ref{fig:Wannier_bands}e.
\\
\paragraph*{Experimental Realizations of the Quantized Quadrupole Moment}
We now describe three systems in which the quadrupole topology can be realized and its signatures can be measured.\\
\underline{\emph{Cold Atoms in Optical Lattices:}}
The quadrupole moment can be realized straightforwardly in a system of ultra-cold atoms in an optical lattice. Our model in Eq.~\ref{eq:quad_hamiltonian} can be constructed from a square lattice Hofstadter model with $\pi$-flux per plaquette with the addition of a superimposed superlattice. For example, in the setup realized in Ref.  \cite{Bloch2013}, a Hofstader model is created by mutually-orthogonal standing optical waves of wavelengths $\lambda_{x,y}=767$ $nm$ and $\lambda_z=844$ $nm$ in which an ultracold gas of $^{87}$Rb atoms is held. A period-2 superlattice (dimerization) in the $x$ and $y$ directions can be induced in the $xy$-plane by two additional orthogonal optical waves at wavelengths $2\lambda_{x,y}$ (Fig.~\ref{fig:cold_atoms}).

Remarkably,  it has already been experimentally demonstrated that such a dimerized lattice in the $xy$-plane can coexist with a uniform effective flux per plaquette that is fully tunable~\cite{Bloch2013}. The effective flux is achieved by inhibiting the tunneling along the horizontal direction via a linear magnetic gradient that generates an equal energy offset of magnitude $\Delta$ between neighboring sites. Then, this inhibited tunneling is restored resonantly using a pair of far-detuned running wave beams with a frequency difference of $\omega=\omega_1-\omega_2=\Delta/\hbar$. The resulting potential from the running wave beams causes an on-site modulation with spatially dependent phases along the direction of the detuning ($x$-direction in Fig.~\ref{fig:cold_atoms}). If the beams are orthogonal, and with a wavevector $k_1\simeq k_2 = 2\pi/\lambda_K$, the phases acquired along the horizontal hoppings are 
$\phi_{m,n} = \pi/\lambda_K(\lambda_x m + \lambda_y n)$ where $(m,n)$ labels the lattice site. Since the phases are acquired only along the horizontal direction, a phase of $\pi$ per plaquette is obtained by fixing the value of $\lambda_K$ and $\lambda_y$ to $\lambda_K = \lambda_y \rightarrow \phi_{plaquette} = \pi.$ While Ref. \cite{Bloch2013} has tuned to a value of $\phi_{plaquette}=\pi/2,$ a value of $\phi_{plaquette}=\pi$ has been achieved in a similar setup in Ref. \cite{miyake2013}.
As such, this system meets all the requirements to realize a the quadrupole model. 

\underline{\emph{Atom-Optics:}}
A second promising direction with cold atoms uses stimulated, two-photon Bragg transitions between free-particle plane-wave momentum states of a Bose-Einstein condensate to mimic electronic tight binding models~\cite{gadway2015,Gadway2016}. Essentially, the local atomic orbitals of a tight binding model are replaced by a plane-wave basis.   Refs. \cite{Gadway2016,meier2016}  demonstrated the creation of effective 1D tight binding models, including a 1D topological chain, in a $^{87}$Rb Bose-Einstein condensate. By programming acousto-optic modulators the amplitude and phases of the laser-induced couplings between momentum states can be precisely controlled. The method can be extended to higher dimensions~\cite{gadway2015}, and the dimerized $\pi$-flux lattice needed for our quadrupole model can be generated similarly  by using the acousto-optic modulators to tune the amplitudes and phases of the appropriate coupling terms.

\begin{figure}[t!]%
\centering
\includegraphics[width=\columnwidth]{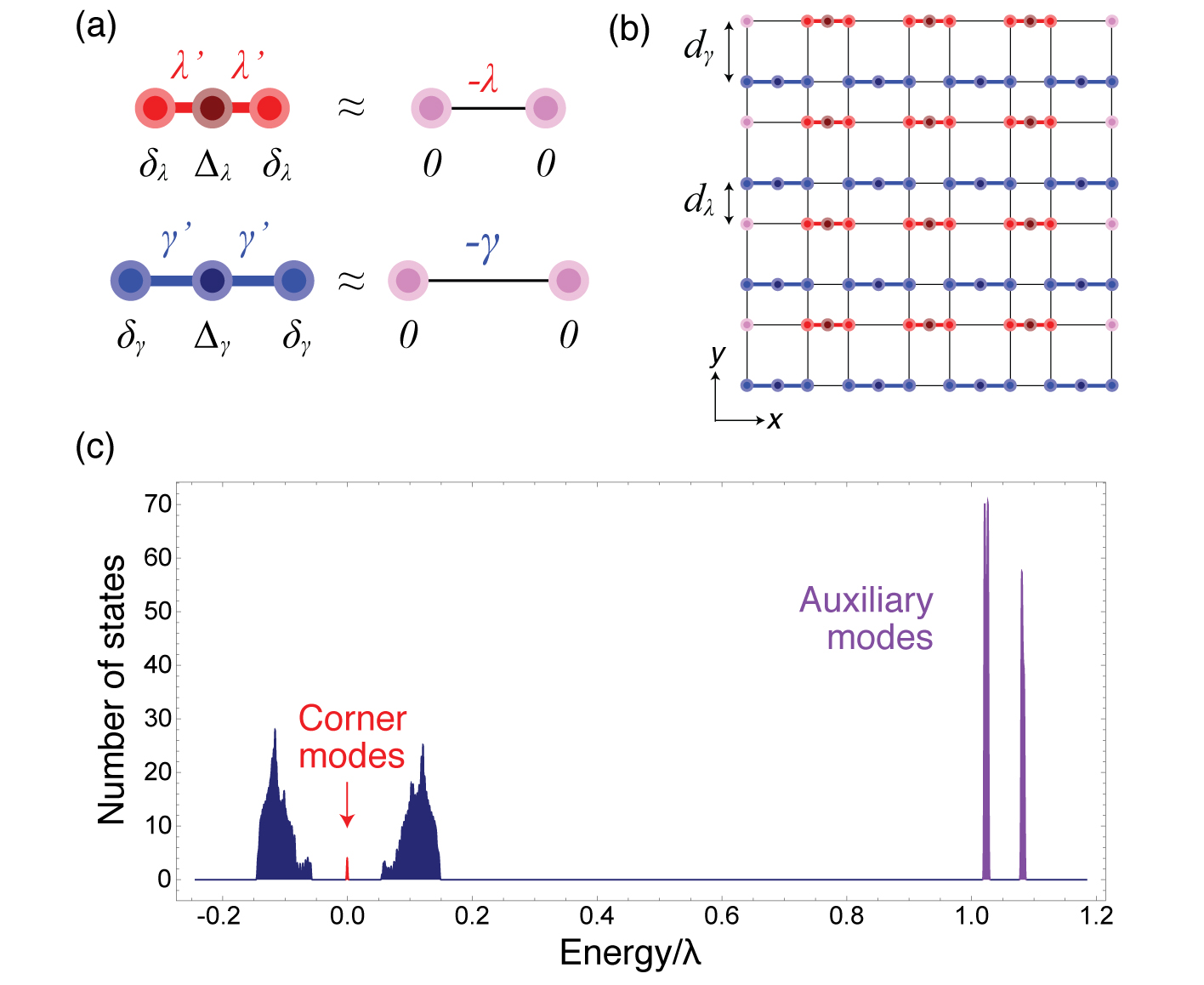}%
\caption{\emph{Proposal for realization of quadrupole in a photonic crystal}. (a) Effective negative hoppings are achieved by introducing auxiliary waveguides (at center) and fine-tunnig the refractive indices on each individual waveguide. (b) Photonic crystal lattice with non-trivial quadurpole moment. (c) Density of states for a lattice as in (b) with $16 \times 16$ unit cells and $d_\lambda=20$ $\mu$m, $d_\gamma = 30$ $\mu$m. For these distances the auxiliary modes are sufficiently detuned and do not affect the quadrupole topological structure at lower energies.}%
\label{fig:photonic_crystal}
\end{figure}

\underline{\emph{Photonic crystals:}}
Our final proposal is for laser-written photonic crystals in fused silica, which have recently been used to simulate a variety of  tightbinding lattice models~\cite{rechtsman2013,rechtsman2013b,plotnik2014}. Until recently our model could not have been easily created in such systems because of the necessity for hoppings with opposite signs that encode the $\pi$-flux. However, Ref. \cite{Szameit2016} proposed, and experimentally demonstrated, the construction of a \textit{negative coupler}. This is accomplished by inserting an additional waveguide between two evanescently coupled waveguides (Fig.~\ref{fig:photonic_crystal}a). Concretely, to induce a negative sign into the hopping amplitude between  two coupled waveguides with coupling amplitude $\lambda$ (that is, to go from $\lambda$ to $-\lambda$), an extra waveguide with on-site energy (which is proportional to the local index of refraction) $\Delta=(\lambda')^2/\lambda-\lambda$ is inserted at equal distance in between the two original waveguides, and an on-site potential $\delta=\lambda$ is added to the two original waveguides (Fig.~\ref{fig:photonic_crystal}a). Here, $\lambda'$ is the hopping amplitude between the extra waveguide and an original waveguide. This coupler approaches an effective two-waveguide coupler with negative hopping in its tight-binding approximation as $\lambda'/\lambda \to \infty$. Using this development, a quadrupole model can be achieved by the dimerized waveguide arrangement shown in Fig.~\ref{fig:photonic_crystal}b. Since the waveguides are evanescently coupled with one another, the hopping amplitudes between two waveguides decreases exponentially with their separation $\lambda \propto e^{-\kappa d_\lambda}$. Thus, the transition between trivial and topological phases is achieved by simply varying the separations $d_\lambda$ and $d_\gamma$, with the transition point occurring at $d_\lambda=d_\gamma$. 

We simulated a $16 \times 16$ unit-cell photonic lattice as in Fig.~\ref{fig:photonic_crystal}b (see SM, Sec. 12 for details). This was done in the tight-binding limit, and considering only nearest neighbor hopping between waveguides. We set $\kappa = 0.125$ $\mu m^{-1}$ (to account for an ambient medium of fused silica with refractive index $n_0=1.45$ and an effective potential of $\Delta n=7\times 10^{-4}$ for a waveguide of radius $R=2.5$ $\mu m$ and incident light with wavelength $633$ $nm$), $d_\lambda=20$ $\mu m$, and $d_\gamma=30$ $\mu m$. Since this leads to $\lambda>\gamma$, the photonic crystal is in the topological phase. This is manifest in the four degenerate, mid-gap energy modes shown in the density of states in Fig.~\ref{fig:photonic_crystal}c, which we verify are exponentially localized at the corners. Importantly, the set of states localized at the auxiliary waveguides are well separated in energy from those that determine the quadrupole topology. To probe this phase experimentally, a corner waveguide can be illuminated and its diffusion into neighboring waveguides can be tracked. In particular, in the trivial phase we expect the initial beam to spread into the bulk, while in the topological phase the beams are expected to remain confined to the corner.\\

\begin{figure}[t!]%
\centering
\includegraphics[width=\columnwidth]{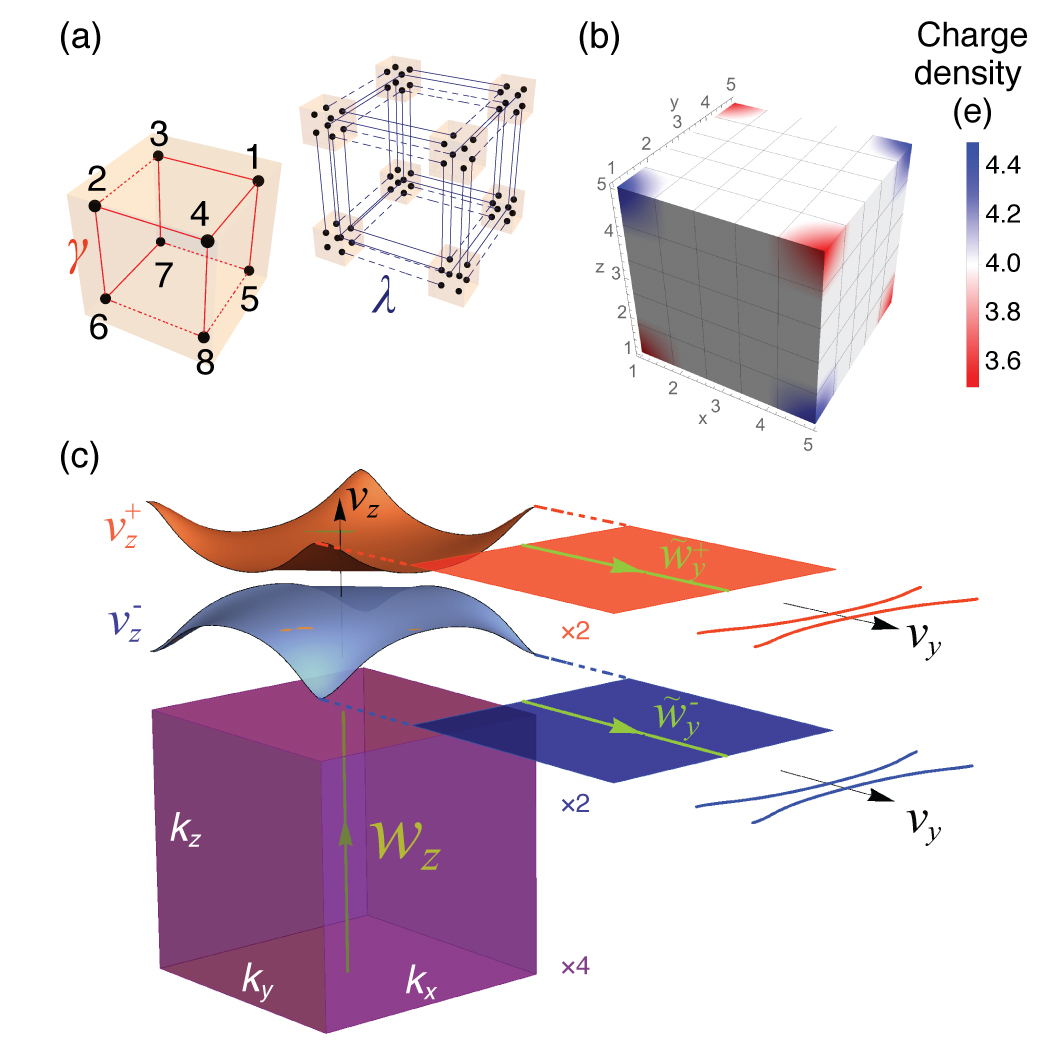}%
\caption{\emph{3D insulator with a quantized octupole moment $o_{xyz}$}. (a) Hopping terms in the Hamiltonian of Eq. \ref{eq:octupole_hamiltonian}. $\gamma, \lambda$ (shown in left, right respectively) represent two hopping strengths, dashed lines represent hopping terms with negative signs. Unit cells are shown in orange boxes. The topological phase occurs when $|\gamma/\lambda|<1$. Numbers indicate basis for $\Gamma'$-matrices. (b) Charge density in the topological phase. Corner charges are $\pm e/2$ relative to background charge. (c) Wannier bands by successively calculating the Wilson loop along $z$, then nested Wilson loop along $y$, and finally along $x$. Each of the 2D Wannier bands determined by $\W_{z,\bf k}$ have a non-trivial quadrupole moment $q_{xy}=\pm e/2$ and each of the 1D Wannier bands determined from $\tilde\W_{y,\bf k}$ have a nontrivial polarization $p_{x}=\pm e/2$.}%
\label{fig:octupole}
\end{figure}

\paragraph*{Model with Quantized Octupole Moment}
A simple, minimal model in 3D with quantized octupole moment has Bloch Hamiltonian (a schematic representation is shown in Fig.~\ref{fig:octupole}a)
\begin{align}
h^o_\delta({\bf k}) &= \lambda \sin(k_y)\Gamma'_1+ \lambda \cos(k_y)\Gamma'_2
+\lambda \sin (k_x)\Gamma'_3\label{eq:octupole_hamiltonian}\\  &+ \lambda \cos (k_x)\Gamma'_4 
+\lambda \sin(k_z)\Gamma'_5 + \lambda \cos(k_z)\Gamma'_6 
+\delta \Gamma'_0,\nonumber
\label{eq:octupole_hamiltonian}
\end{align}
where $\Gamma'_i=\sigma_3 \otimes \Gamma_i$ for $i=0,1,2,3$, $\Gamma'_4=\sigma_1 \otimes I_{4 \times 4}$, $\Gamma'_5=\sigma_2 \otimes I_{4 \times 4}$, and $\Gamma'_6=i \Gamma'_0 \Gamma'_1 \Gamma'_2 \Gamma'_3 \Gamma'_4 \Gamma'_5$. This insulator has two sets of energy bands, each of which is four-fold degenerate. At half filling, and with $\delta = 0$, the insulator has a quantized octupole moment $o_{xyz}=\pm e/2$ when $\gamma/\lambda \in (-1,1).$ The quantized moment is  protected by the presence of all three reflection symmetries, with operators $\hat{m}_x=\sigma_0 \otimes \sigma_x \otimes \sigma_z, \hat{m}_y=\sigma_0 \otimes \sigma_x \otimes \sigma_x,$ and $\hat{m}_z=\sigma_x \otimes \sigma_z \otimes \sigma_0$ (and inversion $\mathcal{I}=\hat{m}_x\hat{m}_y\hat{m}_z$). With $\delta\neq 0$ the cubic symmetry of the crystal is broken down to tetrahedral symmetry, i.e., with no reflections, and the octupole moment loses its quantization, while, importantly, $p_{x,y,z}$ and $q_{xy,yz,xz}$ retain theirs. An infinitesimal value $\delta \neq 0$ is used in the fully-open, cubic geometry of Fig.~\ref{fig:octupole}b to slightly split the degeneracy of the eight, mid-gap, corner localized states. This generates charges $\pm e/2$ localized at the corners in an octupolar pattern. 

To characterize the bulk topology we  calculate the Wilson loop along the $i$-th direction ($i$=x,y, or z), and our model generates gapped 2D Wannier bands in the orthogonal $jk$-plane in momentum space as shown in Fig.~\ref{fig:octupole}c. We find that each Wannier band is two-fold degenerate, and each has a non-trivial quadrupole topology with a quantized $q_{jk}$ moment. We characterize the quadrupole topology by calculating the nested Wilson loop along the $j$-th direction over one set of the 2D Wannier bands. As shown in Fig.~\ref{fig:octupole}c this further splits the two-fold degenerate 2D bands into non-degenerate, gapped 1D Wannier bands. We then calculate a final nested Wilson loop in the $k$-th direction and find a value of $-1$ in the non-trivial octupole phase (see SM, Sec. 13). In the non-trivial phase with $o_{xyz}=\pm e/2$ this result is independent of the order $ijk$.
Physically, this hierarchy of Wilson loops implies that the corner charges are manifestations of bound, surface quadrupole moments. The surface quadrupoles meet at the hinges of the cube giving rise to bound charge polarizations, finally the polarizations converge/diverge from the corners giving the bound charge density. Because the topology is protected by spatial symmetries, these sharp interfaces allow us to generate boundaries on the boundary to uncover the surface topology.\\

\paragraph*{Conclusion}
We have predicted a new class of topological phases with quantized electromagnetic properties. A description of insulators with these higher multipole moments has been a long-standing puzzle~\cite{blount1962}, which we have now solved. We proved that the quadrupole (and octupole) moment in reflection-symmetric crystals is quantized, and can only appear in materials with an even number of occupied bands. In addition, we provided a simple way to compute the topological invariant, and provided three experimental proposals to realize our quadrupole model in cold-atom and photonic systems. These proposals will provide sharp experimental tests for our theoretical predictions. 
\\
\paragraph*{Acknowledgements} We thank C. Fang and A. Soluyanov for useful discussions, and B. Gadway for useful discussions and for pointing us to Ref. \cite{Szameit2016}. WAB and TLH thank the US National Science Foundation under grant DMR 1351895-CAR and the Sloan Foundation for support. BAB acknowledges support from Department of Energy DE-SC0016239, NSF EAGER Award NOA - AWD1004957, Simons Investigator Award, ONR - N00014-14-1-0330, ARO MURI W911NF-12-1-0461, NSF-MRSEC DMR-1420541, Packard Foundation and Schmidt Fund for Innovative Research.


\bibliography{quad_references}
\bibliographystyle{Science}

\end{document}


\title{Quantized Electric Multipole Insulators:\\Supplementary Material} 

\author{Wladimir A. Benalcazar}
\affiliation{Department of Physics and Institute for Condensed Matter Theory, University of Illinois at Urbana-Champaign, IL 61801, USA}
\author{B. Andrei Bernevig}
\affiliation{Department of Physics, Princeton University, Princeton, New Jersey 08544, USA}
\author{Taylor L. Hughes}
\affiliation{Department of Physics and Institute for Condensed Matter Theory, University of Illinois at Urbana-Champaign, IL 61801, USA}

\maketitle

\section{The multipole expansion in the continuum electromagnetic theory}
\label{sec:app_classical_multipoles}
In this appendix we derive the expressions for the charge distribution due to non-zero multipole moments in continuous media. Throughout this appendix, summation is implied over repeated coordinate indices. 

\subsection{The multipole expansion}
\label{sec:multipole_expansion_far_limit}
The potential at position $\vec{r}$ due to a charge distribution over space is
\begin{align}
\phi(\vec{r})=\frac{1}{4\pi \epsilon} \int d^3\vec{r'}  \rho(\vec{r'}) \frac{1}{|\vec{r}-\vec{r'}|},
\label{eq:app_potential_due_to_charge_distribution}
\end{align}
where $\rho(\vec{r})$ is the volume charge density. This expression can be expanded in terms of the Legendre polynomials as
\begin{align}
\phi(\vec{r})=\frac{1}{4\pi \epsilon} \sum_{l=0}^\infty \int d^3\vec{r'}  \rho(\vec{r'}) \frac{r'^l}{r^{l+1}}P_l(n_i n_i'),
\end{align}
where $P_l(x)$ is the $l^\text{th}$ Legendre polynomial and $n_i$ is the $i^\text{th}$ component of the unitary vector of $\vec{r}$, i.e. $r_i = r n_i$, where $r$ is the magnitude of $\vec{r}$. We define the monopole contribution to the potential as that which scales as $1/r$ in the expansion above, the dipole contribution as that which scales as $1/r^2$, etc. In particular, from the general expression above we now read off the contributions up to octupole expansion:
\begin{align}
\phi^0(\vec{r})&=\frac{1}{4 \pi \epsilon} Q \frac{1}{r}\nonumber \\
\phi^1(\vec{r})&=\frac{1}{4 \pi \epsilon} P_i \frac{r_i}{r^3}\nonumber  \\
\phi^2(\vec{r})&=\frac{1}{4 \pi \epsilon} Q_{ij} \frac{3 r_i r_j-r^2 \delta_{ij}}{2r^5}\nonumber  \\
\phi^3(\vec{r})&=\frac{1}{4 \pi \epsilon} O_{ijk} \frac{5 r_i r_j r_k - 3 r^2 \delta_{ij} r_k}{2r^7},
\end{align}
where $Q$, $P_i$, $Q_{ij}$, and $O_{ijk}$ are the monopole, dipole, quadrupole, and octupole moments respectively, defined as
\begin{align}
Q&=\int_v d^3\vec{r'} \rho(\vec{r'})\nonumber \\
P_i&=\int_v d^3\vec{r'} \rho(\vec{r'}) r_i'\nonumber \\
Q_{ij}&=\int_v d^3\vec{r'} \rho(\vec{r'}) r_i' r_j'\nonumber \\
O_{ijk}&=\int_v d^3\vec{r'} \rho(\vec{r'}) r_i' r_j' r_k'.
\label{eq:app_multipole_moments}
\end{align}
Using these multipole contributions, the total potential can be written as
\begin{align}
\phi(\vec{r})=\sum_{i=0}^\infty \phi^i(\vec{r}).
\label{eq:app_potential_due_to_multipole_moments}
\end{align}

\subsection{Dependence of the multipole moments on the choice of reference frame}
\label{sec:app_multipole_translation_invariance}
The multipole moments are in general defined with respect to a particular reference frame. For example, if we shift the coordinate axes used in the definition of the dipole moment in Eq. \ref{eq:app_multipole_moments} such that our new positions are given by $r'_i = r_i + R_i$, the dipole moment is now given by
\begin{align}
P'_i&=\int_v d^3\vec{r'} \rho(\vec{r'}) (r'_i)\nonumber \\
&=\int_v d^3\vec{r} \rho(\vec{r}+\vec{R}) (r_i+R_i)\nonumber \\
&=\int_v d^3\vec{r} \rho(\vec{r}+\vec{R}) r_i+ R_i \int_v d^3\vec{r} \rho(\vec{r}+\vec{R})\nonumber \\
&=P_i+ R_i Q
\end{align}
where $Q$ is the total charge. Notice, however, that the dipole moment is well defined for any reference frame if the total charge $Q$ vanishes. Similarly, a quadrupole moment transforms as
\begin{align}
Q'_{ij}&=\int_v d^3\vec{r'} \rho(\vec{r'}) r'_i r'_j\nonumber \\
&=\int_v d^3\vec{r} \rho(\vec{r}+\vec{R}) (r_i+R_i) (r_j+R_j)\nonumber \\
&=\int_v d^3\vec{r} \rho(\vec{r}+\vec{R}) r_i r_j + 2 R_i \int_v d^3\vec{r} \rho(\vec{r}+\vec{R}) r_j+R_i R_j \int_v d^3\vec{r} \rho(\vec{r}+\vec{R})\nonumber \\
&=Q_{ij} + 2 P_i R_i+ R_i R_j Q
\end{align}
which is not well defined for any reference frame unless both the total charge and the dipole moments vanish. 
In general, for a multipole moment to be independent of the choice of reference frame, all its lower moments must vanish.

\subsection{Multipole moments in macroscopic media and their boundary manifestations}
\begin{figure}
\centering
\includegraphics[width=.25\columnwidth]{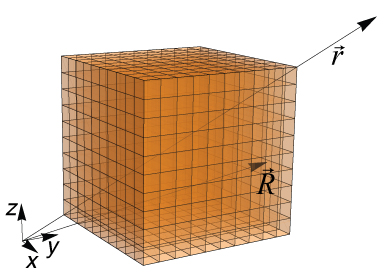}
\caption{\emph{Macroscopic material divided in small voxels over which the multipole moment densities are calculated}. Each voxel is labeled by its center point $\vec{R}$. The observation point at which the potential is calculated is at $\vec{r}$.}
\label{fig:app_macroscopic_multipole}
\end{figure}

We now consider the multipole moments in macroscopic materials.
For this purpose, we divide a macroscopic material (e.g. a macroscopic crystallite) into microscopic voxels, over which multipole moment densities are defined. At macroscopic scales (scales much larger than the voxel size), these densities can be treated as continuous functions of the position. We then embed a finite volume $V$ of this material in the vacuum and ask what observables are present at its boundaries. We start by re-stating that the potential at position $\vec{r}$ due to a charge distribution over space is
\begin{align}
\phi(\vec{r})=\frac{1}{4\pi \epsilon} \int d^3\vec{r'}  \rho(\vec{r'}) \frac{1}{|\vec{r}-\vec{r'}|},
\label{eq:app_potential_macroscopic}
\end{align}
where $\rho(\vec{r})$ is the volume charge density. Now, consider dividing the volume of the macroscopic material into microscopic voxels, as shown in Fig.~S~\ref{fig:app_macroscopic_multipole}.
Then, the expression for the potential can be written as
\begin{align}
\phi(\vec{r})=\frac{1}{4\pi \epsilon} \sum_{\vec{R}} \int_{v(\vec{R})} d^3\vec{r'} \frac{\rho(\vec{r'}+\vec{R})}{|\vec{r}-\vec{R}-\vec{r'}|},
\end{align}
where the integral is over each voxel centered at $\vec{R}$ and with volume $v(\vec{R})$, such that 
\begin{align}
\sum_{\vec{R}} \int_{v(\vec{R})} d^3\vec{r'}=V.
\end{align}
Since the voxels are much smaller than the overall material, we have that $|\vec{r'}| \ll |\vec{r}-\vec{R}|$ \textit{as long as $\vec{r}$ is outside of the material} (notice that this is a weaker condition than in the far limit expansion, Section \ref{sec:multipole_expansion_far_limit}, where we require the distance between the observation point and the object to be much larger than the object's dimensions). Thus, let us expand the potential as follows
\begin{align}
\phi(\vec{r})= \sum_{l=0}^\infty \phi^l(\vec{r})
\end{align}
where
\begin{align}
\phi^l(\vec{r}) = \frac{1}{4\pi \epsilon} \sum_{\vec{R}} \int_{v(\vec{R})} d^3\vec{r'} \rho(\vec{r'}+\vec{R}) \frac{|\vec{r'}|^l}{|\vec{r}-\vec{R}|^{l+1}}P_l\left( \frac{\vec{r}-\vec{R}}{|\vec{r}-\vec{R}|} \cdot \frac{\vec{r'}}{|\vec{r'}|} \right),
\end{align}
and $P_l(x)$ are the Legendre polynomials. Here the contributions to the total potential are, up to octupole moment,
\begin{align}
\phi^0(\vec{r})&=\frac{1}{4 \pi \epsilon} \sum_{\vec{R}} Q(\vec{R}) \frac{1}{|\vec{r}-\vec{R}|}\nonumber \\
\phi^1(\vec{r})&=\frac{1}{4 \pi \epsilon} \sum_{\vec{R}} P_i(\vec{R}) \frac{r_i-R_i}{|\vec{r}-\vec{R}|^3}\nonumber  \\
\phi^2(\vec{r})&=\frac{1}{4 \pi \epsilon} \sum_{\vec{R}} Q_{ij}(\vec{R}) \frac{3 (r_i-R_i) (r_j-R_j)-|\vec{r}-\vec{R}|^2 \delta_{ij}}{2|\vec{r}-\vec{R}|^5}\nonumber  \\
\phi^3(\vec{r})&=\frac{1}{4 \pi \epsilon} \sum_{\vec{R}} O_{ijk}(\vec{R}) \frac{5 (r_i-R_i) (r_j-R_j) (r_k-R_k) - 3 |\vec{r}-\vec{R}|^2 \delta_{ij} (r_k-R_k)}{2|\vec{r}-\vec{R}|^7},
\end{align}
 where
 \begin{align}
 Q(\vec{R})&=\int_{v(\vec{R})} d^3\vec{r'} \rho(\vec{r'}+\vec{R})\nonumber \\
P_i(\vec{R})&=\int_{v(\vec{R})} d^3\vec{r'} \rho(\vec{r'}+\vec{R}) r_i'\nonumber \\
Q_{ij}(\vec{R})&=\int_{v(\vec{R})} d^3\vec{r'} \rho(\vec{r'}+\vec{R}) r_i' r_j'\nonumber \\
O_{ijk}(\vec{R})&=\int_{v(\vec{R})} d^3\vec{r'} \rho(\vec{r'}+\vec{R}) r_i' r_j' r_k'.
 \end{align}
are the charge, dipole, quadrupole and octupole moments at the voxel centered at $\vec{R}$. 
If the voxels are very small compared to the material as a whole, we treat $\vec{R}$ as a continuum, then we define the multipole moment densities
 \begin{align}
 \rho(\vec{R})&=\frac{1}{v(\vec{R})} \int_{v(\vec{R})} d^3\vec{r'} \rho(\vec{r'}+\vec{R})\nonumber \\
p_i(\vec{R})&=\frac{1}{v(\vec{R})} \int_{v(\vec{R})} d^3\vec{r'} \rho(\vec{r'}+\vec{R}) r_i'\nonumber \\
q_{ij}(\vec{R})&=\frac{1}{v(\vec{R})} \int_{v(\vec{R})} d^3\vec{r'} \rho(\vec{r'}+\vec{R}) r_i' r_j'\nonumber \\
o_{ijk}(\vec{R})&=\frac{1}{v(\vec{R})} \int_{v(\vec{R})} d^3\vec{r'} \rho(\vec{r'}+\vec{R}) r_i' r_j' r_k'.
\label{eq:app_multipole_moment_densities}
 \end{align}
to write the potentials as
\begin{align}
\phi^0(\vec{r})&=\frac{1}{4 \pi \epsilon} \int_{V}d^3\vec{R}  \left(\rho(\vec{R}) \frac{1}{|\vec{r}-\vec{R}|} \right)\nonumber \\
\phi^1(\vec{r})&=\frac{1}{4 \pi \epsilon} \int_{V}d^3\vec{R} \left(p_i(\vec{R}) \frac{r_i-R_i}{|\vec{r}-\vec{R}|^3}\right)\nonumber  \\
\phi^2(\vec{r})&=\frac{1}{4 \pi \epsilon} \int_{V}d^3\vec{R} \left(q_{ij}(\vec{R}) \frac{3 (r_i-R_i) (r_j-R_j)-|\vec{r}-\vec{R}|^2 \delta_{ij}}{2|\vec{r}-\vec{R}|^5}\right)\nonumber  \\
\phi^3(\vec{r})&=\frac{1}{4 \pi \epsilon} \int_{V}d^3\vec{R} \left(o_{ijk}(\vec{R}) \frac{5 (r_i-R_i) (r_j-R_j) (r_k-R_k) - 3 |\vec{r}-\vec{R}|^2 (r_k-R_k) \delta_{ij}}{2|\vec{r}-\vec{R}|^7}\right),
\label{eq:app_potential_multipole_moment_densities}
\end{align}
where $V$ is the total volume of the macroscopic material. The potential $\phi^0(\vec{r})$ is due to the free `coarse-grained' charge density of Eq. \ref{eq:app_multipole_moment_densities}. In the limit of $v(\vec{R}) \to 0$, this coarse grained charge density is the original continuous charge density, and we recover the original expression \eqref{eq:app_potential_macroscopic}. In this case, all other multipole contributions identically vanish. We now look at the boundaries manifestations of materials having non-vanishing multipole moments by calculating their contributions to the potential. We do this for each moment density separately.

\subsubsection{Dipole moment}
\begin{figure}
\centering
\includegraphics[width=.2\columnwidth]{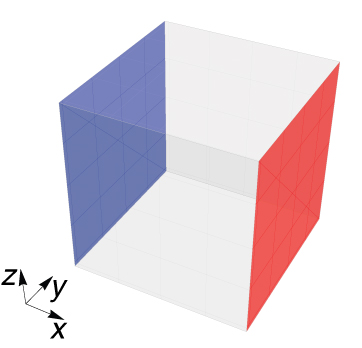}
\caption{\emph{Boundary charge in a material with uniform dipole moment per unit volume $p_x = e$, $p_y=p_z=0$}. Red (blue) color represents positive (negative) charge density per unit area of magnitude $e$.}
\label{fig:app_dipole}
\end{figure}

Let us define the vector
\begin{align}
\vec{\rr} = \vec{r} - \vec{R} 
\end{align}
which points from the a point in the material to the observation point. Then the potential due to a dipole moment density $p_i(\vec{R})$ of Eq. \ref{eq:app_potential_multipole_moment_densities} is given by
\begin{align}
\phi^1(\vec{r}) = \frac{1}{4 \pi \epsilon} \int_V d^3\vec{R}\left(p_i(R)\frac{\rr_i}{\rr^3}\right)
\end{align}
For short, in what follows we refer to the multipole moment densities without their arguments, i.e., we will simply write $p_i$ for $p_i(\vec{R})$, etc. Now, we use
\begin{align}
\frac{\partial}{\partial R_i} \frac{1}{\rr} \equiv \partial_i \frac{1}{\rr}&=\frac{\rr_i}{\rr^3}
\end{align}
to write the potential due to a dipole moment per unit volume $p_i$ as 
\begin{align}
\phi^1(\vec{r}) = \frac{1}{4 \pi \epsilon} \int_V d^3\vec{R}  \left( p_i \partial_i \frac{1}{\rr} \right).
\end{align}
The expression in parenthesis can be decomposed as
\begin{align}
 p_i \left(\partial_i\frac{1}{\rr}\right) = \partial_i \left(p_i \frac{1}{\rr} \right) - \left(\partial_i p_i \right) \frac{1}{\rr},
\end{align}
where $\partial_i$ in $\partial_i p_i$ acts on the arguments of $p_i(\vec{R})$; furthermore, since summation is implied, $\partial_i p_i$ is short notation for the divergence of $\vec{p}(\vec{R})$: $\vec{\nabla} \cdot \vec{p}(\vec{R})$. We use this expression to write the potential as
\begin{align}
\phi^1(\vec{r}) = \frac{1}{4 \pi \epsilon} \int_V d^3\vec{R}  \left[ \partial_i \left(p_i \frac{1}{\rr} \right) - \left(\partial_i p_i \right) \frac{1}{\rr} \right].
\end{align}
Using the divergence theorem on the first term we have
\begin{align}
\phi^1(\vec{r}) = \frac{1}{4 \pi \epsilon} \oint_{dV} d^2\vec{R}  \left( n_i p_i \frac{1}{\rr} \right) + \frac{1}{4 \pi \epsilon} \int_V d^3\vec{R} \left(-\partial_i p_i \frac{1}{\rr} \right).
\end{align}
where $dV$ is the surface of the boundary of the material. To gain clarity, let us rewrite this expression in terms of the original variables
\begin{align}
\phi^1(\vec{r}) = \frac{1}{4 \pi \epsilon} \oint_{dV} d^2\vec{R}  \left( n_i p_i \frac{1}{|\vec{r}-\vec{R}|} \right) + \frac{1}{4 \pi \epsilon} \int_V d^3\vec{R} \left(-\partial_i p_i \frac{1}{|\vec{r}-\vec{R}|} \right).
\end{align}
Since both terms scale as $1/|\vec{r}-\vec{R}|$, where $|\vec{r}-\vec{R}|$ is the distance from a point in the material to the observation point, we can define the charge densities
\begin{align}
\rho &= -\partial_i p_i\nonumber \\
\sigma &= n_i p_i
\label{eq:app_charge_densities_dipole}
\end{align}
The first term is the volume charge density due to a divergence in the polarization, and the second is the area charge density on the boundary of a polarized material. The manifestation of the dipole at the boundary is thus the existence of charge, as shown in Fig.~S~\ref{fig:app_dipole}. 

\subsubsection{Quadrupole moment}
\begin{figure}
\centering
\includegraphics[width=.6\columnwidth]{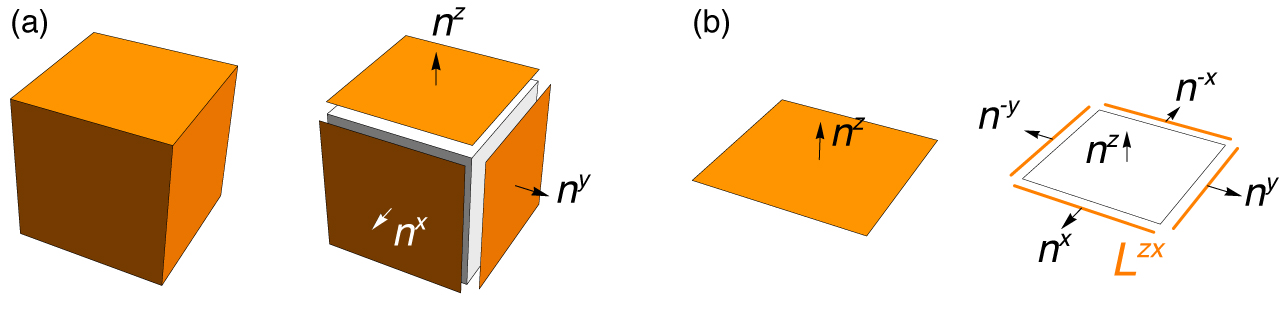}
\caption{\emph{Boundary segmentation for the calculation of quadrupole signatures}. (a) Separation of 2-dimensional boundary into its flat faces. (b) Separation of a 1-dimensional boundary into its straight lines.}
\label{fig:app_boundaries}
\end{figure}

Making the change of variables $\vec{\rr} = \vec{r} - \vec{R}$, as in the previous section, the potential due to a quadrupole moment per unit volume $q_{ij}$ of Eq. \ref{eq:app_potential_multipole_moment_densities} is 
\begin{equation}
\phi^2(\vec{r}) = \frac{1}{4 \pi \epsilon} \int_V d^3\vec{R} \left(q_{ij}\frac{3 \rr_i \rr_j - \rr^2 \delta_{ij}}{2\rr^5} \right).
\end{equation}
We make use of
\begin{align}
\partial_j \partial_i \frac{1}{\rr} &= \frac{3 \rr_i \rr_j - \rr^2\delta_{ij}}{\rr^5}
\end{align}
to write the potential as
\begin{align}
\phi^2(\vec{r}) = \frac{1}{4 \pi \epsilon} \int_V d^3\vec{R}  \left( \frac{1}{2} q_{ij} \partial_i \partial_j \frac{1}{\rr} \right).
\end{align}
Let us rearrange this expression. We use
\begin{align}
q_{ij} \partial_i \partial_j \frac{1}{\rr} &= \partial_i \left( q_{ij} \partial_j \frac{1}{\rr} \right) - \left(\partial_i q_{ij} \right) \partial_i \frac{1}{\rr}\nonumber \\
&= \partial_i \partial_j \left( q_{ij} \frac{1}{\rr} \right) - 2 \partial_i \left[ \left(\partial_j q_{ij} \right) \frac{1}{\rr} \right] + \left( \partial_i \partial_j q_{ij} \right) \frac{1}{\rr}
\end{align}
in the previous expression to get
\begin{equation}
\phi^2(\vec{r}) = \frac{1}{4 \pi \epsilon} \int_V d^3\vec{R} \left[ \frac{1}{2} \partial_i \partial_j \left(  q_{ij} \frac{1}{\rr} \right) - \partial_i \left[ \left(\partial_j q_{ij} \right) \frac{1}{\rr} \right] + \left( \frac{1}{2} \partial_i \partial_j q_{ij} \right) \frac{1}{\rr}  \right].
\end{equation}
Applying the divergence theorem on the first two terms we have
\begin{equation}
\phi^2(\vec{r}) = 
\frac{1}{4 \pi \epsilon} \oint_{dV} d^2\vec{R} \left[ \frac{1}{2} n_i \partial_j \left( q_{ij}  \frac{1}{\rr} \right) \right] 
+\frac{1}{4 \pi \epsilon} \oint_{dV} d^2\vec{R} \left(- n_i \partial_j q_{ij} \right)  \frac{1}{\rr}
+\frac{1}{4 \pi \epsilon} \int_V d^3\vec{R} \left(\frac{1}{2}\partial_j \partial_i q_{ij}\right) \frac{1}{\rr}.
\label{eq:app_quad_potential_1}
\end{equation}
Now consider the boundary of the material, which is a closed surface, as consisting of flat faces (e.g. a cube). At the intersection of different faces, the normal vector is discontinuous. To go around that problem we can break the integral over the entire boundary as a sum over the faces that compose it, as seen in Fig.~S~\ref{fig:app_boundaries}a,
\begin{align}
\oint_{dV} d^2\vec{R}  \left[ \frac{1}{2} n_i \partial_j \left(q_{ji}\frac{1}{\rr} \right) \right] 
&= \sum_\alpha \int_{S_\alpha} d^2\vec{R} \left[ \frac{1}{2} n^\alpha_i\partial_j \left(q_{ji}\frac{1}{\rr} \right) \right].
\end{align}
For the sake of clarity, we have explicitly written the sum over the flat faces $S_\alpha$ with normal vector $n^\alpha_i$. Now, we apply the divergence theorem over the open surfaces $S_\alpha$. We thus have
\begin{align}
\oint_{dV} d^2\vec{R}  \left[ \frac{1}{2} n^\alpha_i\partial_j \left( q_{ji}\frac{1}{\rr} \right) \right] 
&= \sum_{\alpha,\beta} \int_{L_{\alpha,\beta}} d\vec{R} \left(\frac{1}{2} n_j n_i q_{ji} \right) \frac{1}{\rr},
\end{align}
where $L_{\alpha,\beta}$ is the one-dimensional boundary of $S_\alpha$ when it meets $S_\beta$ (see Fig.~S~\ref{fig:app_boundaries}b). Joining the pieces together, the contributions to the potential from a quadrupole moment are
\begin{align}
\phi^2(\vec{r}) &=
\frac{1}{4 \pi \epsilon} \sum_{\alpha,\beta} \int_{L_{\alpha,\beta}} d\vec{R} \left( \frac{1}{2} n_i n_j q_{ij}  \right) \frac{1}{\rr}
+\frac{1}{4 \pi \epsilon} \sum_\alpha \int_{S_\alpha} d^2\vec{R}  \left( - \partial_j n_i q_{ij}  \right) \frac{1}{\rr}\nonumber\\ 
&+\frac{1}{4 \pi \epsilon} \int_V d^3\vec{R} \left( \frac{1}{2}\partial_j \partial_i q_{ij}\right) \frac{1}{\rr},
\label{eq:app_quad_potential_3}
\end{align}
Since all the potentials scale as $1/\rr$, where $\vec{\rr}=\vec{r}-\vec{R}$ is the distance from the point in the material to the observation point, all the expression in parethesis can be interpreted as charge densities, thus, we define the charge densities
\begin{align}
\rho &= \frac{1}{2}\partial_j \partial_i q_{ij}\nonumber \\
\sigma^\alpha &= - \partial_j \left( n^\alpha_i  q_{ij} \right)\nonumber \\
\lambda^{\alpha,\beta} &= \frac{1}{2} \sum_{\substack{\alpha',\beta'=\alpha,\beta \\ \alpha' \neq \beta'}}n^{\alpha'}_i n^{\beta'}_j q_{ij}.
\end{align}
The sums in the last equation amount to permutations of the indices $\alpha$ and $\beta$. There are two contributions to the sum, which are equal, since $q_{xy}=q_{yx}$. Thus, simplifying these contributions we have
\begin{align}
\rho &= \frac{1}{2}\partial_j \partial_i q_{ij}\nonumber \\
\sigma^\alpha &= - \partial_j \left( n^\alpha_i  q_{ij} \right)\nonumber \\
\lambda^{\alpha,\beta} &= n^{\alpha}_i n^{\beta}_j q_{ij}.
\end{align}

\begin{figure}
\centering
\includegraphics[width=.4\columnwidth]{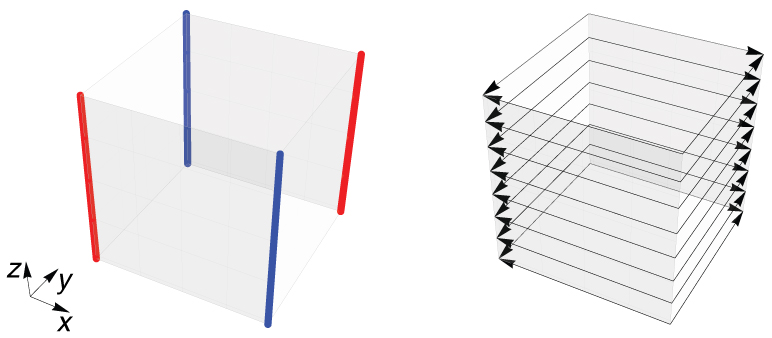}
\caption{\emph{Boundary charge (left) and polarization (right) of a cube with uniform quadrupole moment per unit volume $q_{xy}=e$, $q_{yz}=q_{zx}=0$}. Red (blue) color represents positive (negative) charge densities per unit length of magnitude $e$. Arrows represent boundary polarization per unit area of magnitude $e$.}
\label{fig:app_quadrupole}
\end{figure}

The first term is the direct contribution of the quadrupole moment density to the volume charge density in the bulk of the material. The second term is the area charge density at the boundary surfaces of the material due to a divergence in the quantity $n^\alpha_i q_{ij}$, which exists only at the surfaces of the material perpendicular to the unit vector ${\bf n}^\alpha$. Finally, the third term is the length charge density at the hinges of the material. For a cube or square with constant quadrupole moment $q_{xy}$ the charges are shown in Fig.~S~\ref{fig:app_quadrupole}, as indicated by the expression for $\lambda$. Notice that the expression for the surface charge density $\sigma$ could be written as
\begin{align}
\sigma^\alpha &= - \partial_j p^{S_\alpha}_j,
\end{align} 
where  $p^{S_\alpha}_j = n^\alpha_i  q_{ij}$. This expression resembles the one for the volume charge density $\rho$ in Eq. \ref{eq:app_charge_densities_dipole}. Thus, we interpret $p^{S_\alpha}_j$ as a polarization density (per unit area). Notice, however, that this polarization exists only on the surface of the boundary perpendicular to ${\bf n}^\alpha$. Furthermore, this is a vectorial field that runs parallel to that surface. An illustration of this for a cube with constant quadrupole moment is shown in Fig.~S~\ref{fig:app_quadrupole}.

\subsubsection{Octupole moment}
\begin{figure}%
\centering
\includegraphics[width=.6\columnwidth]{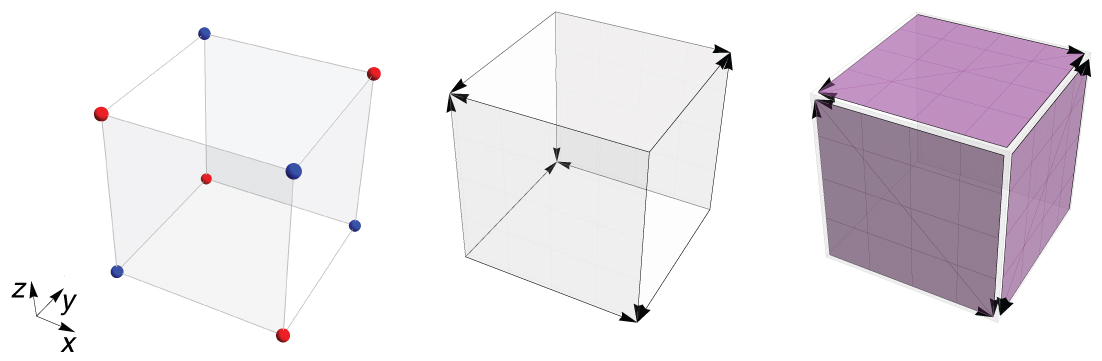}
\caption{\emph{Boundary charge (left), polarization (center), and quadrupole (right) of a cube with uniform octupole moment per unit volume $o_{ijk}=e$ for any permutation of $i,j,k$ such that $i \neq j \neq k$ or zero else}. Red (blue) color represents positive (negative) corner charges of magnitude $e$, arrows represent polarization per unit length of magnitude $e$, and purple squares represent boundary quadrupole moment per unit area of magnitude $e$.}
\label{fig:app_octupole}
\end{figure}

Making the change of variables $\vec{\rr} = \vec{r} - \vec{R}$, as in the previous sections, the potential due to a octupole moment per unit volume $o_{ijk}$ of Eq. \ref{eq:app_potential_multipole_moment_densities} is 
\begin{align}
\phi^3(\vec{r}) = \frac{1}{4 \pi \epsilon} \int_V d^3\vec{R} \left(o_{ijk} \frac{5 \rr_i \rr_j \rr_k - 3 \rr^2 \delta_{ij} \rr_k}{2 \rr^7} \right)
\label{eq:app_potential_multipole_densities}
\end{align}
Using the expression
\begin{align}
\partial_k \partial_j \partial_i \frac{1}{\rr} &= 3\frac{5 \rr_i \rr_j \rr_k - 3 \rr^2 \delta_{ij} \rr_k}{\rr^7} 
\end{align}
we write the potential as
\begin{align}
\phi^3(\vec{r}) = \frac{1}{4 \pi \epsilon} \int_V d^3\vec{R} \left( \frac{1}{6} o_{ijk} \partial_i \partial_j \partial_k \frac{1}{\rr} \right).
\end{align}
If we want to find the potential exclusively as arising from charge distributions, we can proceed as before, the result is
\begin{align}
\phi^3(\vec{r}) &= \frac{1}{4 \pi \epsilon} \int_V d^3\vec{R} \left(-\frac{1}{6} \partial_i \partial_j \partial_k o_{ijk} \right) \frac{1}{\rr}
+ \frac{1}{4 \pi \epsilon} \sum_\alpha \int_{S_\alpha} d^2\vec{R} \left( \frac{1}{2} n^\alpha_i \partial_j \partial_k o_{ijk} \right) \frac{1}{\rr}\nonumber\\
&+ \frac{1}{4 \pi \epsilon} \sum_{\alpha,\beta} \int_{L_{\alpha,\beta}} d\vec{R} \left( -\frac{1}{2} n^\alpha_i n^\beta_j \partial_k o_{ijk} \right) \frac{1}{\rr}
+\frac{1}{4 \pi \epsilon} \sum_{\alpha,\beta,\gamma} \frac{1}{6} n^\alpha_i n^\beta_j n^\gamma_k o_{ijk} \frac{1}{r},
\end{align}
from which we read off the charge densities
\begin{align}
\rho &= -\frac{1}{6} \partial_i \partial_j \partial_k o_{ijk}\nonumber \\
\sigma^\alpha &= \frac{1}{2} n^\alpha_i \partial_j \partial_k o_{ijk}\nonumber \\
\lambda^{\alpha,\beta} &= -\frac{1}{2} \sum_{\substack{\alpha',\beta' = \alpha,\beta \\ \alpha' \neq \beta'}}n^{\alpha'}_i n^{\beta'}_j \partial_k o_{ijk}\nonumber \\
\delta^{\alpha,\beta,\gamma} &= \frac{1}{6} \sum_{\substack{\alpha',\beta',\gamma' = \alpha,\beta,\gamma \\ \alpha' \neq \beta' \neq \gamma'}} n^{\alpha'}_i n^{\beta'}_j n^{\gamma'}_k o_{ijk}.
\end{align}
As before, the sums amount to permutations, which exactly cancel the prefactors. Thus, we are left with
\begin{align}
\rho &= -\frac{1}{6} \partial_i \partial_j \partial_k o_{ijk}\nonumber \\
\sigma^\alpha &= \frac{1}{2} n^\alpha_i \partial_j \partial_k o_{ijk}\nonumber \\
\lambda^{\alpha,\beta} &= -n^{\alpha}_i n^{\beta}_j \partial_k o_{ijk}\nonumber \\
\delta^{\alpha,\beta,\gamma} &= n^{\alpha}_i n^{\beta}_j n^{\gamma}_k o_{ijk}.
\end{align}
Notice the consistency in these expressions when compared with the expressions for dipole and quadrupole moments. We could re-write them as
\begin{align}
\rho &= -\frac{1}{6} \partial_i \partial_j \partial_k o_{ijk}\nonumber \\
\sigma^\alpha &= \frac{1}{2} \partial_j \partial_k q^{S_\alpha}_{jk}\nonumber \\
\lambda^{\alpha,\beta} &= -\partial_k  p^{L_{\alpha,\beta}}_k\nonumber \\
\delta^{\alpha,\beta,\gamma} &= n^{\alpha}_i n^{\beta}_j n^{\gamma}_k o_{ijk},
\end{align}
where 
\begin{align}
q_{jk}^{S_\alpha} &= n^\alpha_i o_{ijk}\nonumber\\
p_k^{L_{\alpha,\beta}} &= n^{\alpha}_i n^{\beta}_j o_{ijk}
\end{align}
are the quadrupole density per unit area on faces perpendicular to ${\bf n}^\alpha$ and the polarization density per unit length on hinges perpendicular to both ${\bf n}^\alpha$ and ${\bf n}^\beta$, respectively. These manifestations at the boundary are illustrated in Fig.~S~\ref{fig:app_octupole} for a cube with uniform octupole moment.

\section{Corner modes in the minimal model with quadrupole moment}
In this section we show that the protected topological corner mode is a simultaneous eigenstate of \emph{both} edge Hamiltonians. Let us begin with our lattice Hamiltonian with $\lambda=1:$
\begin{equation}
H=\sin k_x \Gamma_3+(\gamma_x+\cos k_x)\Gamma_4+\sin k_y\Gamma_1 +(\gamma_y+\cos k_y)\Gamma_2.
\end{equation} One can find an analytic solution for the boundary Hamiltonians even for the lattice model, however to simplify discussion we will solve a continuum version by assuming that $\gamma_x=-1+m_x$ and $\gamma_y=-1+m_y$ for $m_{x/y}$ small and positive (negative) for the topological (trivial) phase. We can take a continuum limit, or equivalently a $k\cdot P$ expansion about $(k_x, k_y)=0,$ to find the Hamiltonian
\begin{equation}
H= k_x \Gamma_3+m_x\Gamma_4 +k_y\Gamma_1+ m_y\Gamma_2.
\end{equation} 

Let us now focus on the upper right corner. We will treat the $x$-edge as a domain wall where $m_x$ steps from positive (inside the topological phase) to negative (outside the topological phase), and the $y$-edge as a domain wall where $m_y$ steps from positive to negative. 
Let us solve for the domain wall low-energy modes on an $x$-edge. We find the equation
\begin{equation}
(-i\partial_x\Gamma_3+m_x(x)\Gamma_4)\Psi(x)=0
\end{equation} and we can use the ansatz $\Psi(x)=\exp (\int_{0}^{x} m_{x}(x')dx')\Phi_{x}$ for some constant spinor $\Phi_{x}.$ The matrix equation can be simplified to $(\mathbb{I}-\tau^{z}\otimes \sigma^z)\Phi_{x}=0$ which is solved by choosing $\Phi_{x}$ to be a positive eigenstate of $\tau^{z}\otimes\sigma^{z},$ i.e., $\Phi_{x1}=(1,0,0,0)$ or $\Phi_{x2}=(0,0,0,1).$ We can project these two degenerate states onto the rest of the Hamiltonian to find the edge Hamiltonian 
\begin{equation}
H_{edge,\hat{x}}=-k_y \mu^y+m_y\mu^x
\end{equation} where $\mu^{a}$ are Pauli matrices in the basis $(\Phi_{x1}, \Phi_{x2}).$

Performing an analogous calculation for an edge with normal vector in the $y$-direction we find the matrix equation
\begin{equation}
(\mathbb{I}-\mathbb{I}\otimes\sigma^z)\Phi_{y}=0
\end{equation} which has solutions which are positive eigenstates of $\mathbb{I}\otimes\sigma^z,$ i.e., $\Phi_{y1}=(1,0,0,0)$ or $\Phi_{y2}=(0,0,1,0).$ To find the edge Hamiltonian we project the remaining bulk terms into this degenerate basis. We find
\begin{equation}
H_{edge,\hat{y}}=-k_x \gamma^y+m_x\gamma^x
\end{equation} where $\gamma^{a}$ are Pauli matrices in the basis $(\Phi_{y1}, \Phi_{y2}).$

Both of these edge Hamiltonians take the form of massive 1+1d Dirac models, i.e., the natural minimal continuum model for a 1+1d topological insulator. Now the key feature we mentioned earlier, i.e., the simultaneous zero mode can be found by considering a corner, i.e., either an $x$-edge with a $y$ domain wall or a $y$-edge with an $x$ domain wall. Using similar calculations to those above we find the following two matrix equations for the former two configurations respectively:
\begin{eqnarray}
(\mathbb{I}-\mu^z)\phi_{x,y}&=&0\\
(\mathbb{I}-\gamma^z)\phi_{y,x}&=&0.
\end{eqnarray}\noindent Hence the corner mode we find for an $x$-edge with a $y$ domain wall is the positive eigenstate of $\mu^z$ while that for the $y$-edge with an $x$ domain wall is the positive eigenstate of $\gamma^z.$ Remarkably we find that these solutions are identical, i.e., in both cases we find the solution is the first basis element which is $\Phi_{x1}=\Phi_{y1}=(1,0,0,0).$ Hence, the corner zero mode is a simultaneous zero mode of both domain wall Hamiltonians, and thus there only has to be a single mode at the corner. Thus, we can conclude that both edges are topological but they only need produce a single zero mode.

\section{Robustness of the corner charge to disorder}
\begin{figure}[h]
\centering
\includegraphics[width=.7\columnwidth]{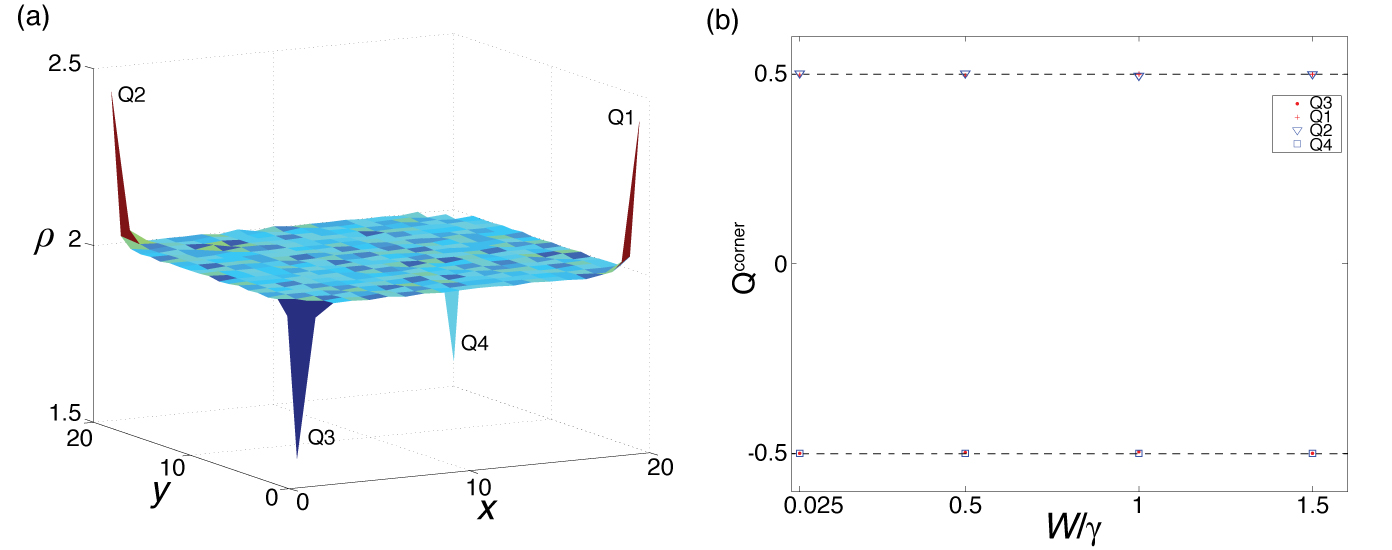}
\caption{\emph{Electron density of quadrupole model with onsite potential disorder having uniform probability distribution}. (a) Electron density distribution $\rho$ for system size $L_x=L_y=20,$ $\gamma/\lambda=1/10$ and $W/\gamma=3/2.$ Density was averaged over 500 disorder realizations.  Quadrupole structure is still clearly apparent, as well as clear density fluctuations in the bulk. (b) Corner charges on the corners $Q1, Q2, Q3,$ and $Q4$ for different values of $W/\gamma.$ Charges still appear sharply quantized.}
\label{fig:disorder}
\end{figure}

The observable properties of the quadrupole are robust, even outside of the pristine crystalline limit. To support this claim, we performed calculations with disorder. We add onsite potential disorder, which naturally breaks both mirror symmetries, and in fact, all symmetries that might protect the quadrupole. We find that even when the bulk and edge are disordered, the corner charges remain sharply quantized. Our calculation shows that our predictions are robust, even when the mirror symmetries are only protected ``on average." For these simulations we used a system size of $L_x=L_y=20$, $\gamma/\lambda=1/10,$ $W/\gamma=1/40, 1/2, 1,$ and $3/2.$ The disorder distribution was uniformly drawn from $[-W/2, W/2].$ In Fig.~S~\ref{fig:disorder}a we show a density plot for the case $W/\gamma=3/2$ averaged over 500 disorder realizations. The quadrupole charge structure is clearly apparent. Furthermore, in Fig.~S~\ref{fig:disorder}b we plot the averaged corner charges for each value of $W$ and find that they remain nearly quantized. All corner charges were calculated averaging over 500 disorder realizations. 

\section{Wannier functions, Wannier centers and polarization in one-dimensional crystals}
\label{sec:app_WF_1d}
The expansion in multipole moments above tells us that one-dimensional crystals admit only the existence of a dipole moment. In insulators, the electronic contribution to the polarization arises from the fact that the electrons are displaced with respect to the atomic positive charges. In this section we diagonalize the electronic projected position operator in one-dimensional crystals and construct the Wannier centers and functions which allows us to calculate the electronic polarization. This follows the presentation in Refs.~\onlinecite{Alexandradinata2014,yu2011}.

The position operator is in a crystal with $N$ unit cells and $N_{orb}$ orbitals per unit cell is
\begin{align}
\hat{x}=\sum_{R,\alpha} c^\dagger_{R,\alpha}\ket{0} e^{-i \Delta_k (R+r_\alpha)}  \bra{0} c_{R,\alpha},
\end{align}
where $\alpha \in 1\ldots N_{orb}$ labels the orbital, $R \in 1 \ldots N$ labels the unit cell, $r_\alpha$ is the position of orbital $\alpha$ relative to a center in the unit cell, and $\Delta_k=2\pi/N$. Consider the discrete Fourier transform
\begin{align}
c_{R,\alpha} &= \frac{1}{\sqrt{N}}\sum_k e^{-i k(R+r_\alpha)}c_{k,\alpha}\nonumber \\
c_{k,\alpha} &= \frac{1}{\sqrt{N}}\sum_R e^{i k(R+r_\alpha)}c_{R,\alpha},
\end{align}
where $k \in \Delta_k(0, 1, \ldots N-1)$. We impose the boundary conditions
\begin{align}
c_{R+N, \alpha}=c_{R,\alpha} \rightarrow c_{k+G,\alpha} = e^{i Gr_\alpha}c_{k,\alpha},
\label{eq:app_creation_operator_boundary_conditions}
\end{align}
where $G$ is a reciprocal lattice vector. In this new basis, we can alternatively write the position operator as
\begin{align}
\hat{x}&=\sum_{k,\alpha} c^\dagger_{k+\Delta_{k},\alpha} \ket{0} \bra{0} c_{k,\alpha},
\end{align}
as well as the second quantized Hamiltonian
\begin{align}
H=\sum_k c^\dagger_{k,\alpha} [h_k]^{\alpha,\beta} c_{k,\beta},
\label{eq:app_Hamiltonian}
\end{align}
where summation is implied over repeated orbital indices. Due to the periodicity \eqref{eq:app_creation_operator_boundary_conditions} the Hamiltonian $h_k$ obeys
\begin{align}
h_{k+G} = V^{-1}(G) h_k V(G),
\label{eq:app_hamiltonian_boundary_conditions}
\end{align}
where
\begin{align}
[V(G)]^{\alpha,\beta}=e^{-iGr_\alpha}\delta_{\alpha,\beta}.
\end{align}
We diagonalize this Hamiltonian as
\begin{align}
[h_k]^{\alpha,\beta} = \sum_n [u^n_k]^\alpha \epsilon_{n,k} [u^{*n}_k]^\beta,
\end{align}
where $[u^n_k]^\alpha$ is the $\alpha$ component of the eigenstate $u^n_k$. To enforce the periodicity \eqref{eq:app_hamiltonian_boundary_conditions}, we impose the periodic gauge
\begin{align}
[u^n_{k+G}]^\alpha = [V^{-1}(G)]^{\alpha,\beta} [u^n_k]^\beta.
\end{align}
This diagonalization allows to write Eq.~\ref{eq:app_Hamiltonian} as
\begin{align}
H = \sum_{n,k} \gamma^\dagger_{n,k} \epsilon_{n,k} \gamma_{n,k},
\end{align}
where 
\begin{align}
\gamma_{n,k} = \sum_\alpha [u^{*n}_k]^\alpha c_{k,\alpha}.
\end{align}
The projection into occupied bands is
\begin{align}
P^{occ}=\sum_{n=1}^{N_{occ}} \sum_k \gamma^\dagger_{n,k}\ket{0} \bra{0}\gamma_{n,k}
\end{align}
where $N_{occ}$ is the number of occupied energy bands. From now on we assume that summations over bands include only occupied bands.
We now proceed to diagonalize the position operator projected into the subspace of occupied bands
\begin{align}
P^{occ} \hat{x} P^{occ} = \sum_{n,k}\sum_{n',k'} \gamma^\dagger_{n,k}\ket{0} \left( \sum_{q,\alpha}  \bra{0}\gamma_{n,k} c^\dagger_{q+\Delta_k,\alpha}\ket{0} \bra{0}c_{q,\alpha} \gamma^\dagger_{n',k'}\ket{0} \right) \bra{0}\gamma_{n',k'}.
\end{align}
From above we have
$\bra{0} \gamma_{n,k}c^\dagger_{q,\alpha}\ket{0} = [u^{*n}_k]^\alpha \delta_{k,q}$, so the projected position operator reduces to
\begin{align}
P^{occ} \hat{x} P^{occ} &= \sum_{m,n=1}^{N_{occ}}\sum_{k} \gamma^\dagger_{m,k+\Delta_k}\ket{0} \braket{u^n_{k+\Delta_k}}{u^m_k} \gamma_{n,k}\bra{0}
\label{eq:app_projected_x}
\end{align}
where we have adopted the notation $\braket{u^n_q}{u^m_k}=\sum_\alpha [u^{*n}_q]^\alpha [u^m_k]^\alpha$ (we warn that $\braket{u^n_k}{u^m_q} \neq \delta_{n,m} \delta_{k,q}$ in general. They do obey $\braket{u^n_k}{u^m_k} = \delta_{n,m}$ though).
The matrix $G_k$ with components $[G_k]^{mn}=\braket{u^n_{k+\Delta_k}}{u^m_k}$ is not unitary due to the discretization of $k$. It is unitary in the thermodynamic limit. To make it unitary for finite $N$, consider the singular value decomposition
\begin{align}
G = U D V^\dagger,
\end{align}
where $D$ is a diagonal matrix. The failure of G to be unitary is manifest in the fact that the (real valued) singular values along the diagonal of $D$ are less than 1. Therefore, we define
\begin{align}
F = U V^\dagger
\end{align}
which is unitary. We refer to $F_k$ as the Wilson line element at $k$. In the thermodynamic limit $[F_k]^{mn}=\braket{u^m_k}{u^n_k}$.
To diagonalize the projected position operator, let us write the eigenvalue problem 
\begin{align}
(P\hat{x}P) \ket{\Psi^{j}} = E^j \ket{\Psi^{j}},
\end{align}
which, in the basis $\gamma_{n,k} \ket{0}$, adopts the following form
\begin{equation}
\left( \begin{array}{ccccc}
0 & 0 & 0 & \ldots & F_{k_N}\\
F_{k_1} & 0 & 0 & \ldots & 0\\
0 & F_{k_2} & 0 & \ldots & 0\\
\vdots & \vdots & \vdots & \ddots & \vdots\\
0 & 0 & 0 & \ldots & 0\\
\end{array} \right)
\left( \begin{array}{c}
\nu_{k_1}\\
\nu_{k_2}\\
\nu_{k_3}\\
\vdots\\
\nu_{k_N}\\
\end{array} \right)^j=
E^j
\left( \begin{array}{c}
\nu_{k_1}\\
\nu_{k_2}\\
\nu_{k_3}\\
\vdots\\
\nu_{k_N}\\
\end{array} \right)^j,
\end{equation}
where $k_1=0$, $k_2 = \Delta_k$, $\dots$, $k_N= \Delta_k (N-1)$, and $j \in 1\ldots N_{occ}$.
Here we have replaced $G_k$ in Eq.~\ref{eq:app_projected_x} by $F_k$ to restore the unitary character of the Wilson line elements. By recursive application of the equations above, one can obtain the relation
\begin{align}
\W_{k_f \leftarrow k_i} \ket{\nu^j_{k_i}} = (E^j)^{(k_f-k_i)/\Delta_k} \ket{\nu^j_{k_f}},
\label{eq:app_parallel_transport_1}
\end{align}
where we are adopting the braket notation $\ket{\nu^j_{k_l}}$ for the vector formed by the collection of values $[\nu^j_{k_l}]^n$, for $n \in 1 \ldots N_{occ}$. We define the discrete Wilson line as
\begin{align}
\W_{k_f \leftarrow k_i} =F_{k_f-\Delta_k} F_{k_f-2\Delta_k} \ldots F_{k_i+\Delta_k} F_{k_i}
\end{align}
For a large Wilson loop, i.e. a Wilson line that goes across the entire Brillouin zone (from now on, by Wilson loop we refer exclusively to large Wilson loops), Eq.~\ref{eq:app_parallel_transport_1} results in the eigenvalue problem
\begin{align}
\W_{k+2\pi \leftarrow k} \ket{\nu^j_k} = (E^j)^N \ket{\nu^j_k}.
\end{align}
Here, the subscript $k$ labels the starting point, or \textit{base point}, of the Wilson loop. While the Wilson-loop eigenstates depend on the base point, its eigenvalues do not. Furthermore, since the Wilson loop is unitary, its eigenvalues are simply phases
\begin{align}
(E^j)^N = e^{i 2\pi \nu^j}
\end{align}
which has $N$ solutions
\begin{align}
E^{j,R} &= e^{i 2\pi \nu/N + i 2\pi R /N}\nonumber\\
&=e^{i \Delta_k (\nu^j + R)}
\end{align}
for $R \in 0 \ldots N-1$. The phases $\nu^j$ are the Wannier centers. They correspond to the positions of the electrons relative to the center of the unit cells. The eigenfunctions of the Wilson loop at different base points are related to each other by the parallel transport equation
\begin{align}
\ket{\nu^j_{k_f}} = e^{-i(k_f-k_i) \nu^j} \W_{k_f \leftarrow k_i} \ket{\nu^j_{k_i}},
\label{eq:app_parallel_transport_2}
\end{align}
which is a restatement of Eq.~\ref{eq:app_parallel_transport_1}. Since $j \in 1 \ldots N_{occ}$ and $R \in 0 \ldots N-1$, there are as many Wilson-loop eigenstates and eigenvalues as there are states in the occupied bands. Given the normalized Wilson-loop eigenstates, the eigenstates of the projected position, which now reads as
\begin{align}
(P^{occ}\hat{x}P^{occ}) \ket{\Psi^j_R} = e^{i \Delta_k (\nu^j + R)}\ket{\Psi^j_R}
\end{align}
are
\begin{align}
\ket{\Psi^j_R} = \frac{1}{\sqrt{N}} \sum_{n=1}^{N_{occ}}\sum_k \gamma^\dagger_{nk}\ket{0} \left[ \nu^j_k \right]^n e^{-i k R}
\label{eq:app_MLWF_1D}
\end{align}
and are known as the \textit{Wannier functions} (WF). Here, $j \in 1 \ldots N_{occ}$ labels the WF and $R \in 0\ldots N-1$ identifies the unit cell. These states obey
\begin{align}
\braket{\Psi^i_{R_1}}{\Psi^j_{R_2}} = \delta_{i,j} \delta_{R_1,R_2},
\end{align}
i.e., they form an orthonormal basis of the Hamiltonian. 

\subsection{Polarization}
As the Wannier centers $\nu^j$ correspond to the positions of the electrons within the unit cell, the electronic contribution to the polarization, measured as the electron charge times the displacement of the electrons from the center of the unit cell is proportional to
\begin{align}
p=\sum_j \nu^j.
\end{align}
This expression is true for any unit cell due to translation invariance, and thus it is a bulk property of the crystal.
Since the Wannier centers are the phases of the eigenvalues of the Wilson loop, we can write the polarization as
\begin{align}
p=-\frac{i}{2\pi} \mbox{Log } \mbox{Det} \left[ \W_{k+2\pi \leftarrow k}\right].
\end{align}
Furthermore, in the thermodynamic limit, if we write the Wilson loop in terms of the potential $[\A_k]^{mn}= -i \bra{u^m_k}\partial_k \ket{u^n_k}$, we have
\begin{align}
p&=-\frac{i}{2\pi} \text{Log } \text{Det} \left[ e^{-i \int_k^{k+2\pi} \A_k dk}\right]\nonumber\\
&=-\frac{1}{2\pi} \int_k^{k+2\pi} \tr [\A_k] dk,
\label{eq:app_polarization_continuum_1d}
\end{align}
which is the well known expression for the polarization in the modern theory of polarization \cite{King-SmithVanderbilt1993}.

\section{Polarization, Wannier bands, and Quadrupole invariant in two-dimensional crystals }
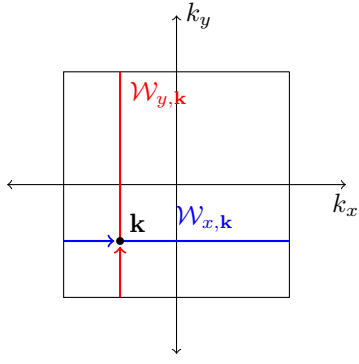
\begin{figure}
\centering
\begin{tikzpicture}[scale=1.5]

\draw[](1,1)--(-1,1)--(-1,-1)--(1,-1)--(1,1);
\draw[<->,black](-1.5,0)--(1.5,0) node[below]{$k_x$};
\draw[<->,black](0,-1.5)--(0,1.5) node[right]{$k_y$};

\draw[->,blue,thick](-1,-0.5)--(-0.55,-0.5);
\draw[blue,thick](-0.5,-0.5)--node[above]{$\W_{x,{\bf k}}$}(1,-0.5);

\draw[red,thick](-0.5,-.5)--(-0.5,1)node[below right]{$\W_{y,\bf k}$};
\draw[->,red,thick](-0.5,-1)--(-0.5,-0.55);

\fill [black] (-0.5,-0.5) circle (1pt) node[above right] {${\bf k}$};

\end{tikzpicture}
\caption{\emph{Wilson loops along $x$ and $y$, both with base point ${\bf k}$.}}
\label{fig:app_Wilson_loops}
\end{figure}

We now explore the topological properties of the WF in two dimensions. We first construct WF along $x$. Repeating the procedure in SM \ref{sec:app_WF_1d} but for two-dimensional crystals, the projected position operator results in 
\begin{align}
P^{occ} \hat{x} P^{occ} &= \sum_{k} \gamma^\dagger_{m,(k_x+\Delta_{k_x}, k_y)}\ket{0} \braket{u^n_{(k_x+\Delta_{k_x},k_y)}}{u^m_{(k_x,k_y)}} \gamma_{n,(k_x,k_y)}\bra{0}
\end{align}
which is similar to Eq.~\ref{eq:app_projected_x}, but with the extra quantum number $k_y$. Importantly, notice that the operator is diagonal in $k_y$. Thus, all the findings in Section \ref{sec:app_WF_1d} follow through in this case too, but with the extra label $k_y$.
The WF along $x$ are
\begin{align}
\ket{\Psi^j_{R_x,k_y}} = \frac{1}{\sqrt{N_x}} \sum_{n=1}^{N_{occ}}\sum_{k_x} \gamma^\dagger_{n,\bf k}\ket{0} \left[ \nu^j_{x,{\bf k}} \right]^n e^{-i k_x R_x}
\label{eq:app_MLWF_2D}
\end{align}
where ${\bf k}=(k_x,k_y)$ is the crystal momentum, with $k_{x,y} = n_{x,y} \Delta_{k_{x,y}}$, for $n_{x,y} \in 0,1,\ldots, N_{x,y}-1$ and $\Delta_{k_{x,y}}=2\pi/N_{x,y}$. 
These functions obey
\begin{align}
\braket{\Psi^j_{R_x,k_y}}{\Psi^{j'}_{R'_x,k'_y}} = \delta_{j,j'} \delta_{R_x,R'_x} \delta_{k_y,k'_y}
\end{align}
i.e., they form an orthonormal basis of the Hamiltonian. 
For the Wilson-loop eigenstates $\ket{\nu^j_{x,{\bf k}}}$, the subscript $x$ specifies the direction of its Wilson loop, and ${\bf k}$ specifies its base point, so, for example,
\begin{align}
\W_{(k_x+2\pi,k_y) \leftarrow (k_x,k_y)} \ket{\nu^j_{x,(k_x,k_y)}} = e^{i 2\pi \nu^j_x(k_y)} \ket{\nu^j_{x,(k_x,k_y)}}.
\end{align}
For short, we write the above expression as
\begin{align}
\W_{x,{\bf k}} \ket{\nu^j_{x,{\bf k}}} = e^{i 2\pi \nu^j_x(k_y)} \ket{\nu^j_{x,{\bf k}}}.
\end{align}
A depiction of these Wilson loops is shown in Fig.~S~\ref{fig:app_Wilson_loops}. Although the phases $\nu^j_x(k_y)$ of the eigenvalues of the Wilson-loop $\W_{x,\bf k}$ do not depend on $k_x$, in general they do depend on $k_y$. Thus, the polarization for one-dimensional crystals translates into polarization as a function of $k_y$ in its two-dimensional counterpart, that is
\begin{align}
p_x(k_y) &=\sum_j \nu^j_x(k_y) = -\frac{i}{2\pi} \text{Log Det}[\W_{x,\bf k}],
\end{align}
and the total polarization along $x$ is
\begin{align}
p_x = \frac{1}{N_y} \sum_{k_y}p_x(k_y).
\end{align}
In the thermodynamic limit, and making $\frac{1}{N_y} \sum_{k_y} \to \frac{1}{2\pi}\int dk_y$, the expression in Eq.~\ref{eq:app_polarization_continuum_1d} for one-dimensional crystals translates into
\begin{align}
p_x &=-\frac{1}{(2\pi)^2} \int_{BZ} \tr [\A_{x,\bf k}] d^2\bf k
\end{align}
for two-dimensional crystals. Here $[\A_{x,\bf k}]^{mn}= -i \bra{u^m_{\bf k}} \partial_{k_x} \ket{u^n_{\bf k}}$ is the non-Abelian Berry potential (where $m,n$ run over occupied energy bands) and $BZ$ is the two-dimensional Brillouin zone. 

\subsection{Wannier bands}
As stated above, the phases of these Wilson-loop eigenvalues $\nu^j_x(k_y)$ in general depend on $k_y$. For example, for an insulator with a single occupied band having Chern number $n$, $\nu_x(k_y)$ winds around $k_y \in (-\pi,\pi]$ an integer $n$ times. A similar result can be proved to time-reversal symmetric topological insulators \cite{yu2011}. Here, however, we focus our attention on the cases in which the Wannier values $\nu^j_x(k_y)$ are gapped across the entire 1-dimensional BZ $k_y \in (-\pi,\pi]$. If that is the case, we can define two \textit{Wannier bands}
\begin{align}
\nu^-_x &= \{\nu^j_x(k_y), \mbox{ s.t. } \nu^j_x(k_y) \mbox{ is below the Wannier gap} \}\nonumber \\
\nu^+_x &= \{\nu^j_x(k_y), \mbox{ s.t. } \nu^j_x(k_y) \mbox{ is above the Wannier gap} \}
\end{align}
We then choose those above or below the gap and form the projector
\begin{align}
P_{\nu_x}&=\sum_{j=1}^{N_W}\sum_{R_x, k_y} \ket{\Psi^j_{R_x,k_y}}\bra{\Psi^j_{R_x,k_y}}\nonumber\\
&=\sum_{j=1}^{N_W} \sum_{n,m=1}^{N_{occ}} \sum_{\bf k} \gamma^\dagger_{n,\bf k}\ket{0}[\nu^j_{x,\bf k}]^n [\nu^{*j}_{x,\bf k}]^m \bra{0} \gamma_{m,\bf k},
\end{align}
where $\sum_j^{N_W}$ is a summatory over all Wannier bands in the sector $\nu_x$, for $\nu_x = \nu^+_x$ or $\nu^-_x$. $N_W$ is the number of Wannier bands in sector $\nu_x$. $R_x \in 0 \ldots N_x-1$ labels the unit cells, and $k_y=\Delta_{k_y} n_y$, for $\Delta_{k_y}=2\pi / N_y$ and $n_y \in 0,1, \ldots, N_y-1$ is the crystal momentum along $y$. We are interested in studying the topological properties of the subspace spanned by $P_{\nu_x}$ across $k_y \in (-\pi,\pi]$. As explained in the main text of this article, the topology over the Wannier sectors is related to the topology of the edge Hamiltonian. In particular, we want to diagonalize the position operator $\hat{y}$ projected into this subspace. 
The position operator along $y$ is
\begin{align}
\hat{y} &= \sum_{{\bf R}, \alpha} c^\dagger_{{\bf R},\alpha} \ket{0}e^{-i \Delta_{k_y} R_y} \bra{0} c_{{\bf R},\alpha}\nonumber \\
&=\sum_{k_x,k_y,\alpha} c^\dagger_{k_x,k_y+\Delta_{k_y},\alpha} \ket{0} \bra{0} c_{k_x,k_y,\alpha}.
\end{align}
We now calculate the position operator $\hat{y}$ projected into the Wannier sector $\nu_x$
\begin{align}
P_{\nu_x} \hat{y} P_{\nu_x} =&\sum_{j,j'=1}^{N_W}\sum_{\bf k} \sum_{n,m,n',m'=1}^{N_{occ}} \gamma^\dagger_{n,{\bf k}+{\bf \Delta_{k_y}}}\ket{0}  \bra{0}\gamma_{n',\bf k}\nonumber\\
&\left([\nu^j_{x,{\bf k}+{\bf \Delta_{k_y}}}]^n [\nu^{*j}_{x,{\bf k}+{\bf \Delta_{k_y}}}]^m \braket{u^m_{{\bf k}+{\bf \Delta_{k_y}}}}{u^{m'}_{\bf k}}[\nu^{j'}_{x,{\bf k}}]^{m'}[\nu^{j'*}_{x,{\bf k}}]^{n'} \right).
\end{align}
To simplify the notation let us define the \textit{Wannier basis}
\begin{align}
\ket{w^j_{x,\bf k}} = \sum_{n=1}^{N_{occ}}\ket{u^n_{\bf k}} [\nu^j_{x,\bf k}]^n
\label{eq:app_Wannier_basis}
\end{align}
for $j \in 1\ldots N_W$. This basis obeys,
\begin{align}
\braket{w^j_{x,\bf k}}{w^{j'}_{x, \bf k}} = \delta_{j,j'}.
\end{align}
However, in general $\braket{w^j_{x,\bf k}}{w^{j'}_{x, \bf q}} \neq \delta_{j,j'} \delta_{\bf k, \bf q}$.
The projected position operator then reduces to
\begin{align}
P_{\nu_x} \hat{y} P_{\nu_x} =&\sum_{j,j'=1}^{N_W}\sum_{\bf k} \sum_{n,n'=1}^{N_{occ}} \gamma^\dagger_{n,{\bf k}+{\bf \Delta_{k_y}}}\ket{0}  \bra{0}\gamma_{n',\bf k}\nonumber\\
&\left([\nu^j_{x,{\bf k}+{\bf \Delta_{k_y}}}]^n \braket{w^j_{x,{\bf k}+{\bf \Delta_{k_y}}}}{w^{j'}_{x,\bf k}}[\nu^{j'*}_{x,{\bf k}}]^{n'} \right),
\end{align}
which has the same structure in $k_y$ as Eq.~\ref{eq:app_projected_x} had in $k$. Notice, however, that the operator is diagonal in $k_x$. Explicitly,
\begin{align}
P_{\nu_x} \hat{y} P_{\nu_x} =&\sum_{k_x,k_y} \sum_{n,n'=1}^{N_{occ}} \gamma^\dagger_{n,(k_x,k_y+ \Delta_{k_y})}\ket{0} [F^{\nu_x}_{y,(k_x,k_y)}]^{n,n'} \bra{0}\gamma_{n',(k_x,k_y)},
\end{align}
where
\begin{align}
 [F^{\nu_x}_{y,(k_x,k_y)}]^{n,n'}=\sum_{j,j'=1}^{N_W}[\nu^j_{x,(k_x,k_y+\Delta_{k_y})}]^n \braket{w^j_{x,(k_x,k_y+\Delta_{k_y})}}{w^{j'}_{x,(k_x,k_y)}}[\nu^{j'*}_{x,{(k_x,k_y)}}]^{n'}.
\end{align}
To diagonalize $P_{\nu_x} \hat{y} P_{\nu_x} $, we calculate the Wilson loop along y
\begin{align}
[{\W}^{\nu_x}_{y, \bf k}]^{n,n'} &= F^{\nu_x}_{y,{\bf k}+N_y {\bf \Delta_{k_y}}} \ldots F^{\nu_x}_{y,{\bf k}+ {\bf \Delta_{k_y}}} F^{\nu_x}_{y,{\bf k}}\nonumber\\
&= [\nu^j_{x,{\bf k}+N_y{\bf \Delta_{k_y}}}]^n [\tilde\W^{\nu_x}_{y,\bf k}]^{j,j'}[\nu^{j'*}_{x,{\bf k}}]^{n'}\nonumber\\
&= [\nu^j_{x,\bf k}]^n [\tilde\W^{\nu_x}_{y,\bf k}]^{j,j'}[\nu^{j'*}_{x,{\bf k}}]^{n'},
\end{align}
$\tilde\W^{\nu_x}_{y,\bf k}$ is the Wilson loop along $y$ over the Wannier sector $\nu_x$ performed over the Wannier basis,
\begin{align}
[\tilde\W^{\nu_x}_{y,\bf k}]^{j,j'} =& \braket{w^j_{x,{\bf k}+N_y{\bf \Delta_{k_y}}}}{w^{r}_{x,{\bf k}+(N_y-1) {\bf \Delta_{k_y}}}} \bra{w^{r}_{x,{\bf k}+(N_y-1) {\bf \Delta_{k_y}}}} \ldots \nonumber\\
&\dots \ket{w^{s}_{x,{\bf k}+{\bf \Delta_{k_y}}}}\braket{w^s_{x,{\bf k}+{\bf \Delta_{k_y}}}}{w^{j'}_{x,\bf k}}.
\label{eq:app_Wilson_loop_Wannier_basis}
\end{align}
In the expression above, summation is implied over repeated indices $r,\ldots,s \in 1 \ldots N_W$ over all Wannier bands in the Wannier sector $\nu_x$. Since $N_W<N_{occ}$, this Wilson loop is done over a subspace \textit{within} the subspace of occupied energy bands. \textbf{In general, we will indicate an operator written in a Wannier basis with a tilde, while no tilde indicates that it is written in the basis of energy bands}. Since we have used $\nu_x$ as the label for the Wannier bands along $x$, we will use the labels $\nu^{\nu_x}_y$ for the eigenvalues and eigenvectors for the Wilson-loop along $y$ over sector $\nu_x$.  This Wilson loop diagonalizes as
\begin{align}
\tilde\W^{\nu_x}_{y,\bf k}  \ket{\nu^{\nu_x,j}_{y,\bf k}} = e^{i 2\pi \nu^{\nu_x,j}_y(k_x)} \ket{\nu^{\nu_x,j}_{y,\bf k}}
\end{align}
for $j \in 1 \ldots N_W$.
The polarization over the Wannier sector $\nu_x$ is then given by the sum of all the $N_W$ phases $\nu^{\nu_x}_y(k_x)$. This can be written as
\begin{align}
p^{\nu_x}_y(k_x) = -\frac{i}{2\pi} \text{Log Det}[\tilde\W^{\nu_x}_{y,\bf k}],
\end{align}
and the total polarization over the Wannier bands $\nu_x$ in the thermodynamic limit is
\begin{align}
p^{\nu_x}_y &=-\frac{1}{(2\pi)^2} \int_{BZ} \tr [\tilde\A^{\nu_x}_{y,\bf k}] d^2\bf k
\label{eq:app_polarization_Wannier_sector}
\end{align}
where $[\tilde\A^{\nu_x}_{y,\bf k}]^{j,j'} = -i \bra{w^j_{x,\bf k}} \partial_{k_y} \ket{w^{j'}_{x,\bf k}}$ is the Berry potential over Wannier bands $\nu_x$, therefore $j,j' \in 1 \ldots N_W$ run over the Wannier bands with value $\nu_x$. This last expression is Eq.~10 in the main text.

\subsection{Relation between Berry potentials}
Let us define the matrix $G_{x,\bf k}$ as
\begin{align}
[G_{x,\bf k}]^{nj} = [\nu^j_{x,\bf k}]^n\nonumber\\
[G^\dagger_{x,\bf k}]^{jn} = [\nu^{j*}_{x,\bf k}]^n
\end{align}
which translates between the basis of energy states and the basis of Wannier states,
\begin{align}
\ket{w^j_{x,\bf k}} = \sum_{n=1}^{N_{occ}}\ket{u^n_{\bf k}} [G_{x,\bf k}]^{nj}.
\end{align}
The $G$ matrix obeys $[G^\dagger G]^{j,j'} = \delta_{j,j'}$, and relates the Wilson loops and Berry potentials as follows
\begin{align}
\W^{\nu_x}_{y,{\bf k}} &= G_{x,\bf k} \tilde\W^{\nu_x}_{y,{\bf k}} G^\dagger_{x,\bf k}\\
\tilde\A^{\nu_x}_{y,\bf k} &= G^\dagger_{x,{\bf k}} \A_{y,\bf k} G_{x,{\bf k}} -i G^\dagger_{x,{\bf k}} \partial_{k_y} G_{x,{\bf k}}.
\end{align}
So, the potential over the $\nu_x$ sector takes the form of a gauge transform of the potential over occupied bands, but, importantly, $G$ is in general not a square matrix, since the components $[G]^{nj}$ exist for $n \in 1 \ldots N_{occ}$ and $j \in 1 \ldots N_W$, where $N_W<N_{occ}$. The polarization over the Wannier sector $\nu_x$ can be written in the form
\begin{align}
p^{\nu_x}_y &=-\frac{1}{(2\pi)^2} \int_{BZ} \tr [P^{\nu_x}_{\bf k} \A_{y,\bf k} + a^{\nu_x}_{y,\bf k}] d^2\bf k,
\end{align}
where
\begin{align}
P^{\nu_x}_{\bf k} &= \sum_{j=1}^{N_W}\ket{\nu^j_{x,\bf k}} \bra{\nu^j_{x,\bf k}}\nonumber\\
&= G_{x,\bf k} G^\dagger_{x,\bf k}
\end{align}
and
\begin{align}
a^{\nu_x}_{y, \bf k}=-i G_{x,\bf k}^\dagger \partial_{k_y} G_{x,\bf k}.
\end{align}
is the Berry potential over the occupied Wannier bands. 

\subsection{Gauge transformations on Wilson loops}
Consider a gauge transformation that expresses  $\ket{u'^n_{\bf k}}$ in terms of $\ket{u^n_{\bf k}}$ by
\begin{align}
\ket{u'^m_{\bf k}} &= \ket{u^n_{\bf k}} \braket{u^n_{\bf k}}{u'^m_{\bf k}}\nonumber \\
&= \ket{u^n_{\bf k}} [U^\dagger_{\bf k}]^{nm},
\end{align}
where $[U^\dagger_{\bf k}]^{nm}=\braket{u^n_{\bf k}}{u'^m_{\bf k}}$ is the unitary matrix that transforms between the two bases. We want to see how the change in $\ket{u^n_{\bf k}}$ for $\ket{u'^n_{\bf k}}$ affects the Wilson loops written in the Wannier basis $\ket{w^i_{\bf k}}$. Using the expression above and its hermitian transpose we have that a Wilson line element transforms as
\begin{align}
\W'_{{\bf k}_2 \leftarrow {\bf k}_1} = U_{{\bf k}_2} \W_{{\bf k}_2 \leftarrow {\bf k}_1} U^\dagger_{{\bf k}_1},
\end{align}
where $\W'_{{\bf k}_2 \leftarrow {\bf k}_1}$ is the Wilson line element in the new basis. It follows that the Wilson loop transforms as
\begin{align}
\W'_{x,{\bf k}} = U_{{\bf k}} \W_{x,{\bf k}} U^\dagger_{{\bf k}}.
\label{eq:app_Wilson_loop_under_gauge_transformation}
\end{align}
Now, consider the Wilson-loop eigenstates
\begin{align}
\W_{x,{\bf k}} \ket{\nu^i_{x,{\bf k}}} = e^{i 2 \pi \nu^i_x(k_y)}\ket{\nu^i_{x,{\bf k}}},
\end{align}
using Eq.~\ref{eq:app_Wilson_loop_under_gauge_transformation} we see that
\begin{align}
\W'_{x,{\bf k}} U_{\bf k} \ket{\nu^i_{x,{\bf k}}} &= U_{\bf k} \W_{x,{\bf k}} \ket{\nu^i_{x,{\bf k}}} \nonumber \\
&= e^{i 2\pi \nu^i_x(k_y)} U_{\bf k}  \ket{\nu^i_{x,{\bf k}}}.
\label{eq:app_gauge_transformation_1}
\end{align}
Thus, $U_{\bf k} \ket{\nu^i_{x,{\bf k}}}$ is an eigenstate of $\W'_{x,{\bf k}}$ with same eigenvalue $e^{i 2\pi \nu^i_x(k_y)}$ as $\ket{\nu^i_{x,\bf k}}$. As such, we can express it as a superposition of the eigenstates of $\W'_{x,{\bf k}}$, as follows:
\begin{align}
U_{\bf k} \ket{\nu^i_{x,{\bf k}}} = \ket{\nu'^j_{x,{\bf k}}} \alpha^{ji}_{{\bf k}},
\label{eq:app_gauge_transformation_2}
\end{align}
where $\alpha^{ji}_{{\bf k}} = \braket{\nu'^j_{x,{\bf k}}}{\nu^i_{x,{\bf k}}}$ is the matrix that connects the Wilson loop eigenstates in the two bases. It has values $\alpha^{ji}_{\bf k} \neq 0$ only if $\nu^j_x(k_y) = \nu^i_x(k_y)$.

Finally, we look at how the Wilson loop in the $\ket{w^i_{x,{\bf k}}}$ basis transforms,

\begin{align}
\ket{w'^i_{x,{\bf k}}} &= \ket{u'^n_{\bf k}} [\nu'^i_{x,{\bf k}}]^n\nonumber \\
&= \ket{u^m_{\bf k}}[U^\dagger]^{mn} [\nu'^i_{x,{\bf k}}]^n\nonumber \\
&= \ket{u^m_{\bf k}} [\nu^j_{x,{\bf k}}]^m [\alpha^\dagger_{\bf k}]^{ji}\nonumber \\
&= \ket{w^j_{x,{\bf k}}} [\alpha^\dagger_{\bf k}]^{ji},
\end{align}
where in the third step we have used Eq.~\ref{eq:app_gauge_transformation_2} in the form $U^\dagger_{\bf k} \ket{\nu'^i_{\bf k}}=\ket{\nu^j_{x,{\bf k}}}[\alpha^\dagger_{\bf k}]^{ji}$. Using this transformation and its hermitian conjugate we see that the Wilson line element transforms as
\begin{align}
\tilde{\W}^{' \nu_x}_{{\bf k}_2 \leftarrow {\bf k}_1} = \alpha_{{\bf k}_2} \tilde{\W}^{\nu_x}_{{\bf k}_2 \leftarrow {\bf k}_1} \alpha^\dagger_{{\bf k}_1}
\end{align}
and, for the Wilson loop
\begin{align}
\tilde{\W}^{'\nu_x}_{\bf k} = \alpha_{\bf k} \tilde{\W}^{\nu_x}_{\bf k} \alpha^\dagger_{\bf k}.
\label{eq:app_Wilson_of_Wilson_under_gauge_transformation}
\end{align}
If the eigenvalues of the Wilson loop are all distinct, $\alpha_{\bf k}$ is a diagonal matrix of phases, and the Wilson loop over the sector $\nu_x$ is gauge invariant and not gauge covariant. This is different from the gauge transformations over the energy bands of Eq.~\ref{eq:app_Wilson_loop_under_gauge_transformation}.

Rather than starting from the energy bands $\ket{u^n_{\bf k}}$, we can perform a gauge transformation directly on the Wilson-loop eigenstates
\begin{align}
\ket{\nu^{'i}_{x,\bf k}} = \ket{\nu^{j}_{x,\bf k}} [\alpha^\dagger_{\bf k}]^{ji},
\end{align}
which also leads to Eq.~\ref{eq:app_Wilson_of_Wilson_under_gauge_transformation}.

\section{Symmetry constraints on Wilson loops}

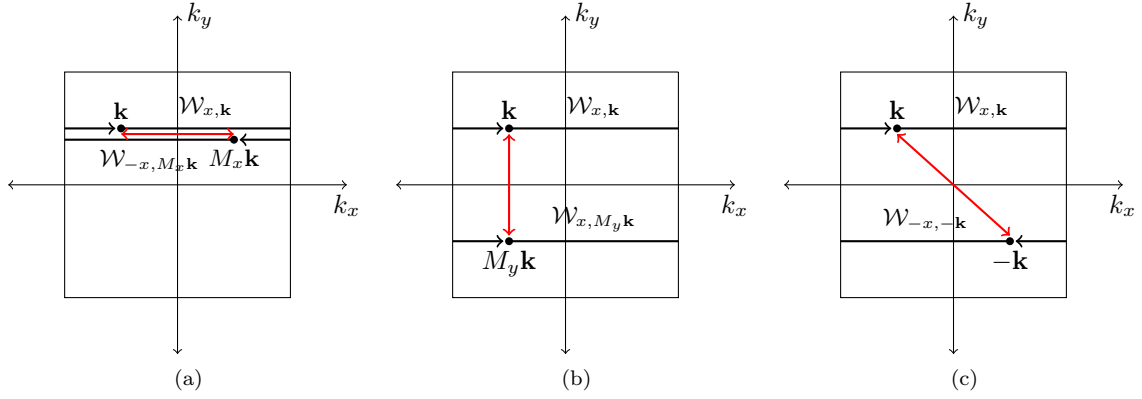
\begin{figure}
\centering
\subfigure[]{
\begin{tikzpicture}[scale=1.5]

\draw[](1,1)--(-1,1)--(-1,-1)--(1,-1)--(1,1);
\draw[<->,black](-1.5,0)--(1.5,0) node[below]{$k_x$};
\draw[<->,black](0,-1.5)--(0,1.5) node[right]{$k_y$};

\fill [black] (-0.5,0.5) circle (1pt) node[above] {\small ${\bf k}$};
\draw[->,black,thick](-1,0.5)--(-0.55,0.5);
\draw[black,thick](-0.5,0.5)--node[above]{\small $\W_{x,{\bf k}}$}(1,0.5);

\fill [black] (0.5,0.4) circle (1pt) node[below] {\small $M_x{\bf k}$};
\draw[black,thick](-1,0.4)--node[below]{\small $\W_{-x,M_x{\bf k}}$}(0.5,0.4);
\draw[<-,black,thick](0.55,0.4)--(1,0.4);

\draw[<->,red, thick](-0.5,0.45)--(0.5,0.45);

\end{tikzpicture}
}
\subfigure[]{
\begin{tikzpicture}[scale=1.5]

\draw[](1,1)--(-1,1)--(-1,-1)--(1,-1)--(1,1);
\draw[<->,black](-1.5,0)--(1.5,0) node[below]{$k_x$};
\draw[<->,black](0,-1.5)--(0,1.5) node[right]{$k_y$};

\fill [black] (-0.5,0.5) circle (1pt) node[above] {${\bf k}$};
\draw[->,black,thick](-1,0.5)--(-0.55,0.5);
\draw[black,thick](-0.5,0.5)--node[above]{\small $\W_{x,{\bf k}}$}(1,0.5);

\fill [black] (-0.5,-0.5) circle (1pt) node[below] {$M_y{\bf k}$};
\draw[->,black,thick](-1,-0.5)--(-0.55,-0.5);
\draw[black,thick](-0.5,-0.5)--node[above]{\small $\W_{x, M_y{\bf k}}$}(1,-0.5);

\draw[<->,red, thick](-0.5,0.45)--(-0.5,-0.45);

\end{tikzpicture}
}
\subfigure[]{
\begin{tikzpicture}[scale=1.5]

\draw[](1,1)--(-1,1)--(-1,-1)--(1,-1)--(1,1);
\draw[<->,black](-1.5,0)--(1.5,0) node[below]{$k_x$};
\draw[<->,black](0,-1.5)--(0,1.5) node[right]{$k_y$};

\fill [black] (-0.5,0.5) circle (1pt) node[above] {${\bf k}$};
\draw[->,black,thick](-1,0.5)--(-0.55,0.5);
\draw[black,thick](-0.5,0.5)--node[above]{\small $\W_{x,{\bf k}}$}(1,0.5);

\fill [black] (0.5,-0.5) circle (1pt) node[below] {$-{\bf k}$};
\draw[black,thick](-1,-0.5)--node[above]{\small $\W_{-x, -{\bf k}}$}(0.5,-0.5);
\draw[<-,black,thick](0.55,-0.5)--(1,-0.5);

\draw[<->,red, thick](-0.5,0.45)--(0.5,-0.45);

\end{tikzpicture}
}
\caption{\emph{Symmetry constraints on Wilson loops}. Relation between Wilson loops along $x$ at base point ${\bf k}$ upon (a) reflection along $x$, (b) reflection along $y$, (c) inversion.}
\label{fig:app_Wx_transformations}
\end{figure}

Insulators with a lattice symmetry obey
\begin{align}
g_{\bf k} h_{\bf k} g_{\bf k}^\dagger = h_{D_g{\bf k}},
\label{eq:app_Hamiltonian_under_symmetry}
\end{align}
where $g_{\bf k}$ is the unitary operator
\begin{align}
g_{\bf k} = e^{-i(D_g {\bf k}). {\bf s}}U_g.
\end{align}
$U_g$ is an $N_{orb} \times N_{orb}$ matrix that acts on the internal degrees of freedom of the unit cell, and $D_g$ is an operator in momentum space sending ${\bf k} \rightarrow D_g {\bf k}$. In real space, on the other hand, we have ${\bf r} \rightarrow D_g{\bf r} + {\bf s}$, for $\bf s=0$ in the case of symmorphic symmetries or fractionary (in unit cell units) in the case of non-symmorphic symmetries. The state $g_{\bf k} \ket{u^n_{\bf k}}$ is an eigenstate of $h_{D_g{\bf k}}$ with energy $\epsilon_{n,{\bf k}}$, as can be seen as follows:

\begin{align}
h_{D_g{\bf k}} g_{\bf k} \ket{u^n_{\bf k}} &= g_{\bf k} h_{\bf k} \ket{u^n_{\bf k}} \nonumber\\
&= \epsilon_{n,{\bf k}} g_{\bf k} \ket{u^n_{\bf k}}.
\end{align}
Therefore, one can expand $g_{\bf k} \ket{u^n_{\bf k}}$ in terms of the eigenbasis of $h_{D_g {\bf k}}$:
\begin{align}
g_{\bf k} \ket{u^n_{\bf k}} &= \ket{u^m_{D_g{\bf k}}} \matrixel{u^m_{D_g{\bf k}}}{g_{\bf k}}{u^n_{\bf k}}\nonumber\\
&=\ket{u^m_{D_g{\bf k}}}B^{mn}_{g,\bf k},
\label{eq:app_sewing_matrix_expansion}
\end{align}
where, from now on, summation is implied for repeated band indices \text{only} over occupied bands.
\begin{align}
B^{mn}_{g,\bf k}=\matrixel{u^m_{D_g{\bf k}}}{g_{\bf k}}{u^n_{\bf k}}
\end{align}
is the unitary sewing matrix that connects states at ${\bf k}$ with those at $D_g{\bf k}$ which have the same energy. 
Using the expansion in Eq. \ref{eq:app_sewing_matrix_expansion}, we can write
\begin{align}
\ket{u^n_{\bf k}}=g_{\bf k}^\dagger \ket{u^m_{D_g {\bf k}}}B^{mn}_{g,\bf k} 
\end{align}
So, an element of a Wilson line from ${\bf k_1}$ to ${\bf k_2}$ is equal to
\begin{align}
\W^{mn}_{{\bf k_2} \leftarrow {\bf k_1}} &= \braket{u^m_{\bf k_2}}{u^n_{\bf k_1}} \nonumber \\
&= B^{\dagger mr}_{g,\bf k_2}\matrixel{u^r_{D_g {\bf k_2}}}{g_{\bf k} g_{\bf k}^\dagger}{u^s_{D_g {\bf k_1}}}B^{sn}_{g,\bf k_1} \nonumber \\
&=  B^{\dagger mr}_{g,\bf k_2} \W^{rs}_{D_g {\bf k_2} \leftarrow D_g {\bf k_1}}B^{sn}_{g,\bf k_1}.
\end{align}
Reording this we have
\begin{align}
B_{g,\bf k_2} \W_{{\bf {\bf k_2}} \leftarrow {\bf k_1}} B^\dagger _{g,\bf k_1} = \W_{D_g{\bf {\bf k_2}} \leftarrow D_g{\bf k_1}}.
\end{align}
In particular, for a Wilson loop at base point ${\bf k}$ we have
\begin{align}
\boxed{B_{g,\bf k} \W_{\C,{\bf k}} B^\dagger_{g,\bf k} = \W_{D_g \C, D_g {\bf k}}}
\label{eq:app_Wilson_loop_under_symmetry}
\end{align}
where $\C$ is the contour along along which the Wilson loop is performed. To simplify notation, from now on we will refer to Wilson loops along the contour $\C =  (k_x,k_y) \rightarrow (k_x + 2 \pi, k_y)$ along increasing (decreasing) values of $k_x$ as $\W_{x,\bf k}$ ($\W_{-x,\bf k}$), where ${\bf k} =(k_x,k_y)$ is the base point of the loop. Similarly, for the path $\C =  (k_x,k_y) \rightarrow (k_x, k_y + 2 \pi)$ along increasing (decreasing) values of $k_y$, we will denote the Wilson loops as $\W_{y,\bf k}$ ($\W_{-y,\bf k}$).
Fig.~S~\ref{fig:app_Wx_transformations} shows how these Wilson loops transform under the three symmetries we will consider here: reflection in $x$, reflection in $y$, and inversion. In what follows, we study the constraints under these symmetries on the Wilson loops over the occupied bands, as well as on the Wilson loops over Wannier sectors.

\subsection{Constraints due to reflection symmetry along $x$}
\label{sec:app_Wilson_loops_under_Mx}
\subsubsection{On the Wilson loop over occupied energy bands}
Under reflection symmetry along $x$, the eigenvalues of Wilson loops along $x$ are constrained to be $+1$, $-1$, or to come in complex-conjugate pairs $e^{\pm i 2 \pi \nu}$ \cite{Alexandradinata2014}, as we will see in this section. Consider a system with reflection symmetry along $x$
\begin{align}
\hat{M}_x h_{\bf k} \hat{M}_x^\dagger = h_{M_x {\bf k}},
\end{align}
where $M_x {\bf k} = M_x (k_x,k_y)=(-k_x, k_y)$. This symmetry allows us to write the expansion
\begin{align}
\ket{u^n_{\bf k}}=\ket{u^m_{M_x {\bf k}}}B^{mn}_{M_x,\bf k},
\end{align}
where 
\begin{align}
B^{mn}_{M_x,\bf k}=\matrixel{u^m_{M_x {\bf k}}}{\hat{M}_x}{u^n_{\bf k}}
\end{align}
is the unitary sewing matrix ($B^\dagger B = B B^\dagger = 1$), which connects states at ${\bf k}$ with states at $M_x {\bf k}$ having the same energy. In particular, $B^{mn}_{M_x,\bf k} \neq 0$ only if $\epsilon_{m, M_x {\bf k}}= \epsilon_{n,{\bf k}}$. 

The relation between Wilson loops in Eq.~\ref{eq:app_Wilson_loop_under_symmetry} for this symmetry is
\begin{align}
B_{M_x,\bf k} \W_{x,{\bf k}} B^\dagger_{M_x,\bf k} = \W_{-x, M_x {\bf k}}=\W^\dagger_{x,M_x{\bf k}}.
\label{eq:app_Wilson_loop_under_Mx}
\end{align}
A schematic of this relation is shown in Fig.~S~\ref{fig:app_Wx_transformations}a. Thus, the Wilson loop at ${\bf k}$ is equivalent (up to a unitary transformation) to the Hermitian conjugate of the Wilson loop at $M_x {\bf k}$. Since the eigenvalues of the Wilson loop along $x$ are $k_x$ independent, this directly imposes a restriction on the allowed Wannier centers at each $k_y$, namely, the set of Wilson-loop eigenvalues must obey
\begin{align}
\left\{ e^{i 2\pi \nu_x^i(k_y)} \right\} = \left\{ e^{-i 2\pi \nu_x^i(k_y)} \right\}.
\label{eq:app_Wilson_loop_eigenvalues_reflection_symmetry}
\end{align}
Thus, the Wannier centers $\nu_x$ are forced to be 0 (centered at unit cell), 1/2 (centered in between unit cells), or to come in pairs $(-\nu,\nu)$ (pairs which are equally displaced from the unit cell but at opposite sides of it). From this it follows that, in order to have a gapped Wannier spectrum, we must have an even number $N_{occ}$ of occupied bands, for if we have an odd $N_{occ}$, at least one of the Wannier centers must have the value $\nu=0$ or $\nu=1/2$.

\subsubsection{On the nested Wilson loop over Wannier sectors}
Now, we focus on the Wilson loop eigenfunctions
\begin{align}
\W_{x,{\bf k}} \ket{\nu^i_{x,\bf k}}=e^{i2\pi \nu_x^i(k_y)}\ket{\nu^i_{x,\bf k}}.
\end{align}
Using Eq.~\ref{eq:app_Wilson_loop_under_Mx}, we have that
\begin{align}
 \W^\dagger_{x,M_x {\bf k}} B_{M_x,\bf k} \ket{\nu^i_{x,\bf k}} &= B_{M_x,\bf k} \W_{x,{\bf k}}  \ket{\nu^i_{x,\bf k}} \nonumber\\
 &= e^{i2\pi \nu_x^i(k_y)} B_{M_x,\bf k} \ket{\nu^i_{x,\bf k}}.
 \label{eq:app_Wilson_loop_eigenstate_under_Mx}
\end{align}
$B_{M_x,\bf k} \ket{\nu^i_{x,\bf k}} $ is hence an eigenfunction of $ \W_{x,M_x {\bf k}}$ with eigenvalue $e^{-i2\pi \nu_x^i(k_y)}$. We now expand this function as
\begin{align}
B_{M_x,\bf k} \ket{\nu^i_{x,\bf k}} = \ket{\nu^j_{x,M_x{\bf k}}}  \alpha^{ji}_{M_x,\bf k},
\label{eq:app_Wilson_loop_eigenstate_expansion_Mx}
\end{align}
where
\begin{align}
 \alpha^{ji}_{M_x,\bf k} = \matrixel{\nu^j_{x,M_x {\bf k}}}{B_{M_x,\bf k}}{\nu^i_{x,\bf k}}
\end{align}
is a unitary sewing matrix that connects Wilson loop eigenstates at base points ${\bf k}$ and $M_x {\bf k}$ having \textit{opposite} Wilson-loop eigenvalues. 
If $ \alpha^{ji}_{M_x,\bf k} \neq 0$, we require that $-\nu_x^j(k_y)=\nu_x^i(k_y)$. In other words, the sewing matrix $\alpha_{M_x,\bf k}$ only connects Wilson-loop eigenstates at base points ${\bf k}$ and $M_x {\bf k}$ having opposite Wannier centers.  This implies that $\alpha_{M_x,\bf k}$ is restricted to be block diagonal in the $\nu=0$ and $\nu=1/2$ sectors and off diagonal \textit{between} the sectors $\nu,-\nu$.

Now, let us act the reflection operator on $\ket{w^j_{x,\bf k}}$, defined in Eq.~\ref{eq:app_Wannier_basis},
\begin{align}
M_x \ket{w^j_{x,\bf k}}&= M_x\ket{u^n_{\bf k}}[\nu^j_{x,\bf k}]^n\nonumber\\
&=\ket{u^m_{M_x\bf k}} \matrixel{u^m_{M_x \bf k}}{M_x}{u^n_{\bf k}}[\nu^j_{x,\bf k}]^n\nonumber\\
&=\ket{u^m_{M_x\bf k}} B^{mn}_{M_x,\bf k} [\nu^j_{x,\bf k}]^n\nonumber\\
&=\ket{u^m_{M_x\bf k}} [\nu^i_{x,M_x\bf k}]^m \alpha^{ij}_{M_x,\bf k}\nonumber\\
&=\ket{w^i_{x,M_x\bf k}} \alpha^{ij}_{M_x,\bf k}.
\label{eq:app_Wannier_basis_under_Mx}
\end{align}
From this relation we can write
\begin{align}
\ket{w^j_{x,\bf k}} &= M_x^\dagger \ket{w^i_{x,M_x \bf k}} \alpha^{ij}_{M_x,\bf k}\nonumber \\
\bra{w^j_{x,\bf k}} &=  [\alpha^\dagger_{M_x,\bf k}]^{ji}  \bra{w^i_{x,M_x {\bf k}}} M_x,
\end{align}
where $\nu_x^i(k_y)=-\nu_x^j(k_y)$ for nonzero $\alpha^{ji}_{M_x,\bf k}$. The Wilson line elements in this basis are related by
\begin{align}
\left[\tilde{\W}^{\nu_x}_{{\bf k}_2 \leftarrow {\bf k}_1}\right]^{ij} &= \braket{w^i_{x,{\bf k}_2}}{w^j_{x,{\bf k}_1}} \nonumber\\
&= [\alpha^\dagger_{M_x,{\bf k}_2}]^{i i'} \braket{w^{i'}_{x,M_x {\bf k}_2}}{w^{j'}_{x,M_x {\bf k}_1}} \alpha^{j' j}_{M_x,{\bf k}_1}\nonumber\\
&= [\alpha^\dagger_{M_x,{\bf k}_2}]^{i i'} \left[\tilde{\W}^{\nu'_x}_{M_x {\bf k}_2 \leftarrow M_x {\bf k}_1}\right]^{i'j'}  \alpha^{j' j}_{M_x,{\bf k}_1}.
\end{align}
In particular, for the Wilson loops along $y$ in the basis $\ket{w^j_{x}}$ we have
\begin{align}
\left[\tilde{\W}^{\nu_x}_{y,{\bf k}} \right]^{ij} &=  \left[\alpha^\dagger_{M_x,\bf k} \right]^{ii'} \left[ \tilde{\W}^{\nu'_x}_{y, M_x {\bf k}} \right]^{i'j'} \left[\alpha_{M_x,\bf k} \right]^{j'j}.
\label{eq:app_Wilson_of_Wilson_under_Mx}
\end{align}
Eq.~\ref{eq:app_Wilson_of_Wilson_under_Mx} implies two things: First, since we have that $\nu^j_x(k_y)=-\nu^i_x(k_y)$ for nonzero $\alpha^{ji}_{M_x,\bf k}$, this expression tells us that Wilson loops along $y$ at base point ${\bf k}$ over Wannier sectors $\nu_x=0$ or $1/2$ are equivalent (up to unitary transformations) to Wilson loops along $y$ over the same Wannier sector at base point $M_x {\bf k}$. Second; suppose that we have gapped Wannier bands $\{\nu_x(k_y),-\nu_x(k_y)\}$ across the entire range $k_y \in (-\pi, \pi]$. This expression also tells us that if we calculate the Wilson loop along $y$ over the Wannier sector $\nu_x(k_y)$ at base point ${\bf k}$, this Wilson loop is equivalent (up to a unitary transformation) to the Wilson loop along $y$ at base point $M_x {\bf k}$ over the sector $-\nu_x(k_y)$.
Thus, for the eigenvalues $\exp[i 2\pi \nu^{\nu_x,j}_y(k_x)]$ of the Wilson loop $\tilde{\W}^{\nu_x}_{y, {\bf k}}$ over the Wannier sector $\nu_x$, reflection symmetry along $x$ implies that
\begin{align}
\left\{e^{i2\pi \nu^{\nu_x,j}_y(k_x)}\right\} = \left\{e^{i2\pi \nu^{-\nu_x,j}_y(-k_x)}\right\}
\end{align}
or
\begin{align}
\{\nu^{\nu_x,j}_y(k_x)\} =\{\nu^{-\nu_x,j}_y(-k_x)\} \;\; \mbox{mod 1},
\end{align}
where $j \in 1\ldots N_{\nu_x}$ labels the eigenvalue, and $N_{\nu_x}$ is the number of Wannier bands in the sector $\nu_x$. The Wannier-sector polarization can be written as
\begin{align}
p^{\nu_x}_y = \frac{1}{N_x} \sum_{k_x} \sum_{j=1}^{N_{\nu_x}} \nu^{\nu_x,j}_y(k_x).
\label{eq:app_Wannier_sector_polarization}
\end{align}
Hence, $M_x$ symmetry implies that
\begin{align}
p^{\nu_x}_y=p^{-\nu_x}_y
\end{align}

\subsection{Constraints due to reflection symmetry along $y$}
\label{sec:app_Wilson_loops_under_My}
\subsubsection{On the Wilson loop over occupied energy bands}
Consider a system that is Wannier-gapped, and which furthermore has reflection symmetry along $y$
\begin{align}
\hat{M}_y h_{\bf k} \hat{M}_y^\dagger = h_{M_y {\bf k}},
\end{align}
where $M_y {\bf k} = M_y (k_x,k_y)=(k_x,-k_y)$. This symmetry allows us to write the expansion
\begin{align}
\ket{u^n_{\bf k}}=\ket{u^m_{M_y {\bf k}}}B^{mn}_{M_y,\bf k},
\end{align}
where 
\begin{align}
B^{mn}_{M_y,\bf k}=\matrixel{u^m_{M_y {\bf k}}}{\hat{M}_y}{u^n_{\bf k}}
\end{align}
is the unitary sewing matrix, which connects states at ${\bf k}$ with states at $M_y {\bf k}$. In particular, $B^{mn}_{M_y,\bf k} \neq 0$ only if $\epsilon_{m,M_y {\bf k}}= \epsilon_{n,{\bf k}}$. 

Under this symmetry, the Wilson loop along $x$ starting at base point ${\bf k}$ obeys
\begin{align}
B_{M_y,\bf k} \W_{x,{\bf k}} B^\dagger_{M_y,\bf k} = \W_{x,M_y {\bf k}}.
\end{align}
A schematic of this relation is shown in Fig.~S~\ref{fig:app_Wx_transformations}b. The Wilson loops at ${\bf k}$ and $M_y {\bf k}$ are equivalent up to a unitary transformation. Hence, the sets of their eigenvalues are the same, namely,
\begin{align}
\left\{ e^{i 2\pi \nu_x^i(k_y)} \right\} = \left\{ e^{i 2\pi \nu_x^i(-k_y)} \right\}
\label{eq:app_Wilson_loop_eigenvalues_reflection_symmetry}
\end{align}

\subsubsection{On the nested Wilson loop over Wannier sectors}
Now, we focus on the Wilson loop eigenfunctions
\begin{align}
\W_{x,{\bf k}} \ket{\nu^i_{x,\bf k}}=e^{i2\pi \nu_x^i(k_y)}\ket{\nu^i_{x,\bf k}}.
\end{align}
Using the symmetry relation for the Wilson loops above as $B_{M_y,\bf k} \W_{x,{\bf k}} = \W_{x,M_y {\bf k}} B_{M_y,\bf k}$, we have that
\begin{align}
 \W_{x,M_y {\bf k}} B_{M_y,\bf k} \ket{\nu^i_{x,\bf k}} &= B_{M_y,\bf k} \W_{x,{\bf k}}  \ket{\nu^i_{x,\bf k}} \nonumber\\
 &= e^{i2\pi \nu_x^i(k_y)} B_{M_y,\bf k} \ket{\nu^i_{x,\bf k}}.
 \label{eq:app_Wilson_loop_eigenstate_under_symmetry}
\end{align}
$B_{M_y,\bf k} \ket{\nu^i_{x,\bf k}} $ is an eigenfunction of $ \W_{x,M_y {\bf k}}$ with eigenvalue $e^{i2\pi \nu_x^i(k_y)}$. We now expand this function as
\begin{align}
B_{M_y,\bf k} \ket{\nu^i_{x,\bf k}} = \ket{\nu^j_{x,M_y{\bf k}}}  \alpha^{ji}_{M_y,\bf k},
\label{eq:app_Wilson_loop_eigenstate_expansion}
\end{align}
where
\begin{align}
 \alpha^{ji}_{M_y,\bf k} = \matrixel{\nu^j_{x,M_y {\bf k}}}{B_{M_y,\bf k}}{\nu^i_{x,\bf k}}
\end{align}
is a sewing matrix that connects Wilson loop eigenstates at base points ${\bf k}$ and $M_y {\bf k}$ having \textit{the same} Wilson-loop eigenvalues. To see this, one can follow a similar procedure as in Section \ref{sec:app_Wilson_loops_under_Mx} to show that if $ \alpha^{ji}_{M_y,\bf k} \neq 0$, we require that $\nu_x^j(-k_y)=\nu_x^i(k_y)$. In other words, the sewing matrix $\alpha_{M_y,\bf k}$ only connects Wilson-loop eigenstates at base points ${\bf k}$ and $M_y {\bf k}$ having the same Wannier center. 
Following the same procedure as in \eqref{eq:app_Wannier_basis_under_Mx} for the Wannier sectors $\ket{w^j_{x,{\bf k}}}$, we have
\begin{align}
M_y \ket{w^j_{x,\bf k}}&=\ket{w^i_{x,M_y\bf k}} \alpha^{ij}_{M_y,\bf k},
\end{align}
from which is follows that
\begin{align}
\ket{w^j_{x,\bf k}} &= M_y^\dagger \ket{w^i_{x,M_y \bf k}} \alpha^{ij}_{M_y,\bf k}
\end{align}
Using these expressions, there is the following relation for a Wilson line element
\begin{align}
\left[\tilde{\W}^{\nu_x}_{{\bf k}_2 \leftarrow {\bf k}_1}\right]^{ij} = [\alpha^\dagger_{M_y,{\bf k}_2}]^{i i'} \left[\tilde{\W}^{\nu'_x}_{M_y {\bf k}_2 \leftarrow M_y {\bf k}_1}\right]^{i'j'}  \alpha^{j' j}_{M_y,{\bf k}_1}.
\end{align}
In particular the Wilson loop along $y$ obeys
\begin{align}
\left[\tilde{\W}^{\nu_x}_{y,{\bf k}} \right]^{ij} &=  \left[\alpha^\dagger_{M_y,\bf k} \right]^{ii'} \left[ \tilde{\W}^{\nu'_x}_{-y, M_y {\bf k}} \right]^{i'j'} \left[\alpha_{M_y,\bf k} \right]^{j'j},
\end{align}
which looks similar to the one in Section \ref{sec:app_Wilson_loops_under_Mx}, but with the important difference in the structure of $\alpha^{ji}_{M_y,\bf k}$, which connects Wilson loop eigenstates such that $\nu^j_x(k_y) = \nu^i_x(-k_y)$. Another important difference is the fact that $M_y$ reverses the loop contour along $y$ and preserves it along $x$. This expression tells us that the Wilson loop eigenvalues are related by
\begin{align}
\left\{e^{i 2 \pi \nu^{\nu_x,j}_y(k_x)}\right\} = \left\{e^{-i 2\pi \nu^{\nu_x,j}_y(k_x)}\right\}
\end{align}
or
\begin{align}
\left\{\nu^{\nu_x,j}_y(k_x) \right\}=\left\{ - \nu^{\nu_x,j}_y(k_x)\right\} \;\; \mbox{mod 1},
\end{align}
from which it follows that $\nu^{\nu_x}_y(k_x)$ is either $0$, $1/2$, or comes in pairs $\nu,-\nu$. Reflection symmetry along $y$ thus implies that the polarization in Eq.~\ref{eq:app_Wannier_sector_polarization} over the Wannier sector $\nu_x$ obeys 
\begin{align}
p^{\nu_x}_y=-p^{\nu_x}_y,
\end{align}
from which it follows that $p^{\nu_x}_y$ is either $0$, $1/2$. In particular, values that come in pairs $\nu,-\nu$ do not contribute to $p^{\nu_x}_y$.

\subsection{Constraints due to inversion symmetry}
\label{sec:app_Wilson_loops_under_inversion}
\subsubsection{On the Wilson loop over occupied energy bands}
Now, we consider the inversion symmetry
\begin{align}
\I h_{\bf k} \I^\dagger = h_{-{\bf k}}
\end{align}
under which the Wilson loop obeys (see Eq. \ref{eq:app_Wilson_loop_under_symmetry})
\begin{align}
B_{\I,\bf k} \W_{x,{\bf k}} B^\dagger_{\I,\bf k} = \W_{-x,-{\bf k}}=\W^\dagger_{x,-{\bf k}}
\end{align}
where
\begin{align}
B^{mn}_{\I,\bf k}=\bra{u^m_{-{\bf k}}}\I \ket{u^n_{\bf k}}
\end{align}
connects energy eigenstates at ${\bf k}$ and $-{\bf k}$ having the same energy. A schematic of this relation is shown in Fig.~S~\ref{fig:app_Wx_transformations}c. 
This relation between Wilson loops implies that the set of eigenvalues follow
\begin{align}
\left\{ e^{i 2\pi \nu_x^i(k_y)} \right\} = \left\{ e^{-i 2\pi \nu_x^i(-k_y)} \right\}
\label{eq:app_Wilson_loop_eigenvalues_inversion_symmetry}
\end{align}
In particular, at $\Pi_y=0,\pi$ we recover the identical condition as for reflection symmetry along $x$. Thus, at these points the $\W_x$ eigenvalues are either $\pm1$ or complex conjugate pairs.

\subsubsection{On the nested Wilson loop over Wannier sectors}
Following the procedures of Sections \ref{sec:app_Wilson_loops_under_Mx} and \ref{sec:app_Wilson_loops_under_My}, one can demonstrate that $B_{\I,\bf k} \ket{\nu^i_{x,\bf k}}$ is an eigenstate of $\W_{x,-{\bf k}}$ with eigenvalue $e^{-i 2\pi \nu_x^i(k_y)}$. Thus, in the expansion
\begin{align}
B_{\I,\bf k} \ket{\nu^i_{x,\bf k}} = \ket{\nu^j_{x,-{\bf k}}}  \alpha^{ji}_{\I,\bf k}
\end{align}
the sewing matrix
\begin{align}
 \alpha^{ji}_{\I,\bf k} = \matrixel{\nu^j_{x,-{\bf k}}}{B_{\I,\bf k}}{\nu^i_{x,\bf k}}
\end{align}
connects Wilson-loop eigenstates at base points ${\bf k}$ and $-{\bf k}$ having opposite Wannier centers [$ \alpha^{ji}_{\bf k} \neq 0$ only if $\nu_x^i(k_y) = -\nu_x^j(-k_y)$]. 
For the Wannier sectors $\ket{w^j_{x,{\bf k}}}$, we have
\begin{align}
\I \ket{w^j_{x,\bf k}}&=\ket{w^i_{x,-\bf k}} \alpha^{ij}_{\I,\bf k},
\end{align}
from which it follows that
\begin{align}
\ket{w^j_{x,\bf k}} &= \I^\dagger \ket{w^i_{x,- {\bf k}}} \alpha^{ij}_{\I,\bf k}.
\end{align}

Using these expressions, there is the following relation for a Wilson line element
\begin{align}
\left[\tilde{\W}^{\nu_x}_{{\bf k}_2 \leftarrow {\bf k}_1}\right]^{ij} = [\alpha^\dagger_{\I,{\bf k}_2}]^{i i'} \left[\tilde{\W}^{\nu'_x}_{- {\bf k}_2 \leftarrow - {\bf k}_1}\right]^{i'j'}  \alpha^{j' j}_{\I,{\bf k}_1}.
\end{align}
In particular the Wilson loop along $y$ obeys
\begin{align}
\left[\tilde{\W}^{\nu_x}_{y,{\bf k}} \right]^{ij} &=  \left[\alpha^\dagger_{\I,\bf k} \right]^{ii'} \left[ \tilde{\W}^{\nu'_x}_{-y, - {\bf k}} \right]^{i'j'} \left[\alpha_{\I,\bf k} \right]^{j'j},
\end{align}
Thus, the Wilson loop eigenvalues are related by
\begin{align}
\left\{e^{i 2 \pi \nu^{\nu_x,j}_y(k_x)}\right\} = \left\{e^{-i 2\pi \nu^{-\nu_x,j}_y(-k_x)}\right\}
\end{align}
or
\begin{align}
\left\{\nu^{\nu_x,j}_y(k_x) \right\}=\left\{ - \nu^{-\nu_x,j}_y(-k_x)\right\}\;\; \mbox{mod 1},
\end{align}
The Wannier-sector polarization \eqref{eq:app_Wannier_sector_polarization}, under inversion symmetry obeys
\begin{align}
p^{\nu_x}_y=-p^{-\nu_x}_y.
\end{align}

\section{Wilson Loop and the Edge Hamiltonian}
\label{sec:app_edge_Hamiltonian}
Being unitary, we can express the Wilson loop as the exponential of a Hermitian matrix,
\begin{align}
\W_{\C,\bf k} \equiv e^{i H_{\W_\C}(\bf k)}.
\end{align}
We interpret $H_{\W_\C}(\bf k)$ as the \textit{edge Hamiltonian} \cite{Klich2011}, whose eigenvalues are defined mod $2\pi$. Alternatively, we will refer to this as the \textit{Wannier Hamiltonian}, for obvious reasons. Notice that in the definition above, the argument $\bf k$ of the edge Hamiltonian is the base point of the Wilson loop. The eigenvalues of $H_{\W_\C}({\bf k})$ only depend on the coordinate of $\bf k$ not spanned by $\C$, e.g., in two-dimensions, the eigenvalues depend on $k_y$ for $\C$ along $k_x$ and viceversa. 

Using the relations
\begin{align}
\W_{-\C,\bf k} = \W^\dagger_{\C,\bf k} = e^{-i H_{\W_\C}(\bf k)},
\end{align}
where $-\C$ is the contour $\C$ but in reverse order, we make the identification
\begin{align}
H_{\W_{-\C}}({\bf k}) = -H_{\W_{\C}}({\bf k}).
\end{align}
The tranformations of the Wilson loops under symmetries shown in \eqref{eq:app_Wilson_loop_under_symmetry} can be written as
\begin{align}
B_{g,\bf k} H_{\W_\C}({\bf k}) B^\dagger_{g,\bf k} = H_{\W_{D_g \C}}(D_g \bf k).
\end{align}
Since this expression is of the same form as Eq.~\ref{eq:app_Hamiltonian_under_symmetry}, we can interpret the sewing matrix $B_{g,\bf k}$ as a symmetry of the edge Hamiltonian. In particular, we have
\begin{align}
B_{M_x,\bf k} H_{\W_x}({\bf k}) B^\dagger_{M_x,\bf k} &= -H_{\W_x}(M_x \bf k)\nonumber \\
B_{M_y,\bf k} H_{\W_x}({\bf k}) B^\dagger_{M_y,\bf k} &= H_{\W_x}(M_y \bf k)\nonumber \\
B_{\I,\bf k} H_{\W_x}({\bf k}) B^\dagger_{\I,\bf k} &= -H_{\W_x}(-\bf k)
\label{eq:app_Wannier_Hamiltonian_symmetries}
\end{align}

\section{No go theorem for quadrupole moments in crystals with commuting reflection operators}
\label{sec:app_no-go_theorem}
In this section, we show that spinless crystals with \emph{commuting} reflection operators cannot realize a quadrupole. This result mathematically formalizes the need for a $\pi$ flux per plaquette in tight binding models that realize a non-vanishing, quantized quadrupole moment. 

For a crystal with $N_{occ}$ occupied energy bands, and for $i,j=x,y$ and $i \neq j$,various cases need to be considered:
\begin{itemize}
\item $N_{occ}$ is odd: In this case, by Eq.~\ref{eq:app_Wilson_loop_eigenvalues_reflection_symmetry} we always have a gapless Wannier spectrum, and thus the quadrupole invariant cannot be defined.
\item $N_{occ}=2$: In this case, the Wannier bands $\nu_i(k_j)$ are necessarily gapless at $k_j=0,\pi$.
\item $N_{occ} = 4$: In this case, the Wannier bands $\nu_i(k_j)$ can be generically gapped, with each Wannier sector being two-dimensional. The two eigenvalues of the Wilson loop over the Wannier sector $\nu_i$, for either $\nu_i = \nu^+_i$ or $\nu^-_i$ in this case come in pairs  $[\nu^{\nu_i}_j(k_i),-\nu^{\nu_i}_j(k_i)]$ at $k_i=0,\pi$. Thus, $p^{\nu_i}_j = 0$.
\item $N_{occ} = 4n$: In this case, we have a generalized version of the $N_{occ}=4$ case.
\item $N_{occ} = 4n + 2$: In this case, the Wannier bands split as in the $4n$ case plus the 2 band case, and the Wannier spectrum is gapless.
\end{itemize}
In what follows, we elaborate in each of the last four cases.

\subsection{$N_{occ}=2$: Gapless Wannier bands}
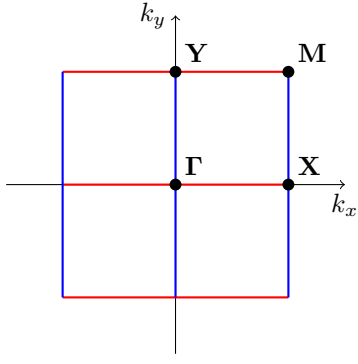
\begin{figure}
\centering
\begin{tikzpicture}[scale=1.5]
 
 		\draw [->,black] (-1.5,0)--(1.5,0) node[below] {$k_x$};
	  	\draw [->,black] (0,-1.5)--(0,1.5) node[left] {$k_y$};		
		
		 \draw [red, thick] (-1,.0)--(1,0);
	 	 \draw [red,thick] (-1,1)--(1,1);
		 \draw [red,thick] (-1,-1)--(1,-1);
		 \draw [blue, thick] (0,-1)--(0,1);
	 	 \draw [blue,thick] (-1,-1)--(-1,1);
	 	 \draw [blue,thick] (1,-1)--(1,1);
	  
		\fill [black] (0,0) circle (1.5pt) node[above right] {${\bf \Gamma}$};
		\fill [black] (1,0) circle (1.5pt) node[above right] {${\bf X}$};
		\fill [black] (0,1) circle (1.5pt) node[above right] {${\bf Y}$};
		\fill [black] (1,1) circle (1.5pt) node[above right] {${\bf M}$};
		
\end{tikzpicture}
\caption{\emph{Reflection invariant points and lines of the Brillouin zone}. The tight binding Hamiltonian at the blue (red) lines commutes with $\hat{M}_x$ ($\hat{M}_y$). At the points $\bf \Gamma, X, Y, M$ the Hamiltonian also commutes with $\I$.}
\label{fig:app_BZ_HSP}
\end{figure}

Consider a spinless crystal with reflection symmetries in $x$ and $y$. For a tight binding Hamiltonian $h(k_x,k_y)$, these symmetries are expressed by
\begin{align}
\hat{M}_x h_{(k_x,k_y)} \hat{M}^{-1}_x = h_{(-k_x,k_y)}\nonumber \\
\hat{M}_y h_{(k_x,k_y)} \hat{M}^{-1}_y = h_{(k_x,-k_y)},
\end{align}
where $\hat{M}_{x,y}$ are the reflection operators. Such a system also has the inversion symmetry
\begin{align}
\I h_{\bf k} \I^\dagger = h_{-\bf k},
\end{align}
where 
\begin{align}
\I = \hat{M}_y \hat{M}_x
\end{align}
is the inversion operator. 
There are special high-symmetry momenta in the Brillouin zone, at which a given operator $\hat{Q}$ (we consider $\hat{Q}=\hat{M}_{x}$, $\hat{M}_{y}$ and $\I$) commutes with the Hamiltonian,
\begin{align}
[h_{\bf k^*},\hat{Q}] = 0.
\end{align}
Those special points are shown in Fig.~S~\ref{fig:app_BZ_HSP} in blue for $\hat{M}_x$, red for $\hat{M}_y$, and black dots for all $\hat{M}_x$, $\hat{M}_y$, and $\I$. 

We are interested in the conditions under which we have gapped Wannier bands along both $x$ and $y$. 
This can be inferred \cite{Alexandradinata2014} from the $\hat{Q}$ eigenvalues at the high-symmetry momenta, as shown in Table~S\ref{tab:app_eigenvalues_relations}.
\begin{table}
\begin{center}
\begin{tabular}{l l l}
Eigenvalues  & Eigenvalues   & Eigenvalues  \\
of $\hat{Q}$ at ${\bf k^*}$ & of $\hat{Q}$ at ${\bf k^*+G/2}$ & of $\W_{{\bf k}^*+{\bf G}
\leftarrow \bf k^*}$\\
\hline\\
$(++)$ & $(++)$ & $(1,1)$\\
$(++)$ & $(+-)$ & $(1,-1)$\\
$(++)$ & $(--)$ & $(-1,-1)$\\
$(+-)$ & $(+-)$ & $(\lambda,\lambda^*)$
\end{tabular}
\end{center}
\caption{\emph{Relation between symmetry eigenvalues and Wilson loop eigenvalues}. $\pm$ are the eigenvalues of reflection or inversion symmetries at high-symmetry momenta $\bf k^*$ and ${\bf k}^*+\bf G/2$. Corresponding Wilson loop eigenvalues are for the Wilson loop in the direction of the reciprocal lattice vector $\bf G$ for two occupied bands \cite{Alexandradinata2014}.}
\label{tab:app_eigenvalues_relations}
\end{table}
There we see that, in order to have gapped Wannier bands $(\lambda, \lambda^*)$ not pinned at $1$ or $-1$ by symmetry, we require that, along each of the blue and red lines of Fig.~S~\ref{fig:app_BZ_HSP}, the reflection eigenvalues come in pairs $(+-)$. To these two conditions (one along blue lines for $\hat{M}_x$ eigenvalues and another one along red lines for $\hat{M}_y$ eigenvalues) we need to add the requirement that the inversion eigenvalues  also come in pairs $(+-)$ at the time-reversal invariant momenta (TRIM) $\bf \Lambda =  \Gamma, X, Y, M$. If the condition is not also met for the inversion eigenvalues, the complex conjugate pairs are then forced to be both $1$ or $-1$ at $k_y=0,\pi$ for Wilson loops along $k_x$, and at $k_x=0,\pi$ for Wilson loops along $k_y$. We will now see that this third condition is not possible to meet in the case of commuting reflection operators. If we have that
\begin{align}
[\hat{M}_x,\hat{M}_y]=0,
\end{align}
it is possible to simultaneously label the energy states at the TRIM by their reflection eigenvalues as $\ket{m_x,m_y}$, where
\begin{align}
\hat{M}_x \ket{m_x,m_y} &= m_x \ket{m_x,m_y}\nonumber \\
\hat{M}_y \ket{m_x,m_y} &= m_y \ket{m_x,m_y},
\end{align}
for $m_{x,y}=\pm$ the eigenvalues of the reflection operators. Since the inversion operator is
$\I = \hat{M}_x \hat{M}_y =\hat{M}_y \hat{M}_x$, the inversion eigenvalues are
\begin{align}
\I \ket{m_x,m_y} = m_x m_y \ket{m_x,m_y}.
\end{align}
The two combinations of states that have $\hat{M}_x$ and $\hat{M}_y$ eigenvalues of $(+-)$ are listed in Table~S\ref{tab:app_states_pm_mx_eigenvalues}. However, those two options do not meet the third condition of having $\I$ eigenvalues $(+-)$ at the TRIM. Instead, the $\I$ eigenvalues at the TRIM are always either $(++)$ or $(--)$, which prevents the possibility that the Wilson loop eigenvalues come in complex conjugate pairs $\lambda, \lambda^*$ with $\lambda \neq +1,-1$. Thus, along the high-symmetry lines (blue and red lines of Fig.~S~\ref{fig:app_BZ_HSP}) we have
\begin{align}
(\lambda,\lambda^*) \rightarrow (1,1) \mbox{ or } (-1,-1)
\end{align}
i.e., at those lines the Wannier bands close the gap.

\begin{table}
\begin{center}
\begin{tabular}{c c c c}
states  &$\hat{M}_x$ eigenvalues &  $\hat{M}_y$ eigenvalues &  $\I$ eigenvalues\\
\hline
$\left(\ket{++},\ket{--}\right)$ & $(+-)$ & $(+-)$ & $(++)$\\
$\left(\ket{+-},\ket{-+}\right)$ & $(+-)$ & $(-+)$ & $(--)$\\
\end{tabular}
\end{center}
\caption{\emph{States with eigenvalues $(+-)$ for $\hat{M}_x$ and $\hat{M}_y$ and their $\I$ eigenvalues at the TRIM.}}
\label{tab:app_states_pm_mx_eigenvalues}
\end{table}

\subsection{$N_{occ}=4$: Gapped Wannier bands with trivial quadrupole}
Unlike the $N_{occ}=2$ case, if four energy bands are occupied, it is possible to meet the conditions of having $\hat{M}_x$, $\hat{M}_y$, and $\I$ eigenvalues that come in pairs $(+-)$ at any TRIM. This occurs only for the choice of states $(\ket{++}, \ket{+-}, \ket{-+}, \ket{--})$. In that case, the Wannier bands at the TRIM are gapped. Using this basis, the sewing matrices for $\hat{M}_x$, $\hat{M}_y$, and $\I$ at the TRIM $\bf \Lambda$ take the forms
\begin{align}
B_{M_x, \bf \Lambda}&=\tau^z\otimes\mathbb{I}\nonumber\\
B_{M_y, \bf \Lambda}&=\mathbb{I}\otimes\sigma^z\nonumber\\
B_{\I, \bf \Lambda}&=\tau^z\otimes\sigma^z.
\end{align}
These matrices are useful because they represent the symmetries that the Wannier Hamiltonian must have at the TRIM (see Eq.~\ref{eq:app_Wannier_Hamiltonian_symmetries}). For example,  $H_{\W_x}({\bf{k}})$ must satisfy 
\begin{align}
[H_{\W_{x}}({\bf \Lambda}),B_{M_y,{\bf \Lambda}}]=\{H_{\W_{x}}({\bf \Lambda}),B_{M_x,{\bf \Lambda}}\}=\{H_{\W_{x}}({\bf \Lambda}),B_{\mathcal{I},{\bf \Lambda}}\}=0.
\end{align}
Similarly, $H_{\W_y}({\bf{k}})$ must satisfy
\begin{align}
[H_{\W_{y}}({\bf \Lambda}),B_{M_x,{\bf \Lambda}}]=\{H_{\W_{y}}({\bf \Lambda}),B_{M_y,{\bf \Lambda}}\}=\{H_{\W_{y}}({\bf \Lambda}),B_{\mathcal{I},{\bf \Lambda}}\}=0.
\end{align}
Imposing these symmetries on all sixteen Hermitian matrices $\tau^i \otimes \sigma^j$, for $i,j =0,x,y,z$ (where $\tau$, $\sigma$ are Pauli matrices and $\tau^0 = \sigma^0 = \mathbb{I}$), the most general form for the Wannier Hamiltonians is
\begin{align}
H_{\W_x}({\bf \Lambda})&=\delta_1\tau^{x}\otimes\sigma^z+\delta_2\tau^{x}\otimes\mathbb{I}+\delta_3\tau^{y}\otimes\sigma^z+\delta_4\tau^{y}\otimes\mathbb{I}\nonumber \\
H_{\W_y}({\bf \Lambda})&=\beta_1\tau^{z}\otimes\sigma^x+\beta_2\mathbb{I}\otimes\sigma^x+\beta_3\tau^{z}\otimes\sigma^y+\beta_4\mathbb{I}\otimes\sigma^y.
\end{align}
where $\delta_i$ and $\beta_i$ for $i=1,2,3,4$, are coefficients which can vary between the different TRIM. The Wannier energies of $H_{\W_x}$ and $H_{\W_y}$ are, respectively, 
\begin{align}
\theta_x&= 2\pi \nu_x = \pm \sqrt{(\delta_1 - \delta_2)^2 + (\delta_3 - \delta_4)^2}, \pm \sqrt{(\delta_1 + \delta_2)^2 + (\delta_3 + \delta_4)^2}\nonumber \\
\theta_y&= 2\pi \nu_y = \pm \sqrt{(\beta_1 - \beta_2)^2 + (\beta_3 - \beta_4)^2}, \pm \sqrt{(\beta_1 + \beta_2)^2 + (\beta_3 + \beta_4)^2}
\end{align}
mod $2\pi$.
By direct computation we find that the eigenstates of the upper (or lower) bands $\nu_x$ of $H_{\W_x}$ have $(+-)$ eigenvalues under $B_{M_y,{\bf \Lambda}}$, for any values of the $\delta$ coefficients. Hence the $\alpha_{M_y,\bf \Lambda}$ sewing matrix at each TRIM has $(+-)$  eigenvalues and thus, the eigenvalues of the Wilson loop over Wannier band $\nu_x$ come in pairs $[\nu^{\nu_x}_y(k_x),-\nu^{\nu_x}_y(k_x)]$ at $k_x=0,\pi$. Now, since it is not possible to continuously deform the bands $[\nu^{\nu_x}_y(k_x),-\nu^{\nu_x}_y(k_x)]$ at $k_x=0,\pi$ to $[0,1/2]$ or $[1/2,0]$ at any other point in $k_x$ without breaking reflection symmetry along $y$, it follows that the eigenvalues of the Wilson loop over Wannier band $\nu_x$  come in pairs $[\nu^{\nu_x}_y(k_x),-\nu^{\nu_x}_y(k_x)]$ at all $k_x \in (-\pi,\pi]$, which results in a vanishing polarization of Eq.~\ref{eq:app_polarization_Wannier_sector}. For $H_{\W_y}$ a similar statement is true. 

\subsection{$N_{occ}=4n$: Generalizing the $N_{occ}=4$ case}
Now let us generalize the previous argument. Suppose we have $4n$ occupied bands and the $M_x$, $M_y$, and $\mathcal{I}$ eigenvalues all come in $(+-)$ pairs at each TRIM. We can arrange the basis of occupied energy bands such that 
\begin{eqnarray}
B_{M_x,{\bf \Lambda}}&=&\tau^z\otimes\mathbb{I}_{2n}\nonumber\\
B_{M_y,{\bf \Lambda}}&=&\mathbb{I}_{2n}\otimes\sigma^z\nonumber\\
B_{\mathcal{I},{\bf \Lambda}}&=&\mu^z\otimes\mathbb{I}_{n}\otimes\sigma^z.
\label{eq:app_sewing_matrices_4n_case}
\end{eqnarray} 
Crucially, each Wannier Hamiltonian at a TRIM has to commute with one reflection sewing matrix, and anticommute with the other since one reflection preserves the contour and the other flips it. Consider $H_{\W_x}(\bf \Lambda).$ It must satisfy $[H_{\W_{x}}({\bf \Lambda}),B_{M_y,{\bf \Lambda}}]=\{H_{\W_{x}}({\bf \Lambda}),B_{M_x,{\bf \Lambda}}\}.$ We can label an eigenstate of $H_{\W_x}({\bf \Lambda})$ as $\ket{\nu^j_x,b_{m_y}}$, where $\nu^j_x$ is its Wannier value and $b_{m_y}$ is the eigenvalue under $B_{M_y,{\bf \Lambda}}$. For each $\ket{\nu^j_x,b_{m_y}}$ we have another state $B_{M_x,{\bf \Lambda}} \ket{\nu^j_x,b_{m_y}}$ which has \emph{opposite} Wannier value, but the \emph{same} $b_{m_y}$. This is because $M_x$ complex-conjugates the Wannier eigenvalue, but since the Wannier Hamiltonian commutes (by assumption) with $M_y$, leaves $b_{m_y}$ invariant.

Now we can see from the form of our sewing matrices in Eq.~\ref{eq:app_sewing_matrices_4n_case} that there are an equal, and \emph{even} number of $\pm$ eigenvalues ($4n$ bands means $2n$ each of $\pm$), which is a necessary and direct result of our need for gapped Wannier bands. This means that each of the gapped Wannier sectors has an equal number of $\pm$ reflection sewing eigenvalues. 

Hence, since the reflection-sewing eigenvalues of a Wannier sector determine its polarization as indicated in Table~S\ref{tab:app_eigenvalues_relations}, we find that the Wannier centers of the projected Wannier sector must come in complex conjugate pairs, and hence its polarization is trivial. 
This result can be applied \emph{mutatis mutandis} for the other Wilson loop Hamiltonian $H_{W_y}.$ Since the Wilson of Wilson must be trivial in both directions the quadrupole is trivial. 

\subsection{$N_{occ}=4n+2$: Gapless Wannier bands}
This case mirrors the $N_{occ}=2$ case. In order to have gapped Wannier bands for any set of $4n+2$ occupied energy bands we must choose an array of occupied states such that there are $2n+1$  eigenvalues $+1$ and $2n+1$ eigenvalues $-1$ of both $M_x$, and $M_y$.  After making this choice we can try to arrange them such that the products of the eigenvalues, i.e., the inversion eigenvalues, also come in $\pm$ pairs, so that the Wannier bands are gapped. To achieve that, we also need $2n+1$ inversion eigenvalues $+1$ and $2n+1$ inversion eigenvalues $-1$. No matter what arrangement we choose, the number of inversion eigenvalues $+1$ and the number of inversion eigenvalues $-1$ is always an \emph{even} number and cannot be $2n+1$. Hence, we cannot ever find exactly matched pairs of $\pm$ inversion eigenvalues. An alternative way of stating this is that we can find exactly matched pairs for $4n$ bands, but the remaining eigenvalues reduce to the $2$ band problem that we have already shown is gapless.

\section{Plaquette flux and its relation to the commutation of reflection operators}
\label{sec:app_symmetries_up_to_gauge}
In this section we study the conditions under which the reflection operators commute or not when non-zero fluxes are added to plaquettes. This is important in our model due to the requirement that reflection operators do not commute in order to have a gapped Wannier bands (see Section \ref{sec:app_no-go_theorem}). Furthermore, the cases in which plaquettes have $0$ or $2\pi$ fluxes are gapless at half-filling, and therefore are useless in the construction of a 2D quadrupole Hamiltonian. On the other hand, plaquettes with $\pi$ flux are gapped at half-filling and obey $[\hat{m}_x,\hat{m}_y] \neq 0$, as we will see. Thus they are useful in the construction of a non-trivial quadrupole. Indeed, our minimal quadrupole model (Eq. 4 of main text) is built using plaquettes with $\pi$ flux.

\begin{figure}[h]
\centering
\includegraphics[width=9cm]{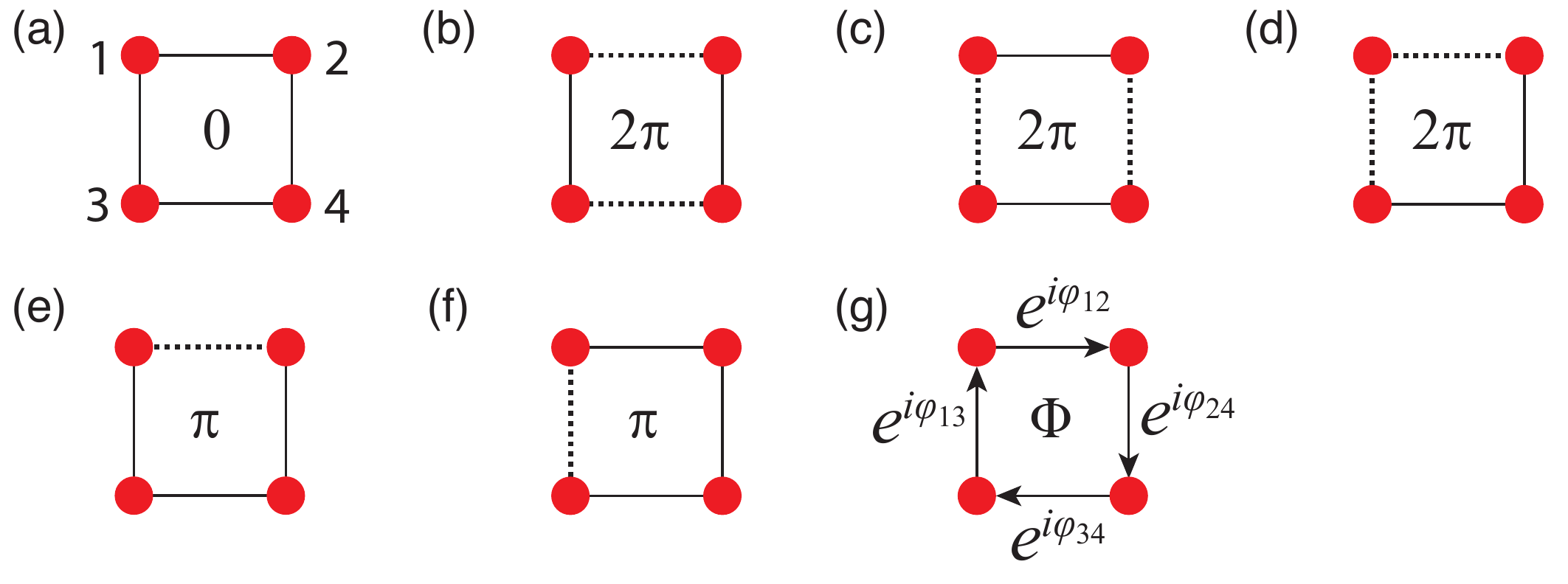}
   \caption{\emph{Hopping configurations on a plaquete with four sites}. Dotted lines indicate a flipped sign compared to solid lines. (a-d) have either $0$ or $2\pi$ flux, while (e),(f) are different configurations with $\pi$ flux. (g) is a generic configuration with flux $\Phi.$}\label{fig:app_square}
\end{figure}

The simple square configuration in Fig.~S~\ref{fig:app_square}a has the Hamiltonian
\begin{equation}
H_0=\left(\begin{matrix} 0 & 1 & 1 &0\\
1 & 0 & 0 &1\\
1& 0 & 0 &1\\
0 & 1 & 1 &0\\\end{matrix}\right).
\end{equation} or, more compactly, $H_0=\mathbb{I}\otimes \sigma^x+\tau^x\otimes\mathbb{I}.$ This model has reflection symmetries that exchange left and right $\hat{m}^{0}_x=\mathbb{I}\otimes\sigma^{x}$ and up and down $\hat{m}^{0}_y=\tau^x\otimes\mathbb{I}.$ These operators multiply to give the inversion operator $\mathcal{I}=\hat{m}^{0}_x \hat{m}^{0}_y=\tau^x\otimes \sigma^x$. In this case we have $[\hat{m}^{0}_x ,\hat{m}^{0}_y]=0.$ Hence $\mathcal{I}^{2}=\hat{m}^{0}_x \hat{m}^{0}_y \hat{m}^{0}_x \hat{m}^{0}_y=(\hat{m}^{0}_x)^2 (\hat{m}^{0}_y)^2 = +1.$ This system has energies $-2, 0, 0, +2$ and therefore is gapless at half filling. 

Now let us consider configurations with $\pi$ flux. When the flipped bond is between sites 1 and 2 (see Fig.~S~\ref{fig:app_square}e) we have
\begin{equation}
H_{12}=-\tau^z\otimes \sigma^x+\tau^x\otimes\mathbb{I}.
\end{equation} The energies of $H_{12}$ are $-1 ,-1, +1, +1$, and hence this system is gapped at half filling. This system has a reflection symmetry in the $x$-direction, but \emph{does not} have an exact reflection symmetry in the $y$-direction, it only has a reflection symmetry times (up to) a gauge transformation. This is because, although the magnetic field is invariant under reflection, the vector potential is not, and we must multiply a pair of the second-quantized operators by a $-1$ in order to recover the symmetry. This $-1$ is the gauge transformation. As such the reflection operator that sends $x \to -x$ does not change, i.e.,  $\hat{m}^{12}_{x}=\mathbb{I}\otimes \sigma^{x}.$ However, the reflection operator in the $y$-direction has opposite signs and we have $\hat{m}^{12}_{y}=\tau^{x}\otimes \sigma^{z}=\hat{m}^{0}_{y}(\mathbb{I}\otimes \sigma^z)$ where $\mathcal{G}=(\mathbb{I}\otimes \sigma^z)$ is one choice for the gauge transformation (the other would be $\mathcal{G}=-(\mathbb{I}\otimes \sigma^z)$). This gauge transformation multiplies either $c^{\dagger}_1$ and $c^{\dagger}_3$ or $c^{\dagger}_2$ and $c^{\dagger}_4$ by a minus sign depending on our choice of $\mathcal{G}$, and leaves the other operators unchanged. 
In this case, the commutation relations have now change to $\{\hat{m}^{12}_x, \hat{m}^{12}_y\}=0$. 

Let us consider another $\pi$ flux configuration such that the flipped bond is between sites 1 and 3, as in Fig.~S~\ref{fig:app_square}f. The Hamiltonian is 
\begin{equation}
H_{13}=\mathbb{I}\otimes \sigma^x- \tau^x\otimes \sigma^z.
\end{equation} This has reflection in $y$, but reflection only up to gauge transformation in the $x$-direction. The gauge transformation in this case is $\mathcal{G}=\tau^z\otimes \mathbb{I}.$ The reflection operators are $\hat{m}^{13}_x=\tau^z\otimes\sigma^x$ and $\hat{m}^{13}_{y}=\tau^{x}\otimes \mathbb{I}.$ These also have a non-vanishing commutator. 

Let us see what happens if we have $2\pi$ flux through the plaquette. If the bonds are arranged as Fig.~S~\ref{fig:app_square}b,c, then the system has reflection symmetries in both the $x$ and $y$ directions with reflection operators $\hat{m}^{0}_x$ and $\hat{m}^{0}_y$ as above, which commute. However, there is another option where the bonds are as in Fig.~S~\ref{fig:app_square}d. In this case both reflection symmetries are only good up to a gauge transformation and the operators are $\hat{m}^{2\pi}_x=\hat{m}^{13}_x$ and $\hat{m}^{2\pi}_y=\hat{m}^{12}_y$ which both have a gauge transformation built in. However, since they both have a gauge transformation it turns out that the two operators commute. Hence, our conclusion is that it is the gauge transformation associated with odd numbers of $\pi$ flux which leads to the strange commutation operators in the spinless case. 

\subsection{General formulation}
Let us now consider the general case. Let us take the general Hamiltonian for a square with flux $\Phi.$
\begin{equation}
H_\Phi=\left(\begin{matrix} 0 & e^{i\varphi_{12}} & e^{-i\varphi_{13}} &0\\
e^{-i\varphi_{12}} & 0 & 0 &e^{i\varphi_{24}}\\
e^{i\varphi_{13}}& 0 & 0 &e^{-i\varphi_{34}}\\
0 & e^{-i\varphi_{24}} & e^{i\varphi_{34}} &0\\\end{matrix}\right).
\end{equation}\noindent where the total flux through a plaquette is $\Phi=\varphi_{12}+\varphi_{24}+\varphi_{34}+\varphi_{13}.$ Now let us consider the reflection operators $\hat{m}_{x}=\hat{m}_{x}^{0}\mathcal{G}_x$ and $\hat{m}_{y}=\hat{m}_{y}^{0}\mathcal{G}_y$ where $\hat{m}_{x,y}^{0}$ are as above, i.e. the reflection operators with vanishing flux and 
\begin{equation}
\mathcal{G}_{x,y}={\rm{diag}}\left[ e^{i\varphi_{1x,y}}, e^{i\varphi_{2x,y}}, e^{i\varphi_{3x,y}}, e^{i\varphi_{4x,y}}\right].
\end{equation}

By brute force evaluation we can check the conditions that $[H_{\phi}, \hat{m}_{x,y}]=0.$ The condition, in both cases, reduces to the constraint $1-e^{2i\Phi}=0$, which is solved by $\Phi=n\pi$ for some integer $n$. This makes physical sense since reflection $M_x$ or $M_y$ flips a magnetic field in the $z$ direction (i.e. threading the plaquette), however, flipping a magnetic flux of $0$, $\pi$ is equivalent to $0$, $-\pi$ through a gauge transformation.

We can now consider the commutator between the reflection operators. By brute force evaluation of the commutator one can show that if $1-e^{i\Phi}=0$ then the commutator vanishes. Otherwise, if $\Phi$ is an odd multiple of $\pi$ we find $[\hat{m}_{x},\hat{m}_{y}]=2\mathcal{I}$ where $\mathcal{I}$ is the inversion operator. Furthermore, one can show that $\mathcal{I}^2=e^{i\alpha}e^{i\Phi}\mathbb{I}$ where $\alpha$ is a global phase that depends on the gauge choice, and $e^{i\Phi}$ is $\pm 1$ for $\Phi$ an even/odd multiple of $\pi.$ We find $\alpha=3\varphi_{12}-\varphi_{13}+\varphi_{24}-\varphi_{34}.$

\section{Robustness of the Wannier-sector polarization to the internal structure of the unit cell}
\begin{figure}[h]
\centering
\includegraphics[width=.5\columnwidth]{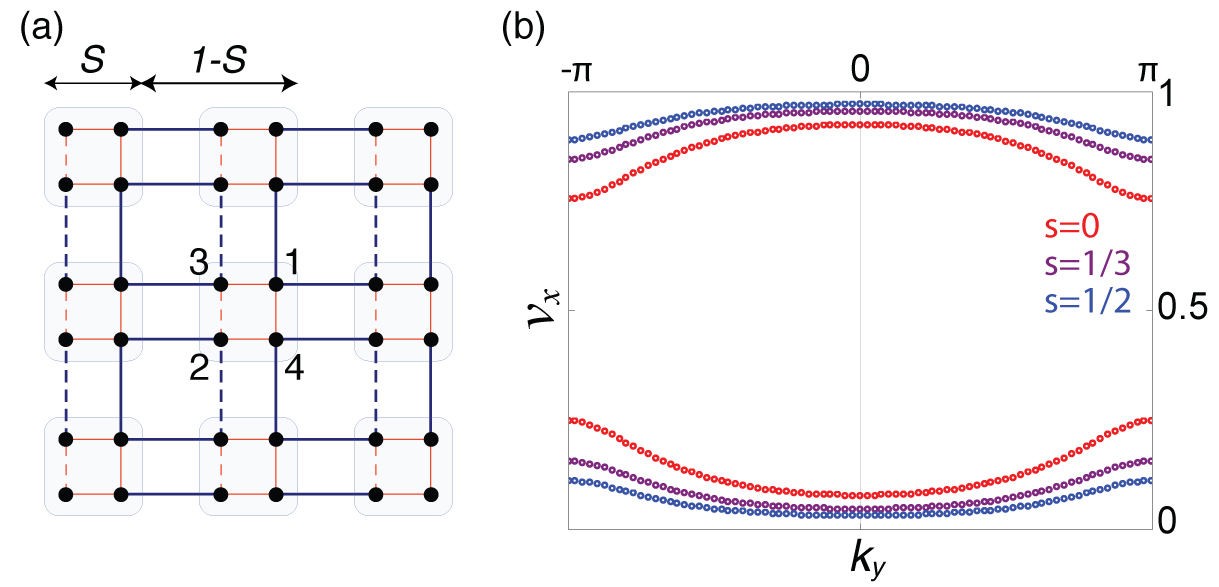}
\caption{\emph{Robustness of Wannier-sector topology to spatial separations of degrees of freedom in the unit cell}. (a) Definition of the parameter $s$ in our quadrupole model as the distance between adjacent sites within the unit cell. (b) Wannier bands for different values of $s$. In these simulations, $\lambda=1$, $\gamma=0.5$.}
\label{fig:quad_s}
\end{figure}

The quantization of the observables in our model is protected by symmetries and thus it is insensitive to the internal structure of the unit cell, as long as this structure preserves the symmetries. One can add in additional position data to the sites, e.g., by appropriately modifying the position operator, but this does not affect our conclusions. 
Although in Figs. 1 and 2 of the Main Text the tight-binding illustrations of the Hamiltonian are schematic, we can adopt the positions of the internal degrees of freedom as real to test how their distance affect the Wannier bands. We characterize this separation by the parameter $s$, shown in Fig.~S~\ref{fig:quad_s}a. 
In general, the Wannier values/functions will be sensitive to $s$ (where $s$ in a broader sense represents the spatial distribution of the degrees of freedom of the unit cell). However, the symmetry-protected topological properties of the Wannier bands are insensitive to $s$. Specifically, due to reflection symmetries, the Wannier centers will always come in pairs $(-\nu,\nu)$, (for each value of momentum) allowing for the subsequent calculation of the quadrupole invariant if the bands do not touch. In Fig.~S~\ref{fig:quad_s}b, we see that this is indeed the case.  The Wannier gap is maintained for $s=0,1/3,1/2$ and the Berry phase of the Wannier bands is unmodified.

\section{Model with corner charges and edge polarization but without quadrupole moment}
\begin{figure}[h]
\centering
\includegraphics[width=.8\columnwidth]{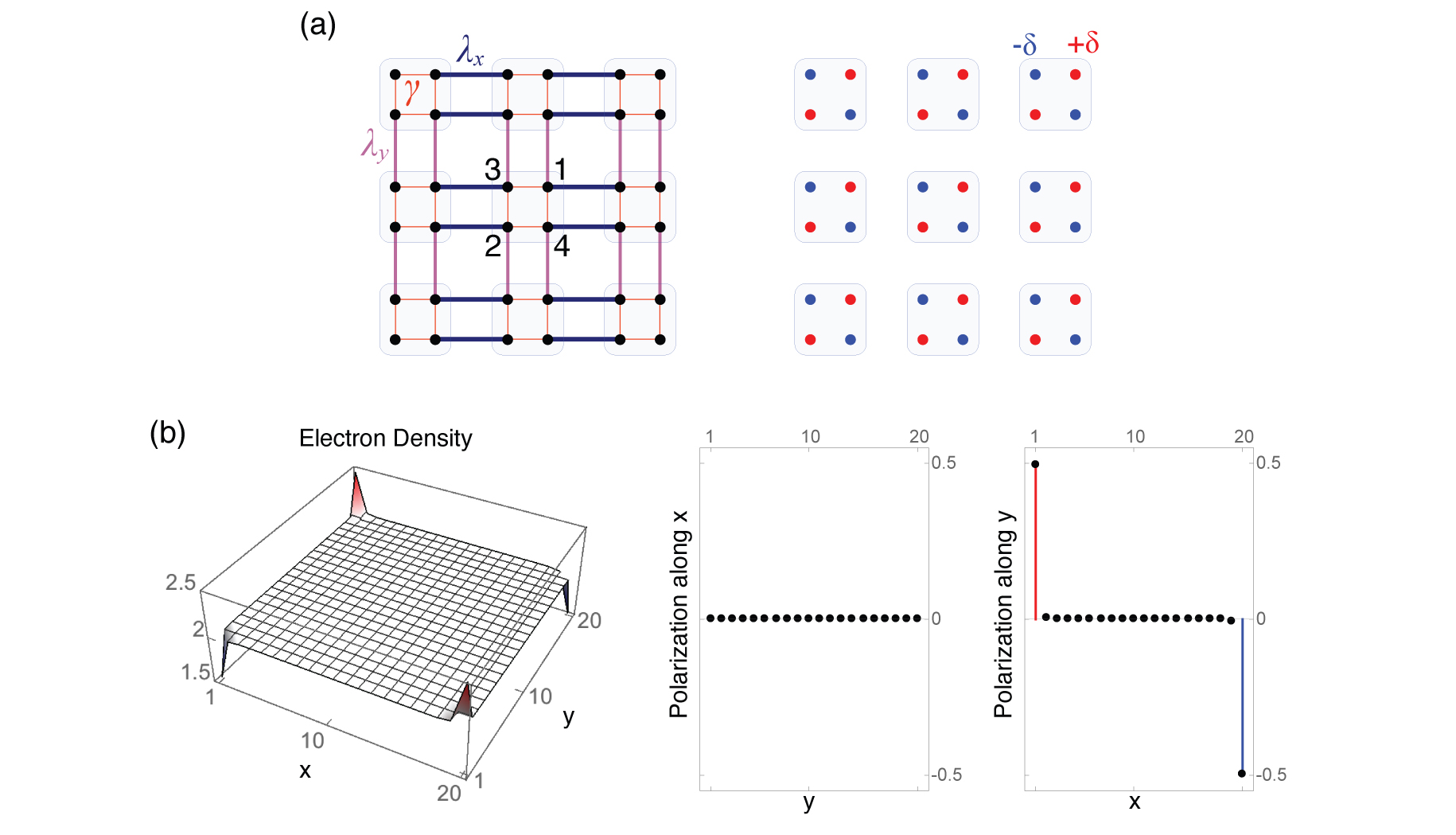}
\caption{Insulator (at half filling) in two dimensions without polarization or quadrupole moment, but with corner charges and edge polarizations. (a) Tight-binding model of insulator: hopping and coupling terms (left) and on-site energy terms (right). It differs from the quadrupole insulator in (1) the requirement for the hopping terms to obey $\lambda_x \neq \lambda_y$ and (2) the absence of $\pi$-flux per plaquette. (b) Electron density and polarizations along x and y for $|\delta| = 10^{-3}$. This system obeys $Q^{corner}=p^{edge}_x + p^{edge}_y$. The parameters for the simulations are $\lambda_x=1$, $\lambda_y=0.5$, $\gamma=0.1$ and the lattice had 20$\times$ 20 unit cells.}
\label{fig:not_quad}
\end{figure}

Corner charges and quadrupole moments are not equivalent. There is a crystalline phase that has half corner charges like a quadrupole, but is \emph{not} a quadrupole. This is shown in Fig.~S~\ref{fig:not_quad}.
This is an insulator at half-filling. If infinitesimal on-site potentials $+\delta$ are put on sites 1, 2 and $-\delta$ are put on sites 3, 4 (see Fig.~S~\ref{fig:not_quad}a right), for $0<\delta \ll \lambda_{x,y}$ (where $\lambda_{x,y}$ is the strength of the hopping terms represented by the horizontal and vertical lines in Fig.~S~\ref{fig:not_quad}a left), we have electron density and polarization as shown in Fig.~S~\ref{fig:not_quad}b. The corner charges are $\pm e/2$. The polarization is identically zero along $x$. Along $y$, on the other hand, the polarization is zero in the bulk but $\pm e/2$ at the edges. Notice that we have $Q^{corner}=p^{edge}_x+p^{edge}_y$. Hence, this model has corner charges but is not a quadrupole moment. We point out that this model has mirror symmetries that commute, and thus its Wannier bands are gapless, which does not allow for the existence of our quadrupole invariant. 

\section{Details on the tight binding simulation of a photonic crystal with quadrupole moment}
\begin{figure}
\centering
\includegraphics[width=.35\columnwidth]{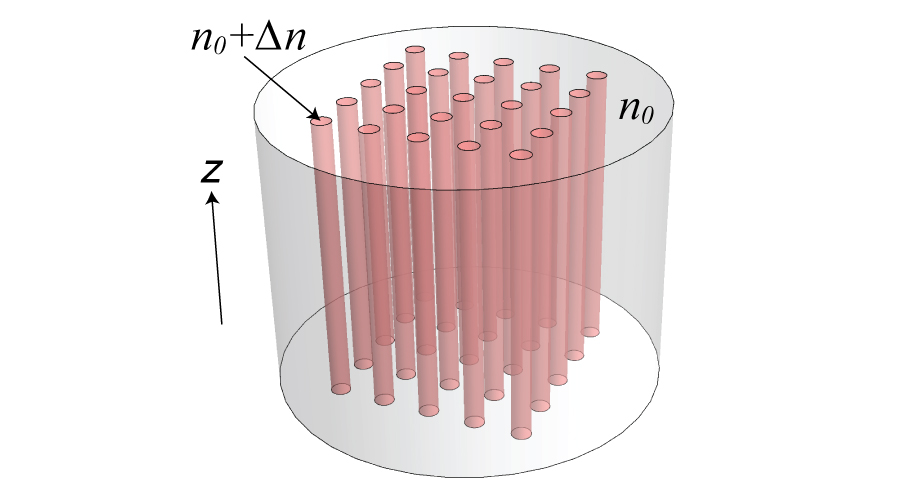}
   \caption{\emph{Schematic of a photonic crystal}. Each waveguide hosts a single mode. Fields propagate along $z$ according to Eq. \ref{eq:app_diffraction}. The waveguide array has crystalline periodicity in the plane transverse to $z$.}\label{fig:app_photonic_crystal}
\end{figure}

In this section we discuss the possibility of realizing the quadrupole moment in a photonic crystal. A photonic crystal consists of a medium of constant refractive index $n_0$ on which cylindrical regions with refractive index $n_0+\Delta n(\vec{r})$ are embeded to highly confine and \textit{guide} the optical waves (see Fig.~S~\ref{fig:app_photonic_crystal}). 
Let us align the axes of the cylinders with the $z$ axis. The waveguides considered here have cross-sectional area small enough to trap a single optical mode, and are arranged to have crystalline periodicity in the transverse plane ${\bf r}=(x,y)$. In the linear regime, the diffraction of light along $z$ through a waveguide array with the characteristics described above is governed by the paraxial Schrodinger equation \cite{Christodoulides1988,Lederer2008}:
\begin{align}
i \partial_z \psi({\bf r},z)= -\frac{1}{2 k_0} \nabla_{\bf r}^2 \psi({\bf r},z) - \frac{k_0 \Delta n({\bf r})}{n_0} \psi({\bf r},z),
\label{eq:app_diffraction}
\end{align}
where $\psi$ is the envelope function of the electric field $E$, as defined by $E({\bf r},z)=\psi({\bf r},z) e^{k_0 z - \omega t} \hat{x}$ (where $\omega=ck/n_0$ and $k$ is the wavenumber within the medium), $\Delta n({\bf r})$ is the refractive index relative to the ambient refractive index $n_0$ which defines the periodic lattice, and $\nabla_{\bf r}^2$ is the Laplacian in the transverse plane ${\bf r}=(x,y)$. Although each of the bound states is confined to its waveguide, the fact that the potential wells given by the contrast in refractive indices $\Delta n$ is finite implies that the bound modes leak into the background with an amplitude that decreases exponentially away from the waveguide. Thus, in a periodic arrangement of waveguides, modes overlap, i.e., they are coupled. If the waveguides sufficiently apart from one another, the coupling arises mostly from the overlap of the exponential tails of the modes. In this case, we can make the tight binding approximation \cite{Garanovich2012}
\begin{align}
i \partial_z \psi_i(z) = - \sum_j h_{ij} \psi_j(z),
\end{align}
where $\psi_i(z)$ is the amplitude of the $i^{\text{th}}$ waveguide and $h_{ij}$ is the coupling constant between waveguides $i$ and $j$. Since the structure is constant along the propagation direction $z$, we make the substitution $\psi_i(z) = \psi_i e^{i \beta z}$, to have
\begin{align}
\beta \psi_i = \sum_j h_{ij} \psi_j.
\end{align}
Notice that the couplings are akin to hopping terms in electronic systems. Additionally, $\beta$ acts as an energy for the Hamiltonian $h$, and time is substituted by the propagation coordinate $z$. Since the bound states at each waveguide are evanescently coupled to one another, the tight binding coupling between waveguides $i,j$ can be written as 
\begin{align}
h_{ij} = C \exp[-\kappa r_{ij}],
\label{eq:app_couplings}
\end{align}
where $r_{ij}$ is the separation between the waveguides and $C$ and $\kappa$ are experimental parameters. The above formulation of the problem and its tight binding simplification is well known in the photonics community, and has been extensively used \cite{Garanovich2012}. We now make use of it to explore the feasibility of realizing the quadrupole in photonic crystals. In our calculations below, we use an effective $\kappa$ and work in the tight binding limit assuming that $\kappa$ is constant across the photonic crystal. Furthermore, due to the exponential decay of the couplings, we only consider nearest neighbor couplings \cite{Garanovich2012}.

\begin{figure}[h!]
\includegraphics[width=.6\columnwidth]{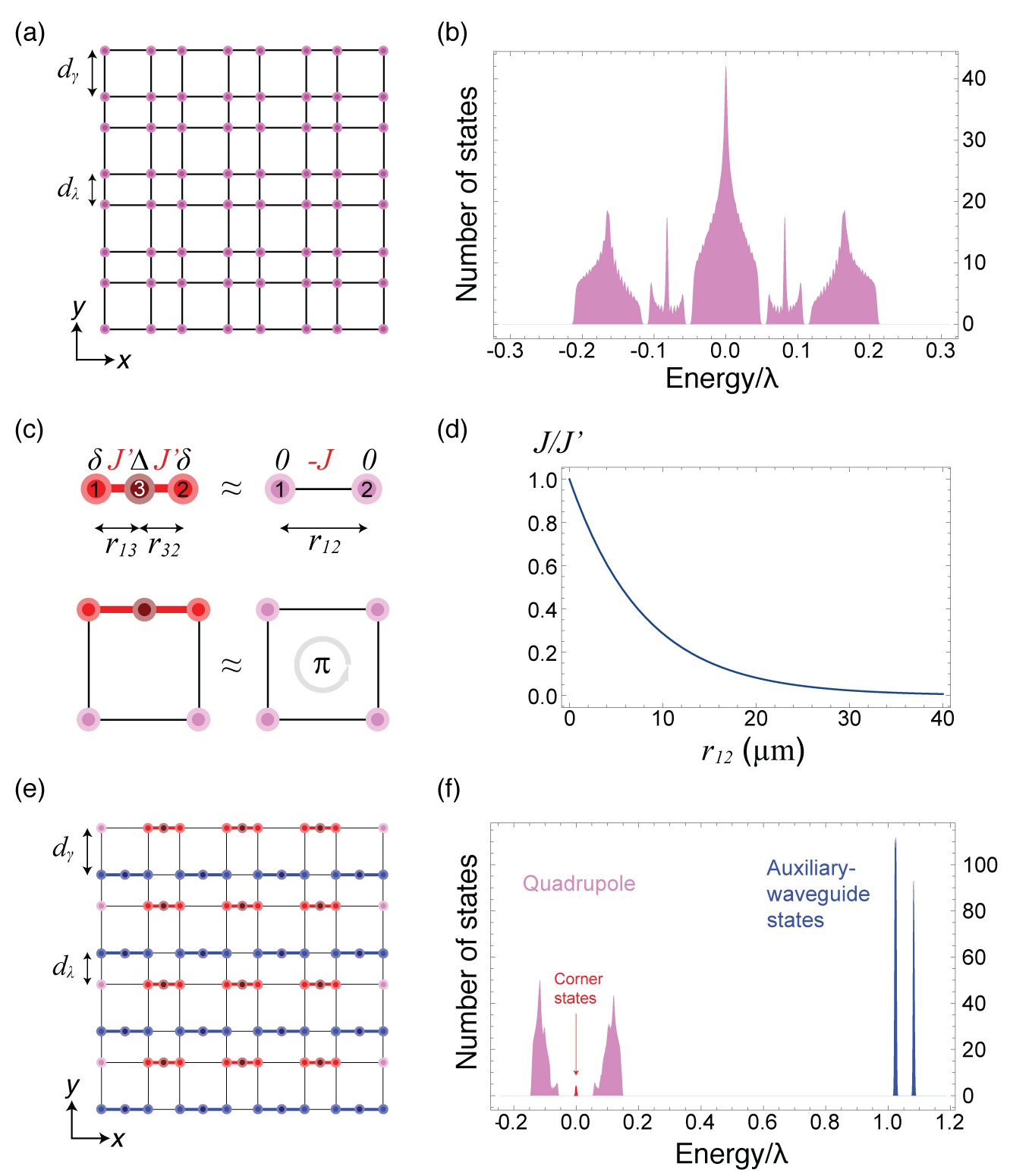}
\centering
   \caption{\emph{Photonic quadrupole}. (a) A failed photonic quadrupole with all positve couplings among waveguides. Lines represent nearest neighbor couplings used in tight binding simulation. (b) Density of states (DOS) for the failed quadrupole in (a) showing a gapless spectrum. (c) Effective negative coupler and effective plaquette with $\pi$ magnetic flux. (d) ratio of $J/J'$ which measures the fidelity in achieving a negative coupler with the configuration in (c). (e) Waveguide lattice arrangement with non-trivial quadrupole. Each plaquette has an effective negative coupling (shown by red and blue lines). (f) DOS for the lattice arangement in (e) with quadrupole region in pink and auxiliary-waveguide states at high-energy region in blue. States localized at corners are shown in red.}
   \label{fig:app_photonic_quadrupole}
\end{figure}

Noticing that our quadrupole has dimerized hoppings in both the horizontal and vertical directions, it would naively follow that the geometric arrangement for a non-trivial photonic quadrupole is as in Fig.~S~\ref{fig:app_photonic_quadrupole}a, where each circle represents a waveguide in the transverse plane spanned by ${\bf r}=(x,y)$. The two characteristic separations $d_\lambda$ and $d_\gamma$ are then responsible for the dimerization, with the corresponding couplings $\lambda$ and $\gamma$ (see Eq. 4 of main text) varying according to Eq~\ref{eq:app_couplings}. 
However, such system is gapless, as shown by the density of states (DOS) in Fig.~S~\ref{fig:app_photonic_quadrupole}b. Furthermore, that arrangement has reflection operators that obey $[M_x,M_y]=0$, but, as shown by the no go theorem in Section \ref{sec:app_no-go_theorem}, the reflection operators $M_x$ and $M_y$ have to obey $[M_x,M_y] \neq 0$ for a non-trivial quadrupole. Both opening an energy gap and having non-commuting reflection operators is achieved by inserting a $\pi$ flux per plaquette, as suggested by our minimal model (Eq. 4 of main text). In photonic lattices, inserting a $\pi$ flux amounts to one of the couplings per plaquette being negative. Interestingly, this was recently developed and demonstrated in Ref.~\onlinecite{Szameit2016}. The idea can be summarized as follows: Let us first define a \textit{regular coupler} as a pair of waveguides with the same on-site refractive index in which the waveguides are evanescently coupled. If we call this coupling $J$, the tight binding eigensystem is 
\begin{align}
\left(
\begin{array}{cc}
0 & J \\
J & 0\\
\end{array}
\right) \ket{u} = \beta \ket{u},
\end{align}
where the energies are measured relative to the on-site energies of each waveguide. This system has the energies and eigenstates: 
\begin{align}
\beta_\pm=\pm J,\quad
\ket{u_\pm}=\frac{1}{\sqrt{2}}(1, \pm 1)^T,
\end{align}
If in this eigensystem we substitute $J \to -J$ to create our hipothetical \emph{negative coupler} the solutions are
\begin{align}
\beta'_\pm=\pm J,\quad
\ket{u'_\pm}=\frac{1}{\sqrt{2}}(1, \mp 1)^T.
\end{align}
Thus, the difference between the regular and negative couplers is manifested in the fact that the eigenstates are flipped with respect to one another. Thus, a mechanism that achieves this flip of eigenstates will mimic a negative coupler. The proposed solution \cite{Szameit2016} consists on the arrangement in Fig.~S~\ref{fig:app_photonic_quadrupole}c. It uses an extra auxiliary waveguide equidistant from the two original ones. Since the separation between waveguides is now half of the original one, the coupling terms, which we denote $J'$ obey $J' > J$. In addition, on-site potentials $\delta$ and $\Delta$ are induced in the waveguides, as shown if Fig.~S~\ref{fig:app_photonic_quadrupole}c (experimentally, these on-site energies are modified by changing the local refractive index at each waveguide). The tight binding eigensystem of this coupler is given by 
\begin{align}
\left(
\begin{array}{ccc}
\delta & 0 & J'\\
0 & \delta & J'\\
J' & J' & \Delta
\end{array}
\right) \ket{v} = \beta \ket{v},
\end{align}
where we have neglected the original coupling $J$, assuming $J' \gg J$.
In that system, the antisymmetric eigenstate $\ket{v_+}=1/\sqrt{2}(1,-1,0)^T$ (which resembles $\ket{u'_+}$ given that the auxiliary waveguide is not occupied) always exists with energy $\beta_-=\delta$, regardless of the values of the other parameters. Thus, setting 
\begin{align}\delta = J
\end{align}
reproduces the energy associated to this state in the negative coupler. It remains to reproduce the state $\ket{u'_-}$ with energy $\beta = -J$. The energy and the symmetry required for $\ket{u'_-}$ are achieved by setting
\begin{align}
\Delta = (J'^2/J)-J.
\label{eq:app_photonic_quadrupole_Delta}
\end{align}
With these values for $\delta$ and $\Delta$ the energies and eigenstates for this three-waveguide coupler are
\begin{align}
\begin{array}{ll}
\beta_+ = +J & \ket{v_+} \propto (1,-1,0)^T\\
\beta_- = -J &\ket{v_-} \propto (1,1,-\frac{2J}{J'})^T\\
\beta_0 = J+J'^2/J & \ket{v_0} \propto (\frac{J}{J'},\frac{J}{J'},1)^T
\end{array}
\end{align}
which approaches the target negative coupler as $J/ J'  \rightarrow 0$. For $r_{13}=r_{32}=r_{12}/2$, where $r_{12}$ is the separation between waveguides in the original coupler (see Fig.~S~\ref{fig:app_photonic_quadrupole}c), we have the ratio $J/ J' = \exp[-\kappa r/2]$, which is shown in Fig.~S~\ref{fig:app_photonic_quadrupole}d for $\kappa = 0.125$ $\mu$m$^{-1}$. In our simulation we use separations of $r=20$ $\mu$m and $r=30$ $\mu$m, which result in ratios $J/ J' =0.08$ and $J/ J' =0.02$. As $J/ J'  \rightarrow 0$, the auxiliary mode $\ket{v_0}$ increasingly resides on the middle waveguide, while the other two modes $\ket{v_\pm}$ on waveguides 1 and 2. Furthermore, in this regime, $|\beta_0| \gg |\beta_\pm|$ indicates a strong detuning of the auxiliary state $\ket{v_0}$ from $\ket{v_\pm}$, thus increasingly isolating it. If we now use one of these negative couplers to build a five waveguide array as in the left of the bottom row of Fig.~S~\ref{fig:app_photonic_quadrupole}c, we effectively create a four-waveguide plaquette with a $\pi$ flux threading it. 

We make use of this development to form the photonic lattice shown in Fig.~S~\ref{fig:app_photonic_quadrupole}e, which is built based on the dimerized arrangement shown in Fig.~S~\ref{fig:app_photonic_quadrupole}a. Since the on-site energies need to be adjusted as function of the couplings to generate the effective negative coupling, and since the couplings vary with distance, two distinct types of negative couplers are necessary in our configuration, one associated with distance $d_\lambda$ and another one with distance $d_\gamma$. These two negative couplers are marked as red and blue in Fig.~S~\ref{fig:app_photonic_quadrupole}e. The configuration accommodates one negative coupler per plaquette, which amounts to adding one $\pi$ flux per plaquette. The density of states for a simulation of this lattice with $16 \times 16$ unit cells is shown in Fig.~S~\ref{fig:app_photonic_quadrupole}f. At low energies, we have the expected gapped DOS proper of an insulator, which confirms that effective negative couplings are achieved. Furthermore, we observe four mid-gap corner-localized states, the observable signature of the non-trivial quadrupole topology. Additional to this, there are 496 states, which localize around the auxiliary waveguides used to achieve negative couplings. However, these states are far detuned from the quadrupole region.

\section{Details on the calculation of the octupole moment}

In this section we discuss in detail the calculation of the octupole moment for our minimal model of Eq. 18  of the main text, which exists in three-dimensional insulators. Classically, the octupole moment manifests at the faces of a three-dimensional material by the existence of surface-bound quadrupole moments (see Section \ref{sec:app_classical_multipoles}). Thus, we expect that in a crystal with non-trivial octupole moment its Wilson loops -and consequently their corresponding Hamiltonians at the two-dimensional boundaries of a material (see Section \ref{sec:app_edge_Hamiltonian})- have a non-trivial quadrupole topology. In this section, we will carry on the calculation that leads to that conclusion.

\begin{figure}[b]
\centering
\includegraphics[width=.4\columnwidth]{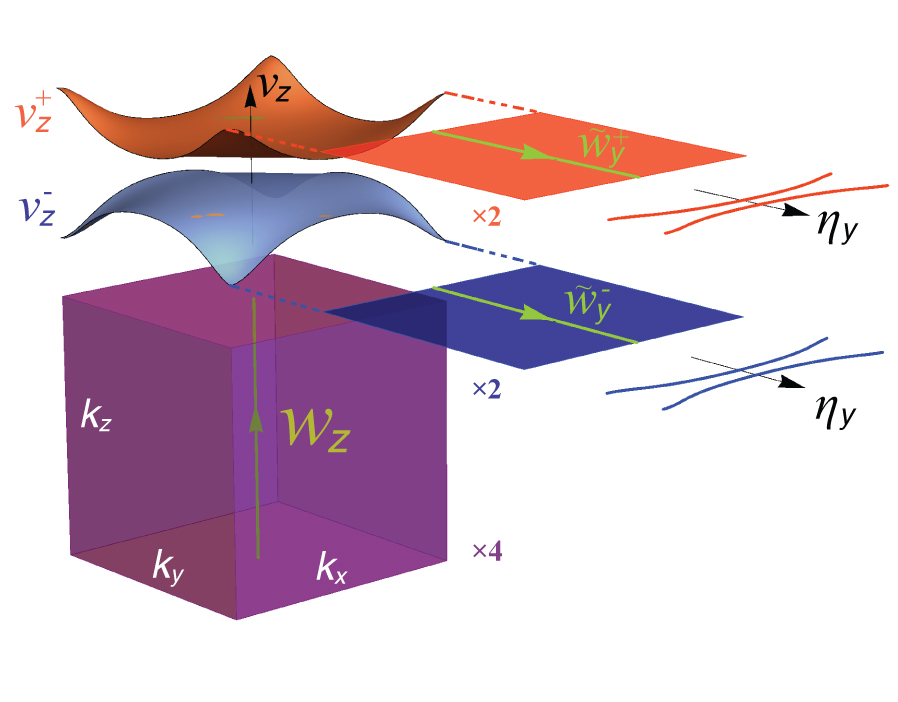}
   \caption{\emph{Schematic of the procedure to determine the topology of an octupole moment}. A Wilson loop along $z$ over the three-dimensional BZ (purple cube) divides it in two sectors, according to its Wannier value $\nu^\pm_z$ (red and light blue plots over the cube). Each sector has 2 bands (represented by the red and blue squares) and has quadrupole topology. This can be verified by calculating Wilson loops along $y$ over each sector, which renders two Wannier sectors $\eta^\pm_y$ (red or blue pair of symmetric lines), each of them having a Berry phase of $\pi$ $(0)$ in its Wilson loop along $x$ in the $\lambda>\gamma$ ($\lambda<\gamma$) regime. }
   \label{fig:app_octupole_invariant}
\end{figure}

The Hamiltonians at the two-dimensional boundaries of the three-dimensional material can be adiabatically connected to its Wilson loops. Without loss of generality, for example, we can make the identification
\begin{align}
\W_{z,\bf k}= e^{-i H_{surface}({\bf k})}
\end{align}
where $H_{surface}({\bf k})$ has the same topology of the Hamiltonian at the surface of the material in the $xy$ plane [we can similarly assign Wilson loops along $x$ ($y$) to Hamiltonians on the surface $yz$ ($zx$)]. To characterize the boundary topology we diagonalize this Wilson loop,
\begin{align}
\W_{z,\bf k} \ket{\nu^j_{z,\bf k}}= e^{i 2\pi \nu^j_z({\bf k_\perp})} \ket{\nu^j_{z,\bf k}}.
\label{eq:app_octupole_1}
\end{align}
Here ${\bf k_\perp}=(k_x,k_y)$. Since the minimal model has 8 bands, and the octupole occurs at half-filling, the Wilson loop along $z$ is a $4 \times 4$ matrix, which has eigenstates $\ket{\nu^j_{z,\bf k}}$ for $j=1,2,3,4$. There is, however,  only 2 two-fold degenerate Wannier bands, which we denote as $\nu^\pm_z({\bf k_\perp})$. These are shown in red and light blue in Fig.~S~\ref{fig:app_octupole_invariant}. Crucially, these Wannier bands are gapped, which is the first signature of the surface Hamiltonian having a quadrupole topology, since the minimal quadrupole occurs at four-band gapped insulators at half filling. We then split the Wilson loop into two sectors across the entire BZ. Let us then re-write Eq.~\ref{eq:app_octupole_1} as
\begin{align}
\W_{z,\bf k} \ket{\nu^{\pm,j}_{z,\bf k}}= e^{i 2\pi \nu^\pm_z({\bf k_\perp})} \ket{\nu^{\pm,j}_{z,\bf k}}.
\end{align}
for $j=1,2$. The subspace of occupied states of $H_{surface}$ associated with each Wannier sector $\nu^\pm_z$ is two-dimensional, as required for a quadrupole. We choose the sector $\nu^+_z$ and characterize its topology. If the theory of multipole moments in crystalline insulators is to follow the hierarchy of the theory in the continuum classical theory this topology should be that of a quadrupole. To probe the topology of sector $\nu^+_z$, let us construct the Wannier states
\begin{align}
\ket{w^{+,j}_{z,\bf k}} = \sum_{n=1}^{N_{occ}}\ket{u^n_{\bf k}} [\nu^{+,j}_{z,\bf k}]^n
\end{align}
for $j=1,2$. We use these Wannier states to calculate the nested Wilson loop along $y$
\begin{align}
[\tilde\W^+_{y,\bf k}]^{j,j'} =& \braket{w^{+,j}_{z,{\bf k}+N_y{\bf \Delta_{k_y}}}}{w^{+,r}_{z,{\bf k}+(N_y-1) {\bf \Delta_{k_y}}}} \bra{w^{+,r}_{z,{\bf k}+(N_y-1) {\bf \Delta_{k_y}}}} \ldots \nonumber\\
&\dots \ket{w^{+,s}_{z,{\bf k}+{\bf \Delta_{k_y}}}}\braket{w^{+,s}_{z,{\bf k}+{\bf \Delta_{k_y}}}}{w^{+,j'}_{z,\bf k}}.
\label{eq:app_Wilson_loop_Wannier_basis}
\end{align}
where ${\bf \Delta_{k_y}} = (0, 2\pi/N_y,0)$. Notice that, since $j,r,\ldots,s,j' = 1,2$, this nested Wilson loop also is non-Abelian. [This Wilson loop was defined in in Eq.~\ref{eq:app_Wilson_loop_Wannier_basis} for two-dimensional crystals, but we reproduce it here in its obvious extension to three dimensions]. The nested Wilson loop has an associated Hamiltonian $H_{hinge}({\bf k})$ given by
\begin{align}
\tilde{\W}^+_{y,\bf k} = e^{-i H_{hinge}({\bf k})}
\end{align}
which has the same topological properties as the Hamiltonian at the one-dimensional boundaries of the two-dimensional surface $xy$ of the material (i.e., we are now looking into the boundary of the boundary). To characterize the topology at the one-dimensional boundary, we diagonalize the Wilson loop
\begin{align}
\tilde{\W}^+_{y,\bf k} \ket{\eta^{\pm}_{y,\bf k}}= e^{i 2\pi \eta^{\pm}_y(k_x)} \ket{\eta^{\pm}_{y,\bf k}}.
\end{align}
The Wannier bands $\eta^\pm_y(k_x)$ are gapped, which implies that the hinge Hamiltonian is gapped. This topology should be that of a one-dimensional TI. To see that, we once again define the two Wannier sectors 
\begin{align}
\ket{w^{\pm}_{y,\bf k}} = \sum_{n=1}^{N_{occ}}\ket{u^n_{\bf k}} [\eta^{\pm}_{y,\bf k}]^n
\end{align}
and use the $\eta^+_y$ sector to calculate the nested Wilson loop along $x$
\begin{align}
\tilde\W^+_{x,\bf k} =& \braket{w^{+}_{y,{\bf k}+N_x{\bf \Delta_{k_x}}}}{w^{+}_{y,{\bf k}+(N_x-1) {\bf \Delta_{k_x}}}} \bra{w^{+}_{y,{\bf k}+(N_x-1) {\bf \Delta_{k_x}}}} \ldots \nonumber\\
&\dots \ket{w^{+}_{y,{\bf k}+{\bf \Delta_{k_x}}}}\braket{w^{+}_{y,{\bf k}+{\bf \Delta_{k_x}}}}{w^{+}_{y,\bf k}}
\label{eq:app_Wilson_loop_Wannier_basis}
\end{align}
which results in the Wannier-sector polarization
\begin{align}
p^+_x=-\frac{i}{2\pi}\frac{1}{N_y N_z} \sum_{k_y,k_z} \mbox{Log} \left[\tilde\W^+_{x,\bf k}\right] = \left\{\begin{array}{ll}
0 & \gamma > \lambda\\
1/2 & \gamma < \lambda
\end{array} \right.
\end{align}
i.e., at hinges we have a one-dimensional TI. From this, it follows that the topology of $H_{surface}$ is that of a quadrupole, and the topology of the entire Hamiltonian is that of an octupole.

In this calculation the order of the nested Wilson loops $\W_z \to \W_y \to \W_x$ was arbitrary. Same results are obtained for any order of Wilson loop nesting. Our model has reflection symmetries $M_{x,y,z}$ (up to a gauge transformation, see Section \ref{sec:app_symmetries_up_to_gauge}), all of which send a classical octupole $o_{xyz}$ to $-o_{xyz}$. In the classical theory, this admits only the solution $o_{xyz}=0$, but the ambiguity in the position of the electronics due to the introduction of the lattice also allows the solution $o_{xyz}=e/2$ mod $e$. Since the minimal quadrupole requires 4-bands at half filling, and an octupole has quadrupole topology in each of its Wannier sectors, we conclude that the minimal octupole requires 8 bands at half filling. Thus, our model is the minimal model with octupole moment.

\bibliography{quad_references}
\bibliographystyle{Science}